\documentclass[useAMS,usenatbib,usegraphicx,onecolumn]{mn2e}
\usepackage{multirow}

\title[Finite Source Effects in Strong Lensing]
{Finite Source Effects in Strong Lensing: Implications for the
Substructure Mass Scale}
\author[Gregory Dobler \& Charles R. Keeton]
{Gregory Dobler$^1$ \& Charles R. Keeton$^2$ \\
$^1$Department of Physics and Astronomy, University of Pennsylvania,
209 S.\ 33rd Street, Philadelphia, PA 19104 USA \\
$^2$Department of Physics and Astronomy, Rutgers University,
136 Frelinghuysen Road, Piscataway, NJ 08854 USA
}

\newcommand\mm{\mathcal{M}}
\newcommand\Msun{\mbox{ M}_{\odot}}
\newcommand\vect[1]{{\textbf{\em #1}}}
\newcommand\refsec[1]{\S \ref{sec:#1}}

\begin{document}

\maketitle

\begin{abstract}
Flux ratio `anomalies' in quadruply-imaged gravitational lenses
can be explained with galactic substructure of the sort predicted
by $\Lambda$CDM, but the strength and uniqueness of that hypothesis
needs to be further assessed.  A good way to do that is to use
the physical scale associated with the size of the source quasar,
and its dependence on wavelength.  We develop a toy model to study
finite source effects in substructure lensing.  Treating substructure
as a Singular Isothermal Sphere allows us to compute the images of
a finite source analytically, and then to explore how the image
configurations and magnifications depend on source position and
size.  Although simplified, our model yields instructive general
principles: image positions and magnifications are basically
independent of source size until the source is large enough to
intersect a substructure caustic; even sources that are much
larger than the substructure Einstein radius can be perturbed at
a detectable level; and most importantly, there is a tremendous
amount to be learned from comparing image positions and
magnifications at wavelengths that correspond to different source
sizes.

In a separate analysis, we carefully study four observed radio
lenses to determine which of the images are anomalous.  In
B0712+472, the evidence for a radio flux ratio anomaly is marginal,
but if the anomaly is real then image C is probably the culprit.
In B1422+231, the anomaly is in image A.  Interestingly, B2045+265
and B1555+375 both appear to have \emph{two} anomalous images.
Coincidentally, in each system one of the anomalies is in image C,
and the other is in either image A or image B (both possibilities
lead to acceptable models).  It remains to be seen whether
$\Lambda$CDM predicts enough substructure to explain multiple
anomalies in multiple lenses.  When we finally join our modeling
results and substructure theory, we obtain lower bounds on the
masses of the substructures responsible for the observed anomalies.
The mass bounds are broadly consistent with expectations for
$\Lambda$CDM.  Perhaps more importantly, we outline various
systematic effects in the mass bounds; poor knowledge of whether
the substructure lies within the main lens galaxy or elsewhere
along the line of sight appears to be the dominant systematic.
\end{abstract}


\section{Introduction}

While the $\Lambda$CDM cosmological scenario has been quite successful
in describing measurements of cosmological structures on large scales
\citep[e.g.,][]{te,do,WMAP}, it seems to overpredict the number of
galactic satellites by about an order of magnitude \citep{klypin,moore}.
The discrepancy could be resolved by modifying dark matter -- making
it warm, self-interacting, or otherwise exotic to reduce the predicted
amount of small-scale structure \citep[e.g.,][]{WDM,SIDM}.  Another
possibility is that star formation in low-mass haloes is suppressed by
photoionization \citep[e.g.,][]{BKW,somerville,benson}, which would
mean that many small haloes are present by dark.  The latter hypothesis
is readily tested with gravitational lensing, which is sensitive to the
distribution of both luminous and dark matter over a range of scales in
galaxy haloes.

Strong gravitational lensing has long been known to probe galaxy mass
distributions on kiloparsec scales \citep[e.g.,][]{refsdal,young,csk91},
and even to be sensitive to the fine graininess of stellar mass
distributions \citep[microlensing; e.g.,][]{chang,wamb}.  More
recently, \citet{MS} pointed out that lensing is also sensitive to
structure on intermediate scales ($\ell \sim \mbox{few pc}$,
$M \sim 10^{6}\Msun$).  This effect, sometimes termed
`millilensing,' could solve a long-standing problem in lens
modeling.  Detailed mass models of quadruply-imaged lenses are quite
successful at matching the relative positions of the images, but
often fail to reproduce the relative fluxes.  \citeauthor{MS} pointed
out that intermediate-scale substructure could nicely explain the
`anomalous' flux ratios in radio lenses, taking the troublesome
lens B1422+231 as their example.  Later, \citet{me} connected this
idea to the predictions of $\Lambda$CDM and suggested that the
statistical distribution of lens flux ratios could be used to test
the hypothesis that galaxies contain significant substructure.
\citet{DK} carried out the statistical test for seven quadruply-imaged
quasars to infer that the fraction of galactic mass in substructure
is $2.0_{-1.4}^{+5.0}$ per cent (90 per cent confidence), which seemed
to match the amount of substructure expected for $\Lambda$CDM, and to
rule out modified dark matter models.

Connecting observed flux ratio anomalies to inferences about dark
matter requires a fairly long chain of logic, whose strength is
still being assessed.  The very first link is the identification of
flux ratio anomalies.  Careful analysis of the lens mapping reveals
model-independent relations between certain images in 4-image lenses
with `cusp' or `fold' configurations, relations which can only
be violated if the lens galaxy contains significant small-scale
structure \citep{cuspreln,foldreln}.  As valuable as that analysis
is, it only reveals which lens \emph{systems} contain anomalies;
it does not pinpoint which individual \emph{images} are anomalous.
This issue is crucial because lens theory predicts fundamental
differences in how positive- and negative-parity images are affected
by small-scale structure \citep{SW,analytics}, which offers a key
test of the substructure hypothesis that may rule out competing
explanations of flux ratio anomalies \citep[see][]{KD}.  Furthermore,
the large number of substructures implied by $\Lambda$CDM 
\citep[e.g.,][]{coorayandsheth, shethandjain} suggests the possibility
that more than one image could be perturbed, but previous analyses
have not determined whether that actually occurs.  One goal of our
analysis is to revisit models of flux ratio anomaly lenses to see
if we can figure out which of the images are affected by substructure.

Another link in the chain of logic involves determining the length
or mass scale associated with flux ratio anomalies.  According to
the substructure hypothesis, the anomalies are caused by mass clumps
in the range $M \sim 10^{6}$--$10^{8}\Msun$, corresponding to
a length scale of a few to tens of parsecs.  However, violations
of the cusp and fold relations really only indicate structure on
the scale of the separation between images \citep[see][]{cuspreln},
which is typically no smaller than a few tenths of an arcsecond, or
hundreds of parsecs.  The difference in scale means that substructure
cannot yet be established as the only viable explanation for flux
ratio anomalies (e.g., \citealt{EW}; but see \citealt{KD}).  Moreover,
even within the substructure hypothesis, comparisons between the
predicted and inferred amount of substructure are very sensitive to
scale.

The size of the source quasar brings an additional scale into the
problem.  Heuristically, a source `feels' lensing structure only
on scales larger than itself.  Combining conventional wisdom about
structure in lens galaxies with the standard model of quasars
\citep[e.g.,][]{peterson,krolik}, it is believed that quasar
optical continuum light is very sensitive to both microlensing and
millilensing; that the optical broad emission lines are certainly
sensitive to millilensing and may or may not be affected by
microlensing \citep[see][]{abajas,LI,richards}; that the radio and
mid-IR light can only be affected by millilensing; and that the
optical narrow emission lines should not be affected by any
small-scale structure.  Measuring the flux ratios associated with
several different source sizes could therefore provide a way to
determine the substructure scale \citep{MM}.  \citet{DK} used these
ideas in a general way, selecting radio lenses in order to focus on
millilensing (and ignore microlensing).  \citet{0435b} compared the
optical continuum and broad line flux ratios for HE~0435$-$1223 to
infer that there must be microlensing in that system, and maybe some
millilensing as well.  \citet{met} compared the optical narrow line
flux ratios with the radio and mid-IR flux ratios for Q2237$+$0305
to find evidence for millilensing and place limits on the substructure
mass scale.

Despite this evidence for the value of working with different source
sizes, there has been no general study of source size effects in
millilensing.  The second main goal of our paper is to present a
semi-analytic toy model that allows us to examine a wide range of
finite source effects.  Assuming that any given flux ratio anomaly
is caused by a single, isolated clump that can be modeled as an
isothermal sphere is admittedly a toy model -- but in the best sense
of the term: a tool that not only reveals, but also elucidates, some
interesting general principles.  As we completed our work, we learned
that \citet{inoue} recently considered the same toy model and derived
analytic approximations for the millilensing magnification in the
limit of a large source.  Our work complements theirs by presenting
exact results for a large range of source sizes, by considering some
of the effects in more details, and also by applying the general
theory to four specific observed lenses.

Thus our paper has two main goals: to better understand the flux
ratio anomalies in four observed radio lenses; and to study finite
source effects in millilensing.  The two parts are independent of
each other, although we do combine them in the end to place
constraints on the substructures required to produce the observed
flux ratio anomalies.  Pedagogically, it makes sense to begin with
the study of finite source effects.  In \S 2 we develop our toy
model for millilensing and use it to study the image configurations
and magnifications for different source sizes and positions; the
discussion goes into some depth, so we offer a review of the main
points in \refsec{comments}.  In \S 3 we introduce a method for
using our millilensing theory can to place lower bounds on the
masses of substructures responsible for flux ratio anomalies.  In
\S 4 we turn to the analysis of real lens systems; we first use
lens models to determine which images are anomalous, and then apply
our millilensing theory to derive the substructure mass bounds.
We summarize our results and conclusions in \S 5.  Throughout the
paper we assume a cosmology with $\Omega_M = 0.3$,
$\Omega_\Lambda = 0.7$, and $h = 0.7$.

\section{SIS Images of a Finite Source}

This section presents our toy model for studying the effects of a
finite source size.  We define the model and then consider the image
configurations and magnifications as a function of the position and
size of the source.  Although the focus is millilensing, our results
have some broader implications that are discussed in \refsec{comments}.

\subsection{Macromodel}

In lens modeling, it is common to begin with a smooth `macromodel'
that reproduces the number and positions of the lensed images.
Smooth models generally fail to fit the observed flux ratios, so
small clumps are introduced that modify the flux ratios enough to
fit the data.  (The clumps may modify the image positions as well,
but not by much more than current measurement errors; see
\refsec{pos}.)  To understand the effects of a clump near one of
the lensed images, we zoom in and consider the lens mapping only in
the vicinity of the clump.  The clump is small compared with the
galaxy ($R_{\rm clump}/R_{\rm gal} \sim 10^{-3}$), so on this scale
the macromodel can be approximated as a constant convergence and
shear ($\kappa$ and $\gamma$, respectively).  The image magnification
predicted by the macromodel is
\begin{equation}
  \mu_0 = \left| (1-\Gamma)^{-1} \right|
  = \frac{1}{(1-\kappa)^2-\gamma^2}\ ,
\label{eq:macromagnification}
\end{equation}
where $1-\Gamma$ is the local lens mapping (in coordinates aligned
with the local shear for simplicity),
\begin{equation}
  1 - \Gamma = \left[\begin{array}{cc}
    1-\kappa-\gamma & 0 \\
    0 & 1-\kappa+\gamma \end{array}\right].
\end{equation}
We can distinguish between three types of images based on the
eigenvalues of this matrix.  Positive parity images have
$1-\kappa+\gamma > 1-\kappa-\gamma > 0$.  Negative parity images
have $1-\kappa+\gamma > 0 > 1-\kappa-\gamma$, so they are parity
reversed in one direction.  Double negative parity images have
$1-\kappa-\gamma < 1-\kappa+\gamma < 0$, so they are parity reversed
in both directions; however, images of this type are faint, rarely
observed, and of relatively little importance for substructure lensing
\citep[e.g.,][]{wi}.  For the special case of an isothermal ellipsoid
macromodel, $\kappa=\gamma$ everywhere.

Many studies of millilensing have assumed that the clump lies in the
halo of the main lens galaxy, but \citet{analytics} and \citet{metcalfLOS}
have pointed out that a clump elsewhere along the line of sight could
still have a significant effect.  While one can invoke statistical
arguments about whether a clump is more likely to lie in the galaxy
or along the line of sight \citep[e.g.,][]{chen}, strictly speaking
the clump redshift is unknown and that may lead to a systematic
uncertainty in a millilensing analysis.  Fortunately, this effect is
easily accommodated in our formalism.  \citet{analytics} showed that
if the clump does lie at a different redshift than the lens galaxy,
the macromodel can still be treated as a simple convergence and shear,
but with effective values
\begin{eqnarray}
  \kappa_{\rm eff} &=& \frac{(1-\beta)[\kappa - \beta (\kappa^2 - \gamma^2)]}
    {(1-\beta\kappa)^2 - (\beta\gamma)^2}\ , \nonumber\\
  \gamma_{\rm eff} &=& \frac{(1-\beta)\gamma}
    {(1-\beta\kappa)^2 - (\beta\gamma)^2}\ , \label{eq:LOS}
\end{eqnarray}
where $\beta = (D_{cl} D_{os})/(D_{ol} D_{cs})$ for a foreground
clump ($z_c < z_l$), while $\beta = (D_{lc} D_{os})/(D_{oc} D_{ls})$
for a background clump ($z_l < z_c < z_s$).  In what follows we
simply use $\kappa$ and $\gamma$ to denote the macromodel, bearing
in mind that we should use the effective values if we want to consider
a line-of-sight clump.

Another possible systematic effect arises from the `mass sheet
degeneracy' in the macromodel.  Adding a uniform mass sheet (and
rescaling the galaxy mass appropriately) leaves the image positions
and flux ratios unchanged \citep{GFS,saha}.  Turning the problem
around, lens models cannot detect the presence of a mass sheet,
which can bias the conclusions drawn from the models \citep{KZ}.
We should therefore consider how a mass sheet might affect a
millilensing analysis.  For our purposes, adding a mass sheet of
density $\kappa_{\rm sheet}$ is equivalent to a simple rescaling
of the macromodel:
\begin{eqnarray}
  \kappa' &=& (1-\kappa_{\rm sheet}) \kappa + \kappa_{\rm sheet}\,, \nonumber\\
  \gamma' &=& (1-\kappa_{\rm sheet}) \gamma\,. \label{eq:sheet}
\end{eqnarray}
This rescaling is the same no matter whether the clump lies in the
lens galaxy or along the line of sight.

\subsection{Micromodel}

We model the mass clump as a singular isothermal sphere (SIS).  
One of the advantages of the SIS is that its $\rho \propto r^{-2}$
density profile yields a simple form for the deflection angle,
\begin{equation}
  \alpha(\vect{x}) = b \frac{\vect{x}}{|\vect{x}|}\ ,
\end{equation}
where the Einstein radius $b$ is
\begin{equation}
  b = 4 \pi \left(\frac{\sigma}{c}\right)^2 \frac{D_{ls}}{D_{os}}\ .
\end{equation}
Here, $\sigma$ is the velocity dispersion of the SIS, while $D_{os}$
and $D_{ls}$ are angular diameter distances from the observer or lens
to the source.  Although N-body simulations predict a different form
for the density profile of dark matter structures \citep[e.g.,][]{NFW},
the SIS has been used for modeling substructures in previous studies
\citep{me,DK}, and its simplicity makes it an attractive choice for
a toy model whose purpose is to yield general insights.  The mass of
the SIS increases linearly with radius, and the projected 2-D mass
within the Einstein radius is
\begin{equation}
  M(b) = \frac{c^2}{4G}\ \frac{D_{os}}{D_{ol}D_{ls}}\ b^2
  = \pi \Sigma_{\rm cr} b^2\,,
\end{equation}
where $\Sigma_{\rm cr} = c^2 D_{os}/(4\pi GD_{ol} D_{ls})$ is the
critical surface density for lensing.

The clump + macromodel system is governed by the lens equation
\begin{eqnarray}
  u &=& (1-\kappa-\gamma) r\cos\theta - b\cos\theta\,, \nonumber\\
  v &=& (1-\kappa+\gamma) r\sin\theta - b\sin\theta\,, \label{eq:lenseq}
\end{eqnarray}
where $\vect{u} = (u,v)$ are coordinates in the source plane and
$\vect{x} = (r\cos\theta,r\sin\theta)$ are coordinates in the image
plane (centred on the clump).  In substructure lensing, solutions
of this lens equation represent `micro-images' that are not
separately resolved but combine to form the observed macro-image.
For a point source, the individual micro-image magnifications are
given by
\begin{equation}
  \mu^{-1} = \left|\frac{\partial\vect{u}}{\partial\vect{x}}\right|
  = \mu_0^{-1} - \frac{b}{r} (1-\kappa-\gamma\cos2\theta)\,.
  \label{eq:mageq}
\end{equation}
The tangential critical curve for the lens system can be found by taking
$\mu \rightarrow \infty$, which yields
\begin{equation}
  r_{\rm crit}(\theta) = \frac{b~(1-\kappa-\gamma\cos2\theta)}
    {(1-\kappa)^2 - \gamma^2}\ .
\end{equation}
Plugging this into the lens equation then gives a parametric equation
for the tangential caustic.  The radial pseudo-caustic is the curve
in the source plane that maps to the origin in the image plane; from
eq.~(\ref{eq:lenseq}), it can be written parametrically as
\begin{eqnarray}
  u_p &=& -b\cos\theta\,, \\
  v_p &=& -b\sin\theta\,.
\end{eqnarray}

It can be seen from equations (\ref{eq:lenseq}) and (\ref{eq:mageq})
that the positions and magnifications of the images of a point
source depend on the perturber strength $b$, the position of the
source relative to the perturber, and $\kappa$ and $\gamma$ from
the macromodel.  There is no general analytic solution to the lens
equation even for a simple SIS perturber.  Nevertheless, it is
possible to find an analytic solution for a source of \emph{finite}
size at an arbitrary position.  \citet{finch} showed how to compute
the area enclosed by the caustics of an SIS lens in an external shear
field, and in the following sections we extend their method to find
the positions, shapes, and magnifications of the images of a finite
source lensed by an SIS in a convergence and shear field.

\subsection{Analytic solution for the images of a finite source}

First, we consider a circular source and parametrize its boundary:
\begin{eqnarray}
  u &=& u_0 + a  \cos(\lambda)\,, \\
  v &=& v_0 + a  \sin(\lambda)\,,
\end{eqnarray}
where $u_0$ and $v_0$ are the coordinates of the centre of the source,
$a$ is the source size, and $\lambda$ varies from $0$ to $2\pi$.  
(\citeauthor{finch} considered the special case $u_0 =v_0 = 0$ and
$a = b$.)  Plugging the source boundary into the lens equation
(\ref{eq:lenseq}) yields
\begin{eqnarray}
  a \cos(\lambda) &=& (1-\kappa-\gamma) r\cos\theta - b\cos\theta - u_0\,, \\
  a \sin(\lambda) &=& (1-\kappa+\gamma) r\sin\theta - b\sin\theta - v_0\,.
\end{eqnarray}
We can eliminate $\lambda$ by squaring and adding these two equations to
obtain
\begin{eqnarray}
  0 &=& r^2\,\Gamma_{-}^2\,\cos^2\theta + r^2\,\Gamma_{+}^2\,\sin^2\theta
    - 2\,u_0\,\Gamma_{-}\,r\cos\theta - 2 v_0\,\Gamma_{+}\,r\sin\theta
    - 2\,\Gamma_{-}\,r\,b\cos^2\theta - 2\,\Gamma_{+}\,r\,b\sin^2\theta
    \nonumber\\
  && + u_0^2 + v_0^2 + b^2 - a^2 + 2\,u_0\,b\cos\theta + 2\,v_0\,b\sin\theta\,,
\end{eqnarray}
where $\Gamma_{\pm} = (1-\kappa\pm\gamma)$.  This is a quadratic equation
for $r(\theta)$ whose solution yields the boundary of the image(s):
\begin{equation}
  r_{\pm}(\theta) = \frac{A \pm \sqrt{B}}{C}\ , \label{eq:boundary}
\end{equation}
where
\begin{eqnarray*}
  A &=& -2 u_0 (\gamma+\kappa-1) \cos\theta - 2 b (\gamma\cos 2\theta +\kappa-1)
        +2 v_0 (\gamma-\kappa+1) \sin\theta\,, \\
  B &=& -4\left[\gamma^2+(\kappa-1)^2+2\gamma(\kappa-1)\cos 2\theta\right]
        \left[ b^2 - a^2 + u_0^2 + v_0^2 + 2\,b\,u_0\,\cos\theta
        + 2\,b\,v_0\,\sin\theta\right]\\
  &&\qquad +4\left[u_0(\kappa+\gamma-1)\cos\theta+b(\kappa+\gamma\cos2\theta-1)
        +v_0(\kappa-\gamma-1)\sin\theta\right]^2 , \\
  C &=& 2\left[\gamma^2+(\kappa-1)^2+2\gamma(\kappa-1)\cos 2\theta\right] .
\end{eqnarray*}
Though complicated, this is a completely analytic mapping of the boundary
of the source to the boundary of the image(s).

While eq.~(\ref{eq:boundary}) completely describes the image boundary, it
is important to note that only solutions with $r_{\pm}(\theta)$ real and
positive are physical.  For some parameter combinations, $B$ can be negative
which implies that $\sqrt{B}$, and hence $r_{\pm}(\theta)$, is complex.  
In particular, for given values of $(\gamma, \kappa, a, b, u_0, v_0)$,
$B$ may or may not be negative for a particular $\theta$.  The range of
$\theta$ for which $B\geq0$ defines the azimuthal extent of the image(s).
In addition, there are also parameter combinations for which
$r_{\pm}(\theta) < 0$.  Such solutions are not physical and form the
boundaries of an `artefact' image.

\begin{figure}
 \begin{center}
\includegraphics[width=0.3\textwidth]{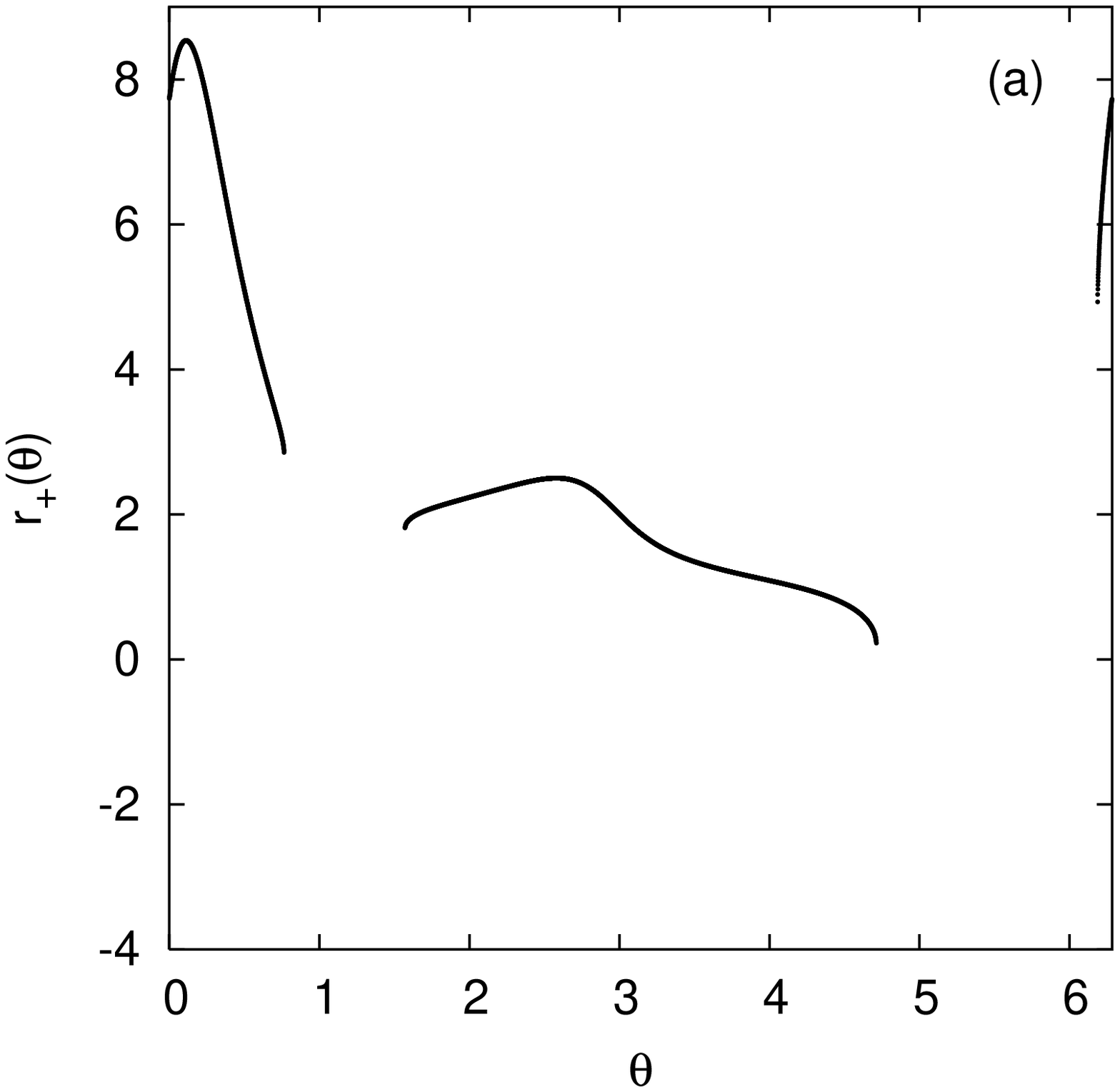}
\includegraphics[width=0.3\textwidth]{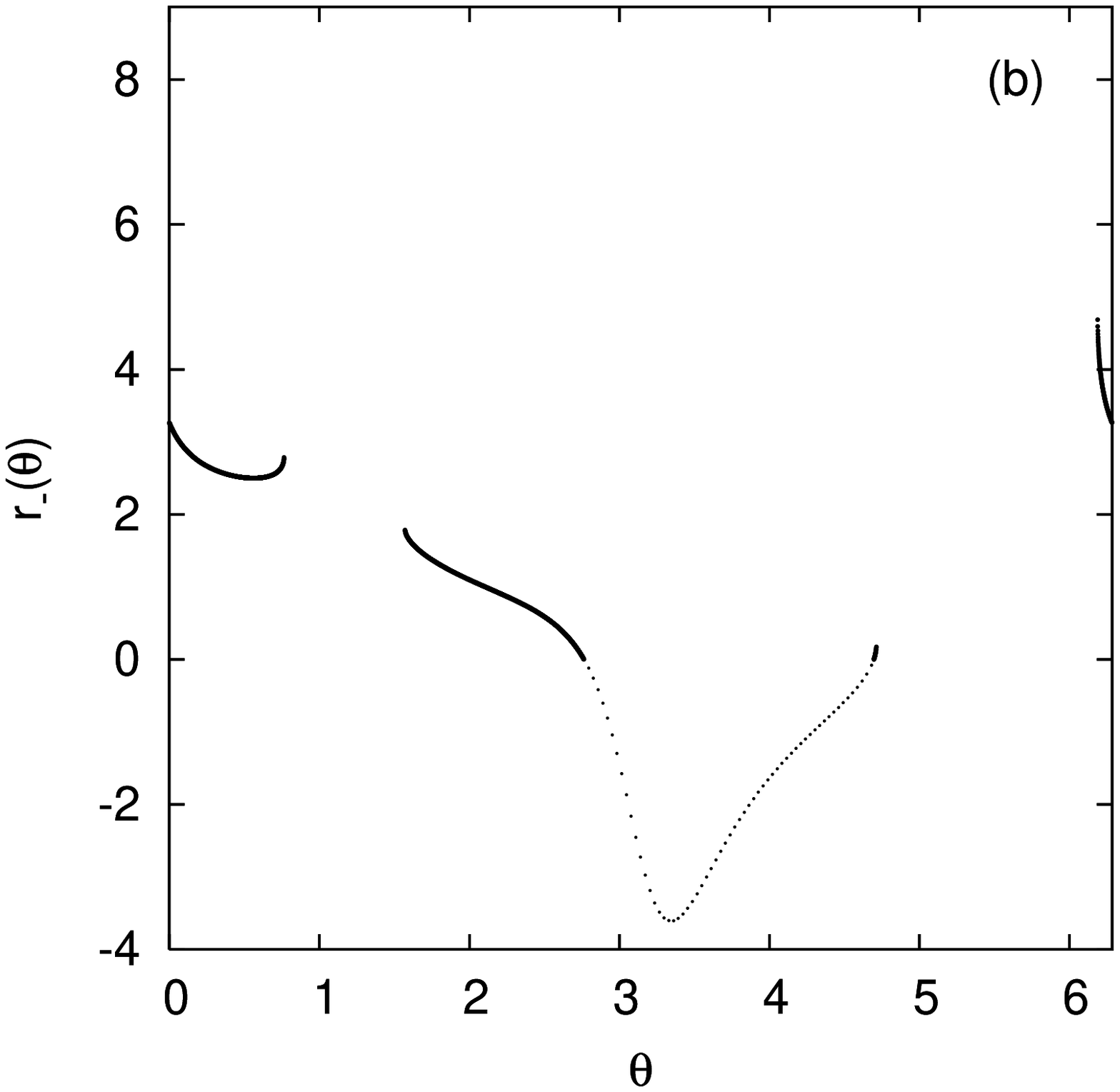}
\linebreak
\includegraphics[width=0.4\textwidth]{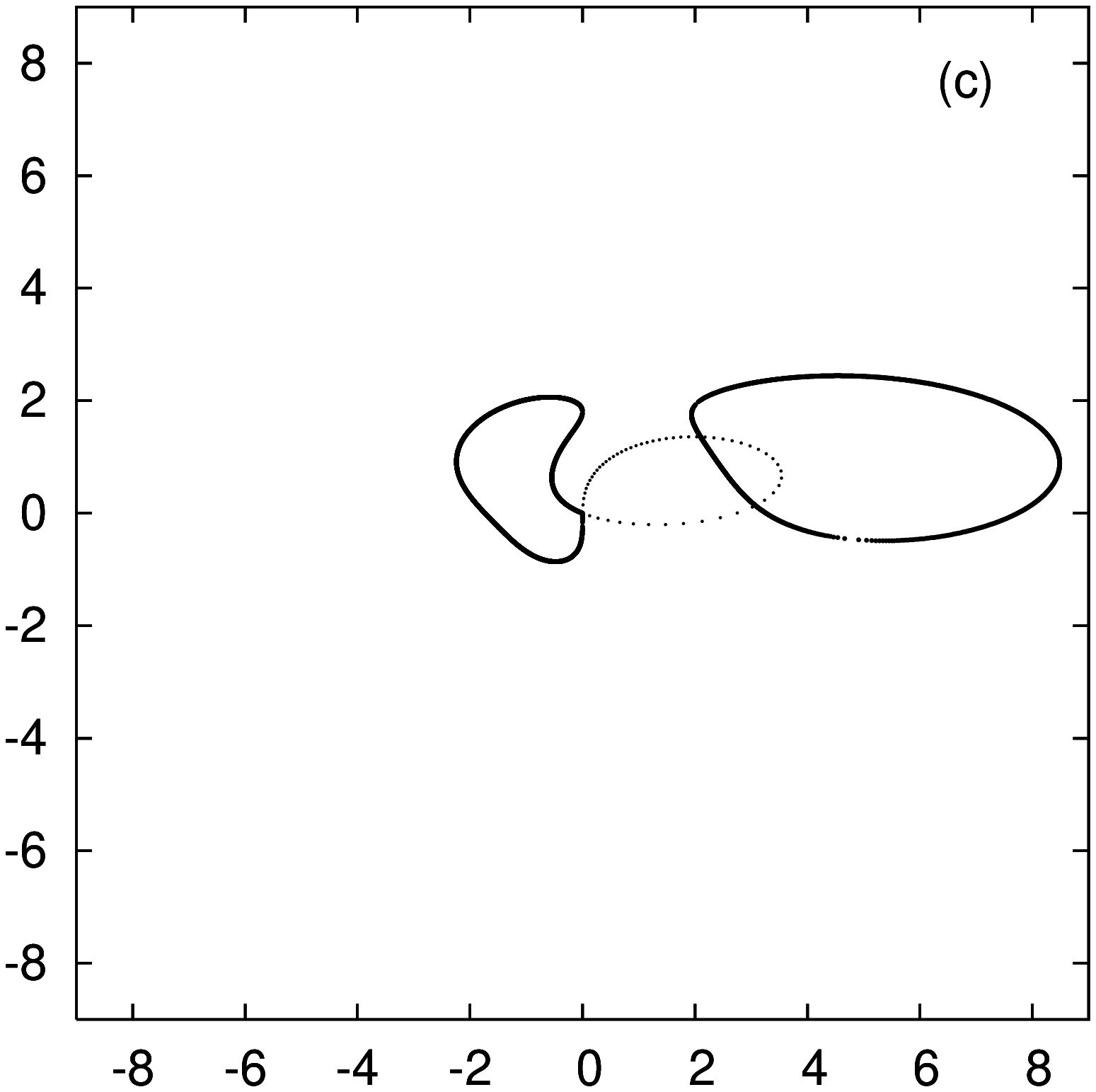}
 \caption{
Analytic image boundary solutions for $\kappa=\gamma=0.3$,
$(u_0,v_0)=(1.2,0.8)$, and $(a,b)=(1.2,1.0)$.
(a) $r_{+}(\theta)$ versus $\theta$.  The region of $\theta$ where
$r_{+}(\theta)$ is defined gives the azimuthal extent of the images.
(b) $r_{-}(\theta)$ versus $\theta$.  The dotted line indicates
where $r_{-}(\theta)$ is real but negative, which corresponds to an
unphysical `artefact' image.
(c) Image boundaries in the $(x,y)$ plane.  Solid lines show the
physical solutions, while the dotted line shows the artefact image
with $r_{-}(\theta)<0$.
}\label{fig:artefactimage}
 \end{center}
\end{figure}

The $r_{\pm}(\theta)$ solutions shown in Fig.~\ref{fig:artefactimage}
exhibit both of these features.  For this example, $\kappa=\gamma=0.3$ so
the unperturbed image has positive parity and magnification $\mu_0 = 2.5$.
Panels (a) and (b) show $r_{\pm}(\theta)$ which are real for only certain
values of $\theta$.  Physically this implies that the images have a finite
azimuthal extent, as can be seen in panel (c).  Panel (b) also shows that
there is a range of $\theta$ where $r_{-}(\theta) < 0$.  This corresponds
to the unphysical artefact image shown by the dotted line in panel (c).

\begin{figure}
 \begin{center}

\includegraphics[width=0.2\textwidth]{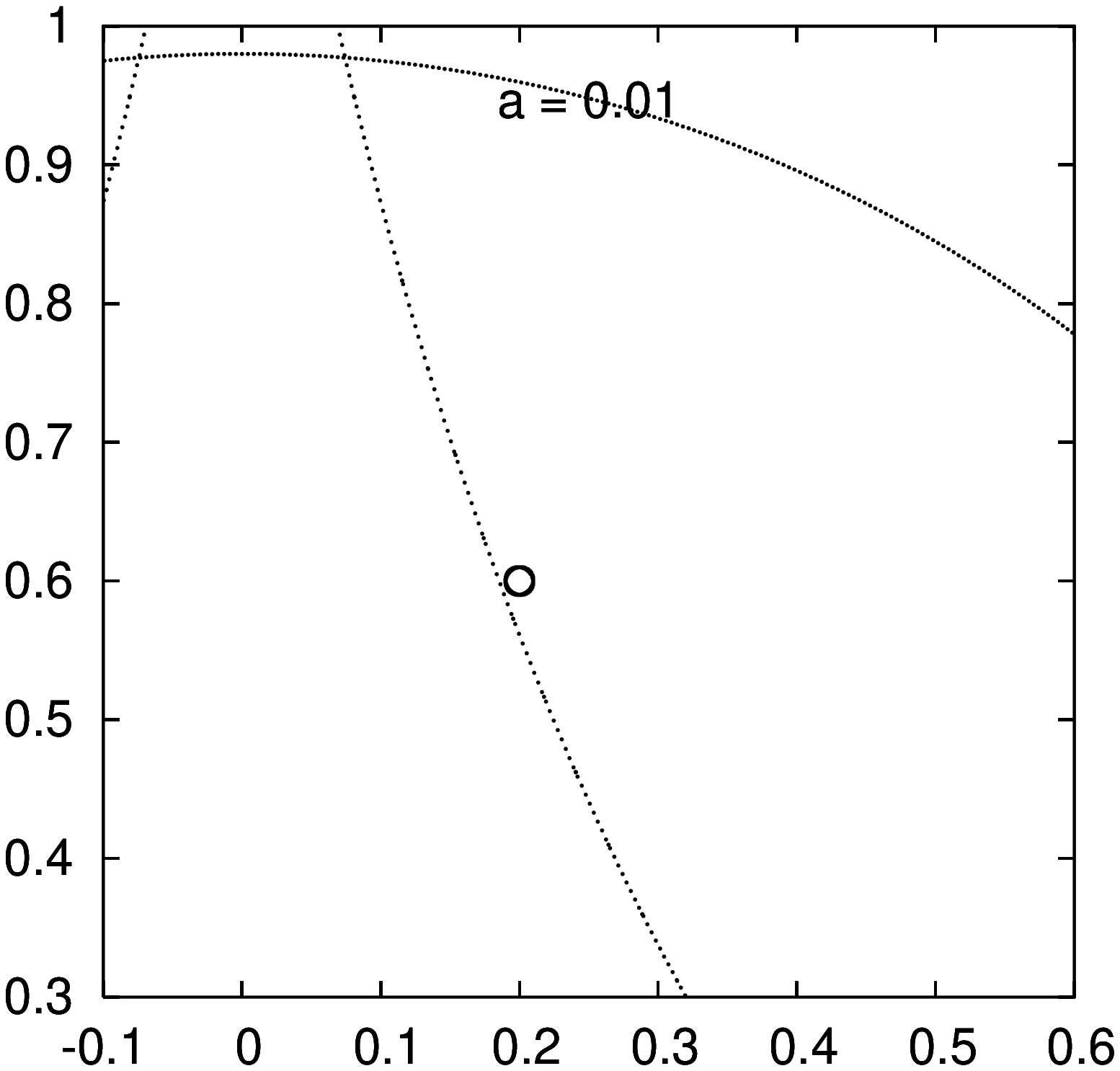}
\includegraphics[width=0.2\textwidth]{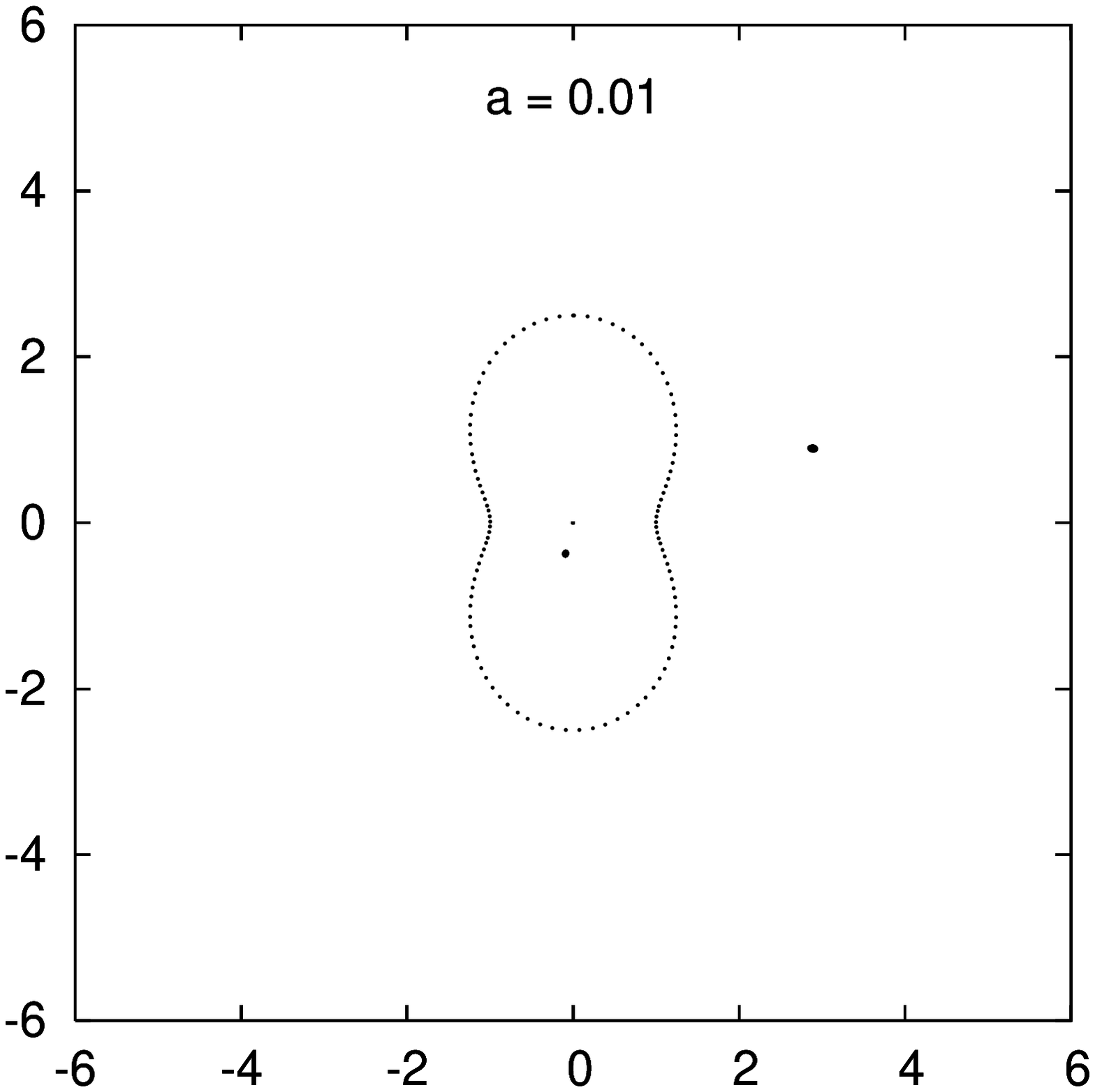}

\includegraphics[width=0.2\textwidth]{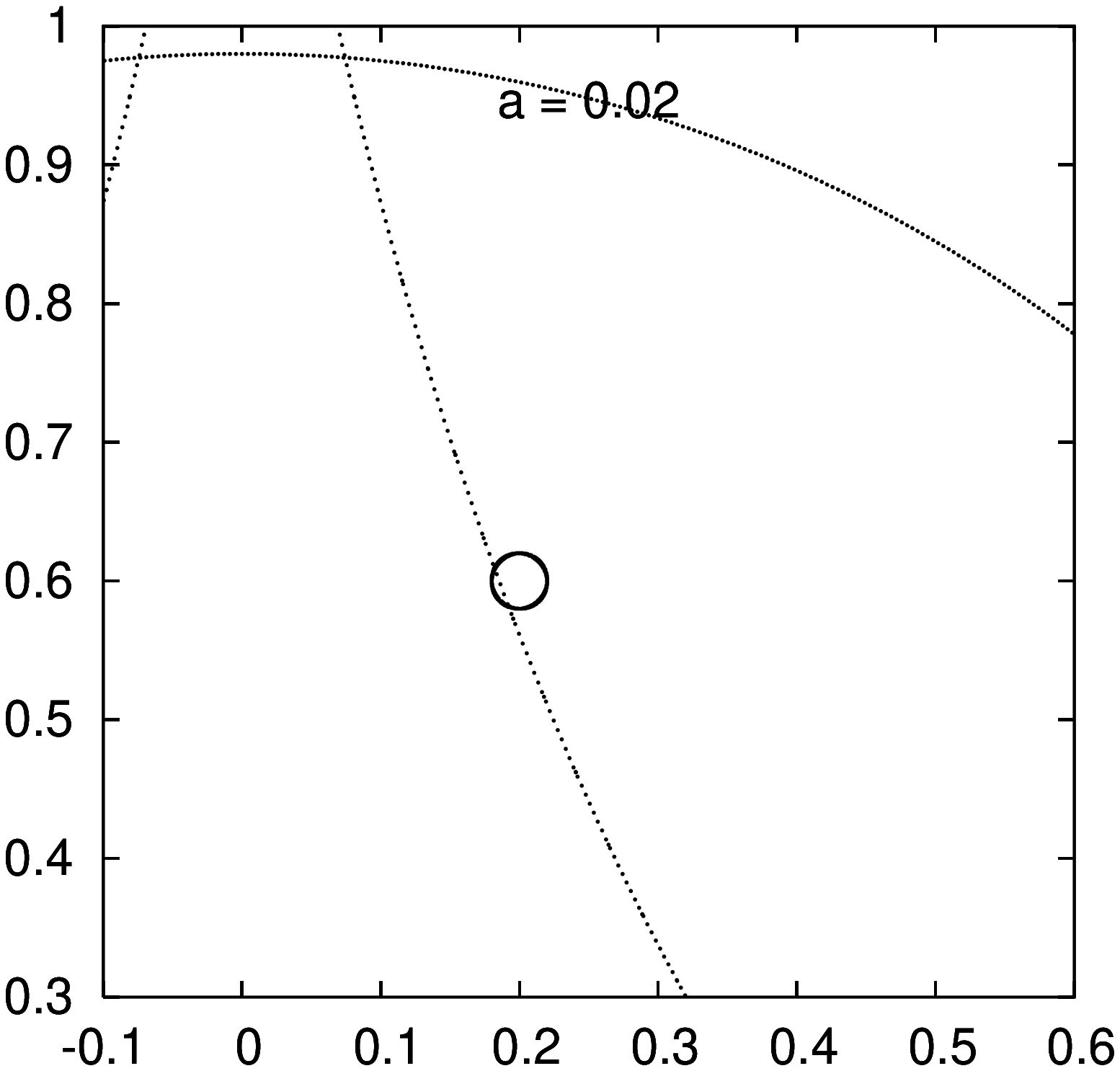}
\includegraphics[width=0.2\textwidth]{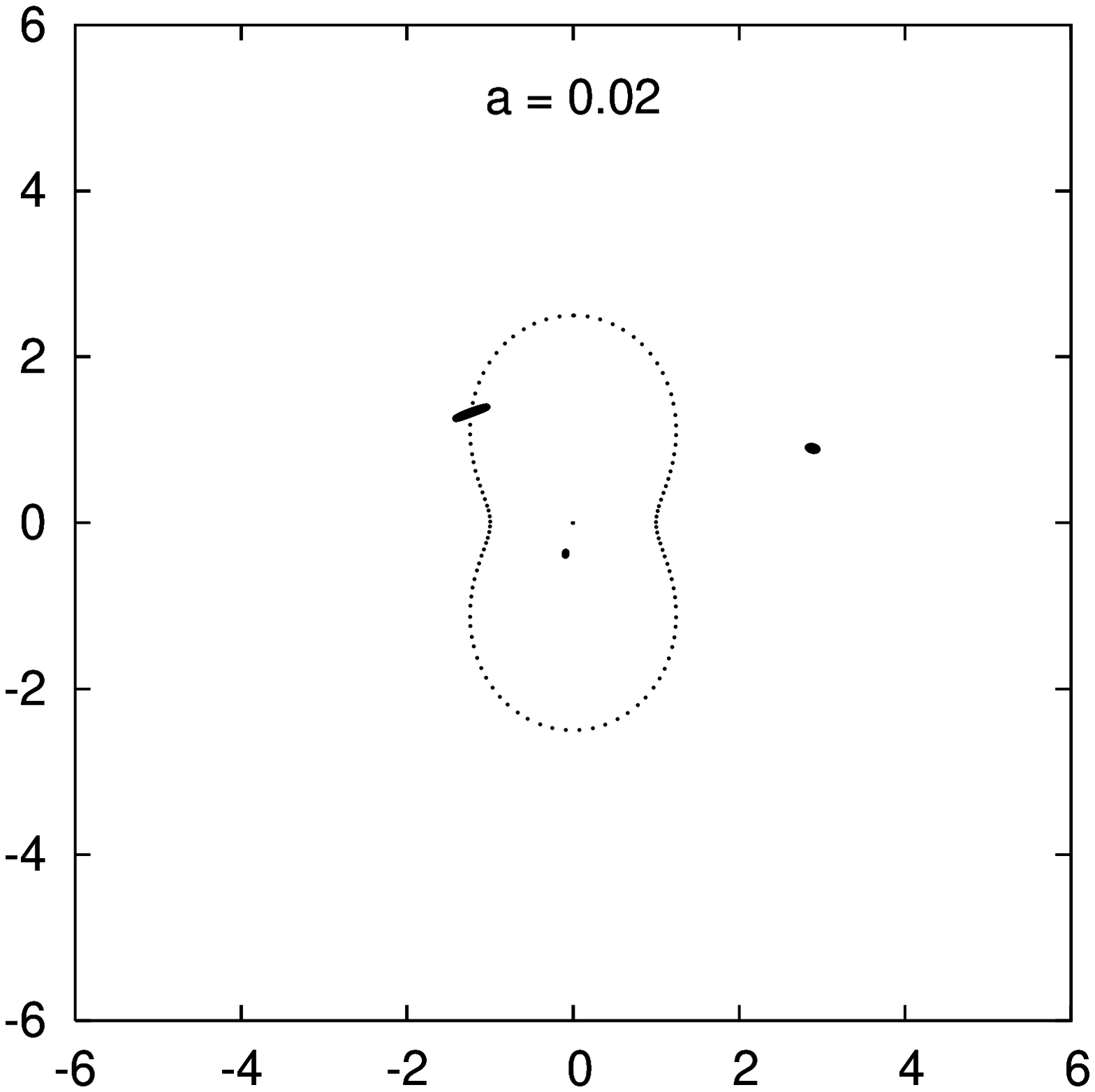}

\includegraphics[width=0.2\textwidth]{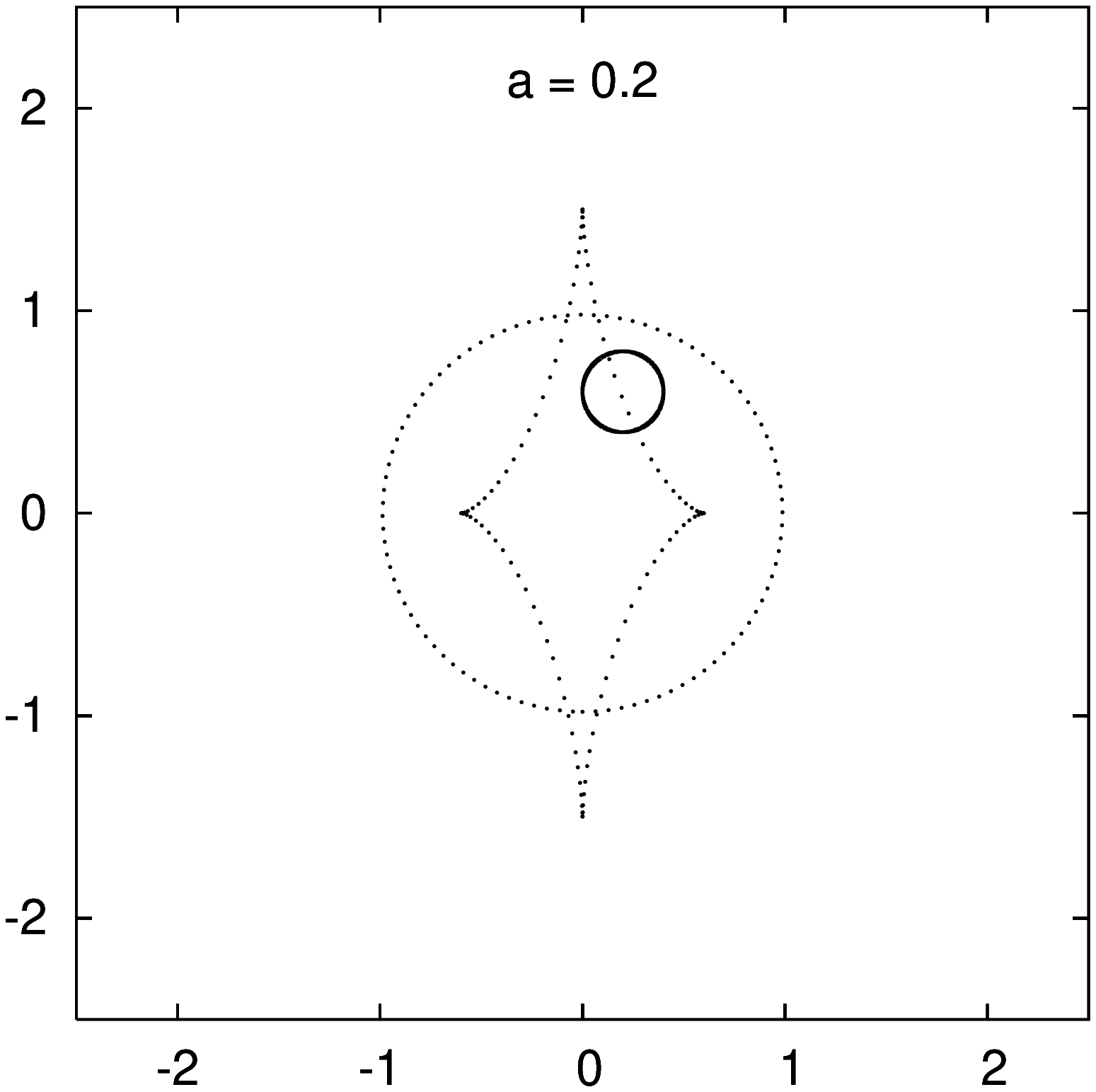}
\includegraphics[width=0.2\textwidth]{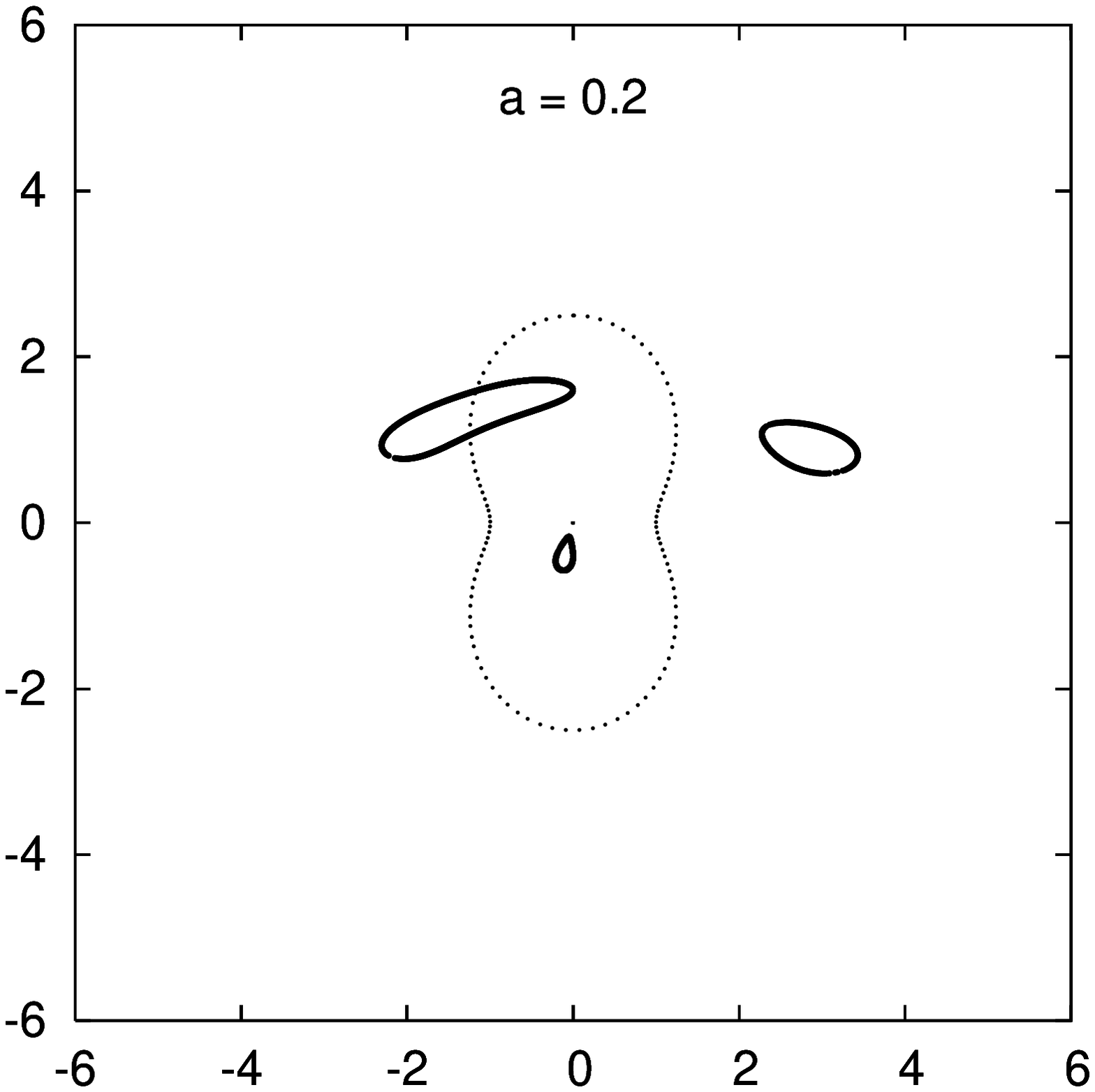}

\includegraphics[width=0.2\textwidth]{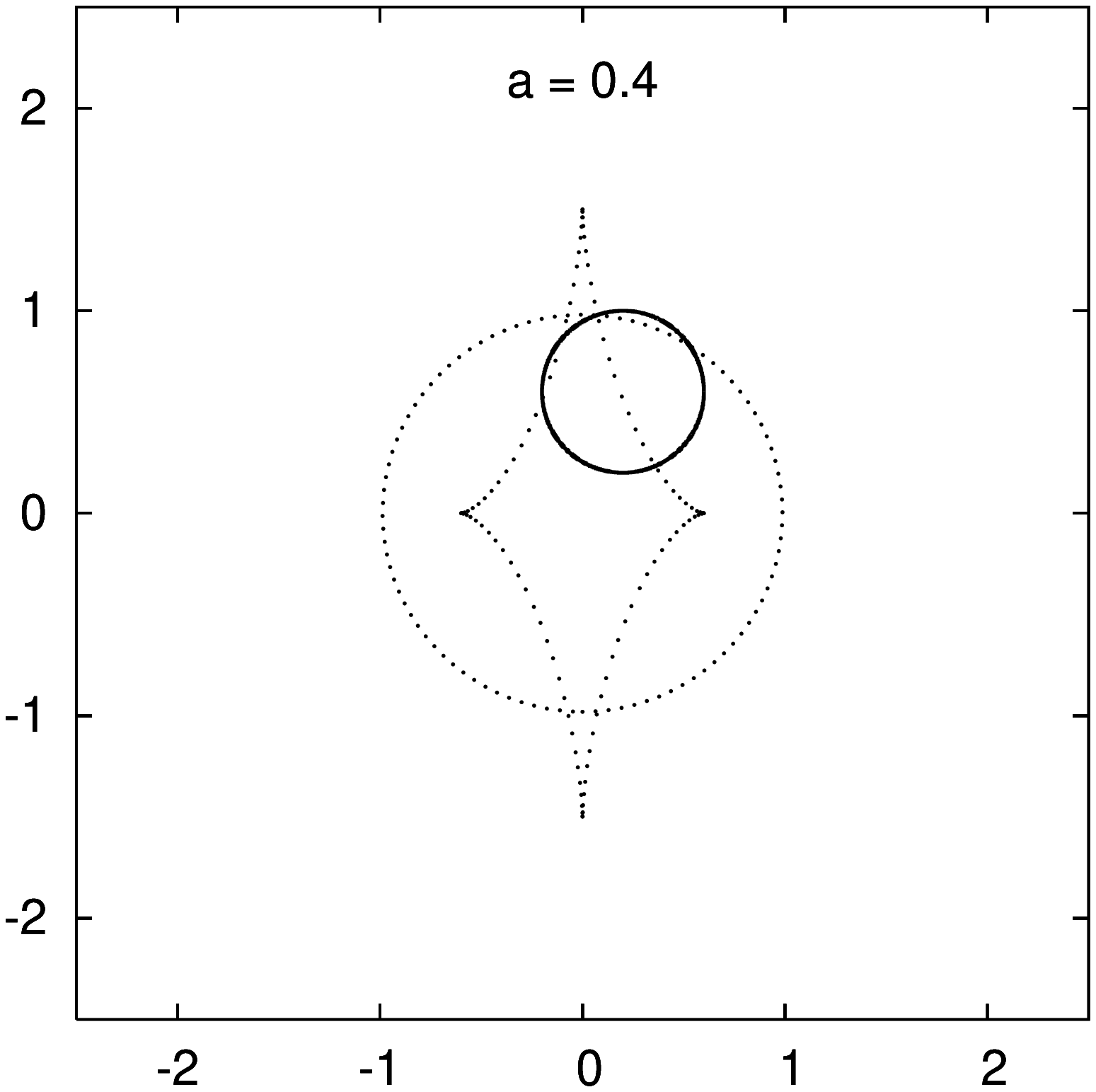}
\includegraphics[width=0.2\textwidth]{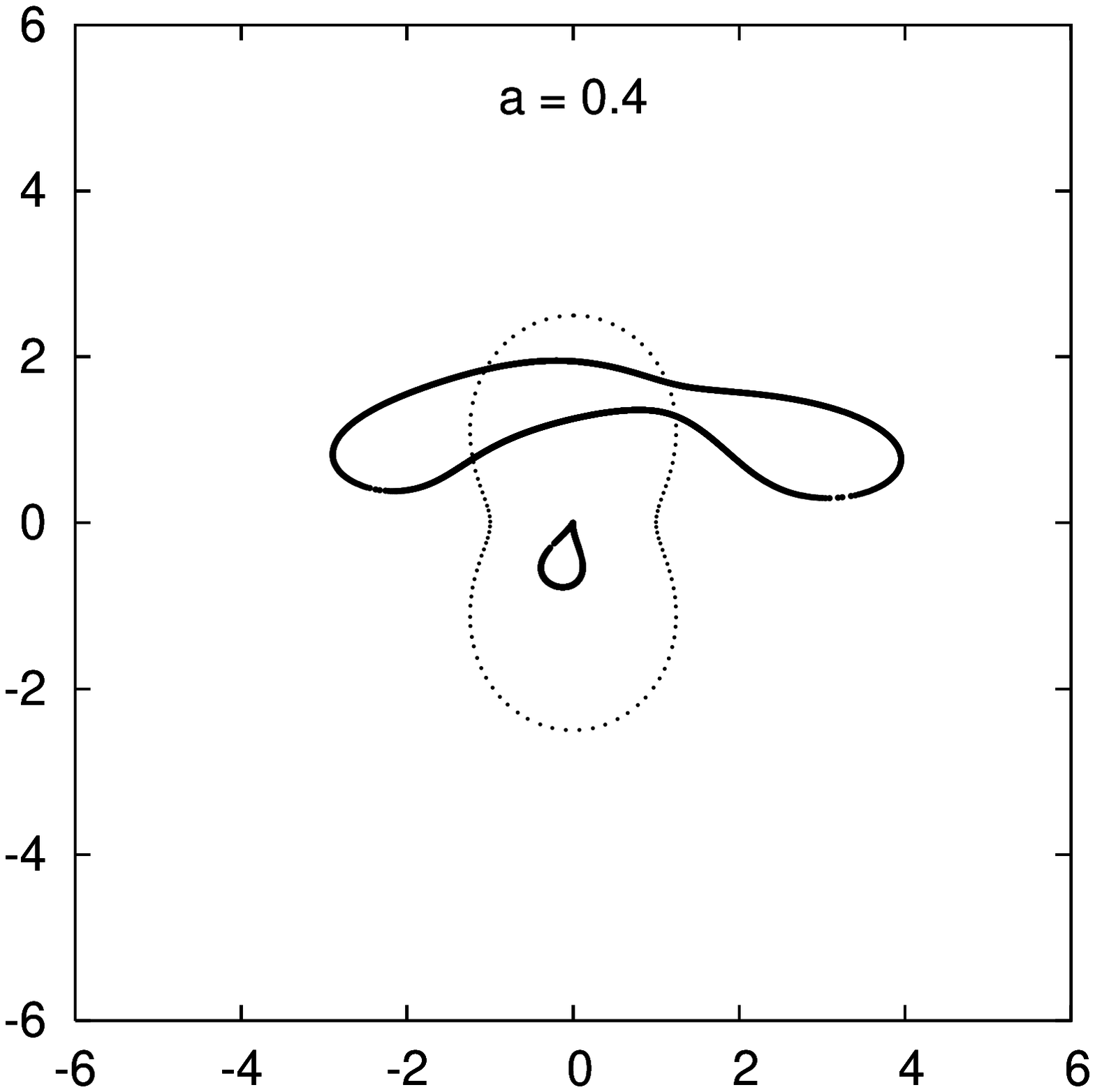}

\includegraphics[width=0.2\textwidth]{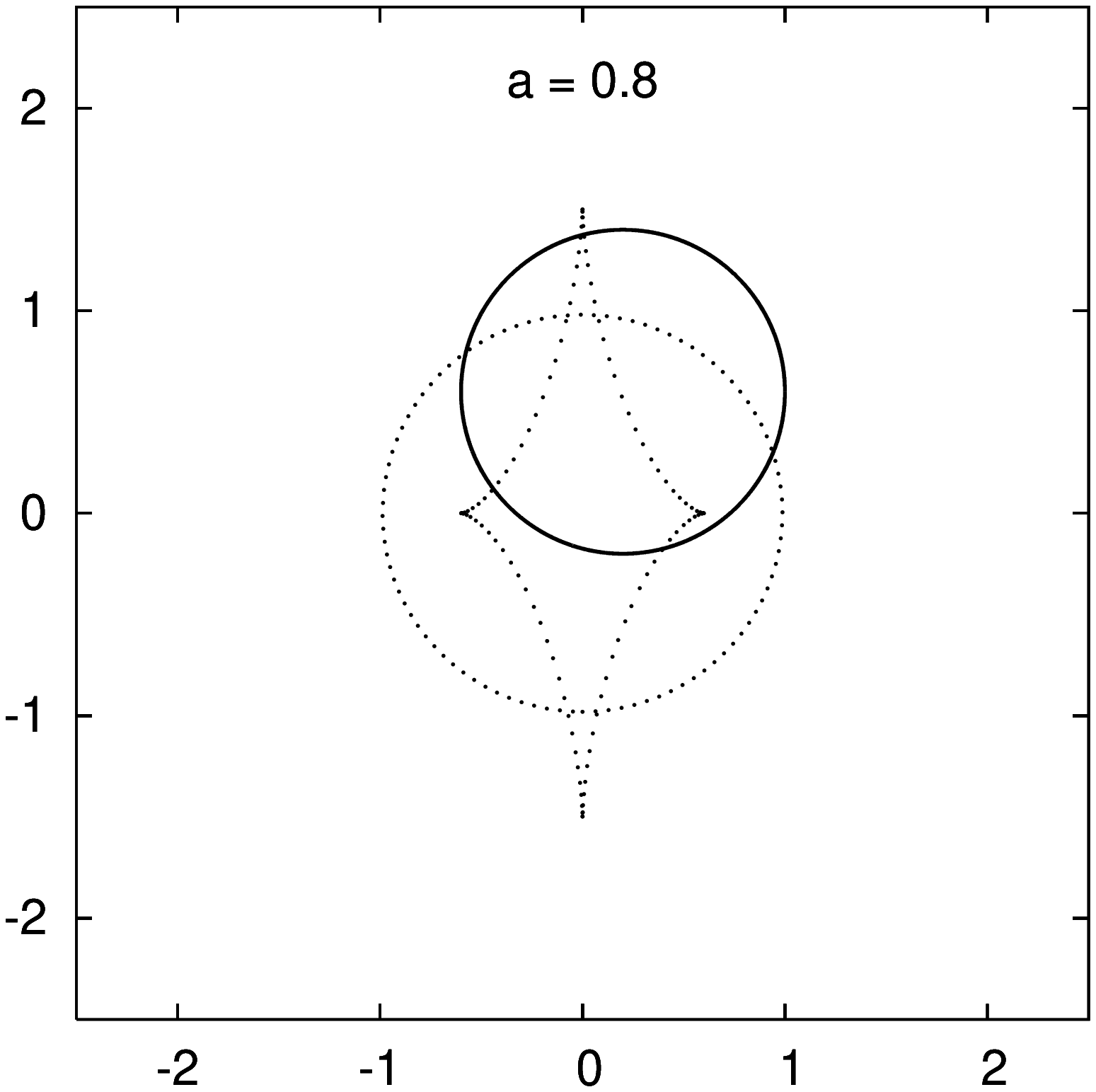}
\includegraphics[width=0.2\textwidth]{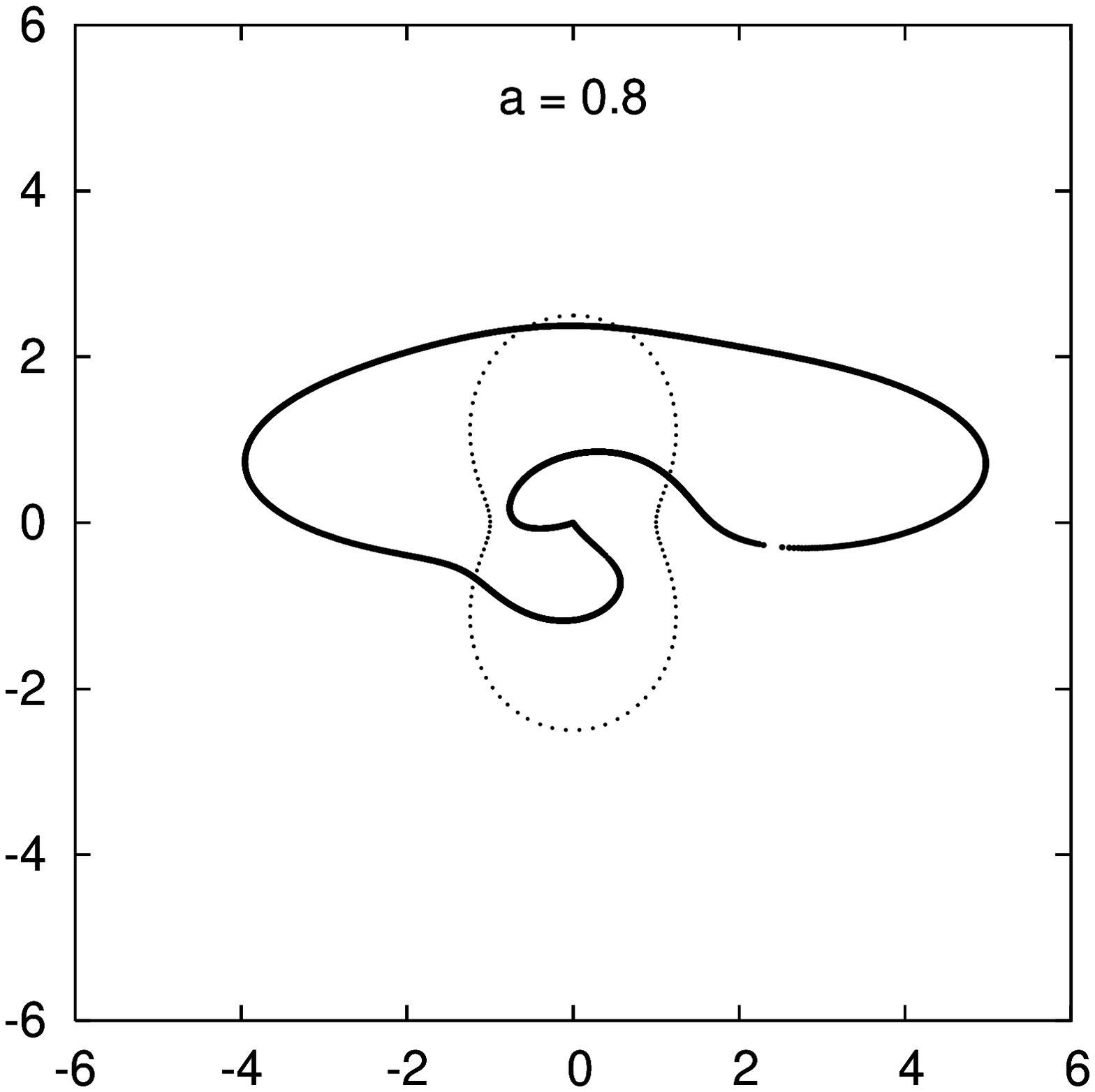}

\includegraphics[width=0.2\textwidth]{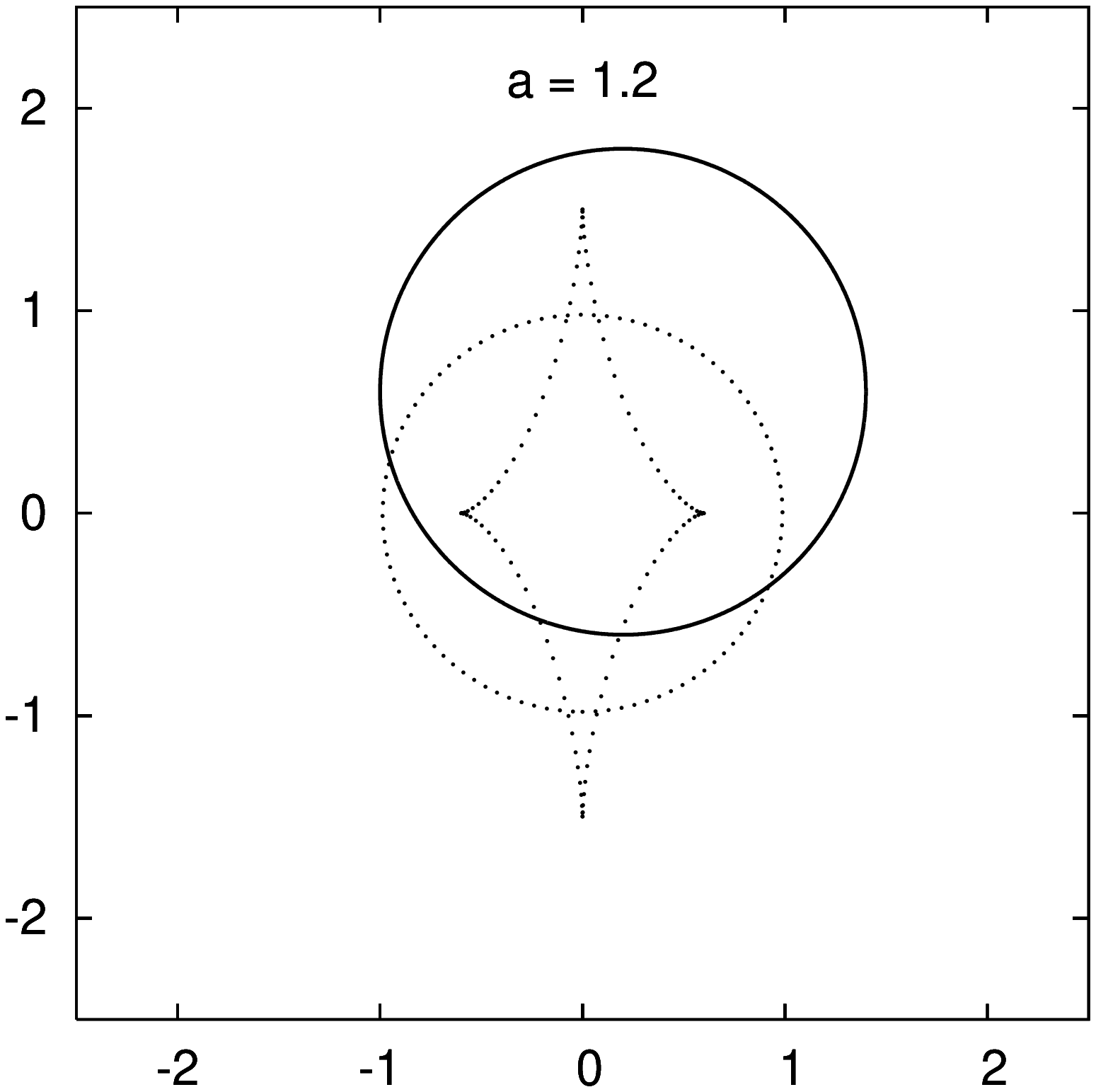}
\includegraphics[width=0.2\textwidth]{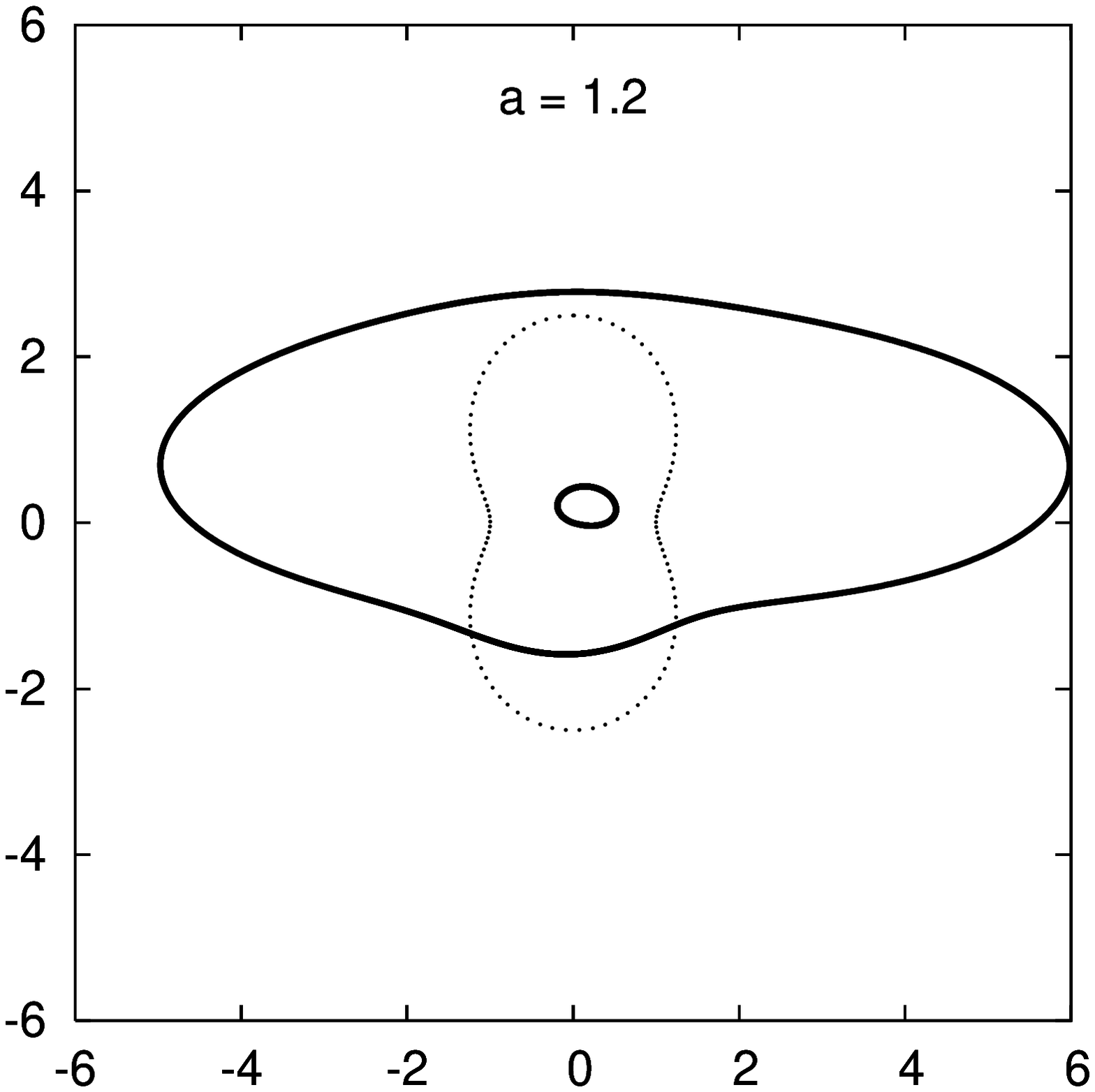}

\caption{
Image configurations for different source sizes.  The left column shows
the source boundaries and caustics, while the right column shows the image
boundaries and critical curves.  In these examples, $\kappa=\gamma=0.3$,
$(u_0,v_0)=(0.2,0.6)$, $b=1$, and $a$ is increased from $0.01$ to $1.2$.
At large source size ($a/b > 1$) there are still significant deviations
from the ellipse image expected for a pure convergence and shear field.
}\label{fig:positiveimages}
 \end{center}
\end{figure}

Fig.~\ref{fig:positiveimages} shows how image configurations change as
the source size $a$ is increased.  The left column shows the source and
the caustics while the right column shows the images and the critical
curves; without loss of generality, we work in units with $b=1$.  For
the top row, the source with $a=0.01$ has been placed near a fold caustic
but has not intersected it.  The source lies completely within a two-image
region and the $r(\theta)$ solution does in fact give two images.  In the
second row, the source has doubled in size and now crosses the fold caustic.
In this configuration, part of the source lies in the two-image region and
another part lies in the four-image region.  The $r(\theta)$ solution shows
that the initial two images have grown slightly in size and a third image
(which is actually a merged image pair) has appeared in the upper left.  
As the source size is increased further, it crosses more and more of the
caustics yielding complex image solutions which consist of merging and
growing images.  By $a=1.2$, the source covers most of the caustic and
the resulting image is clearly becoming the ellipse that one would
expect for a simple convergence and shear field.  Nevertheless, even at
this large source size there are still significant deviations from the
unperturbed image.

\begin{figure}
 \begin{center}

\includegraphics[width=0.2\textwidth]{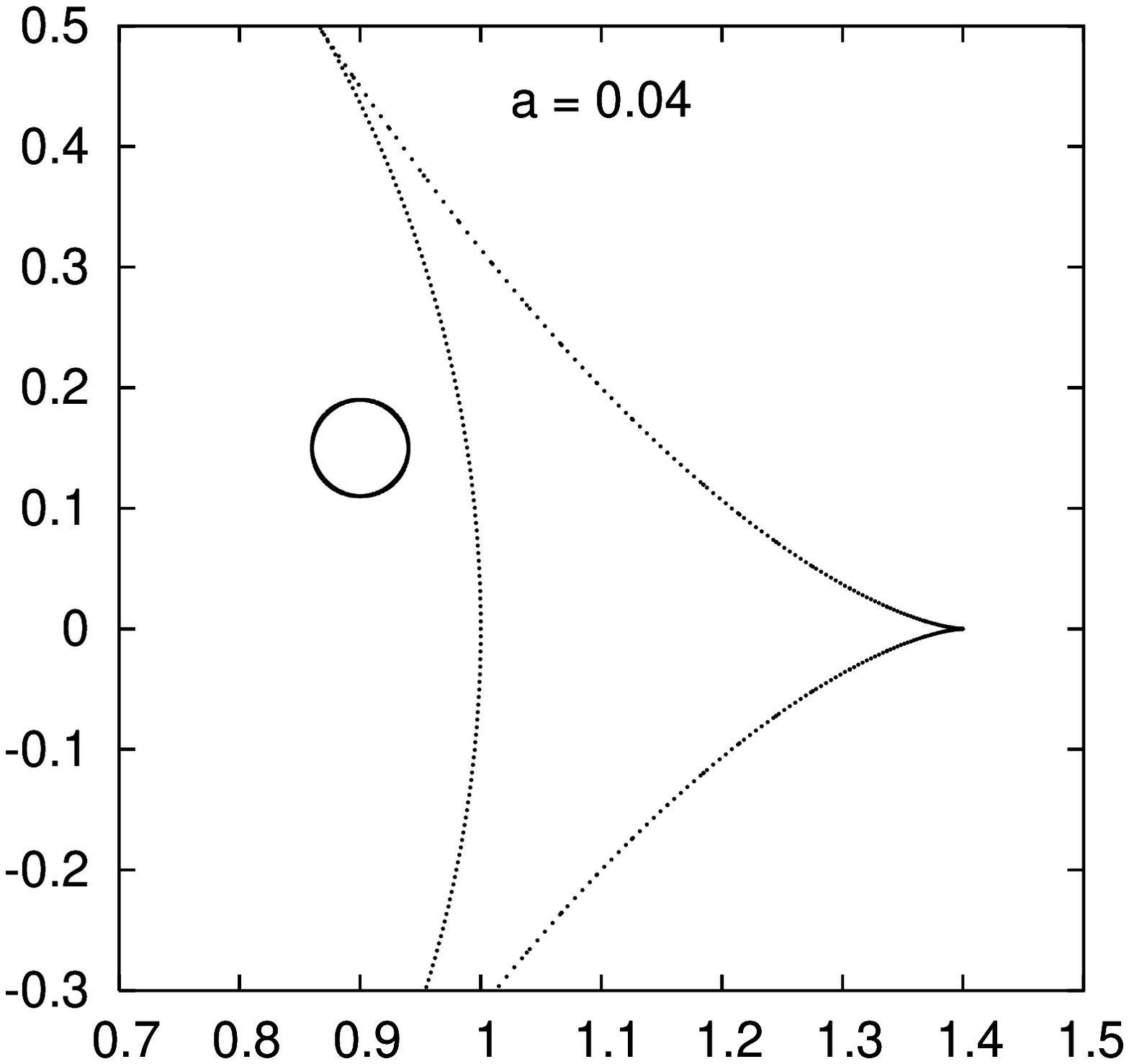}
\includegraphics[width=0.2\textwidth]{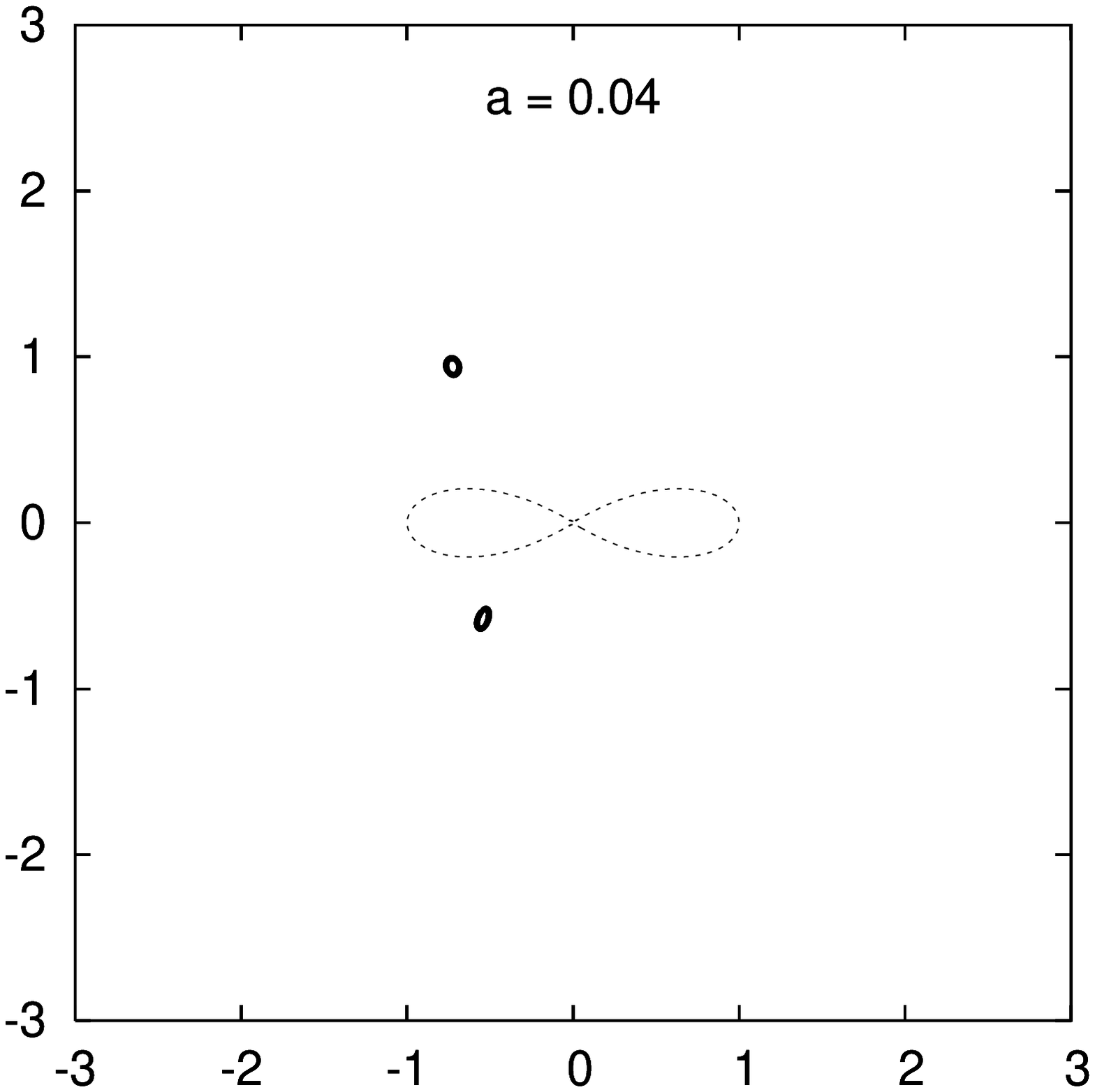}

\includegraphics[width=0.2\textwidth]{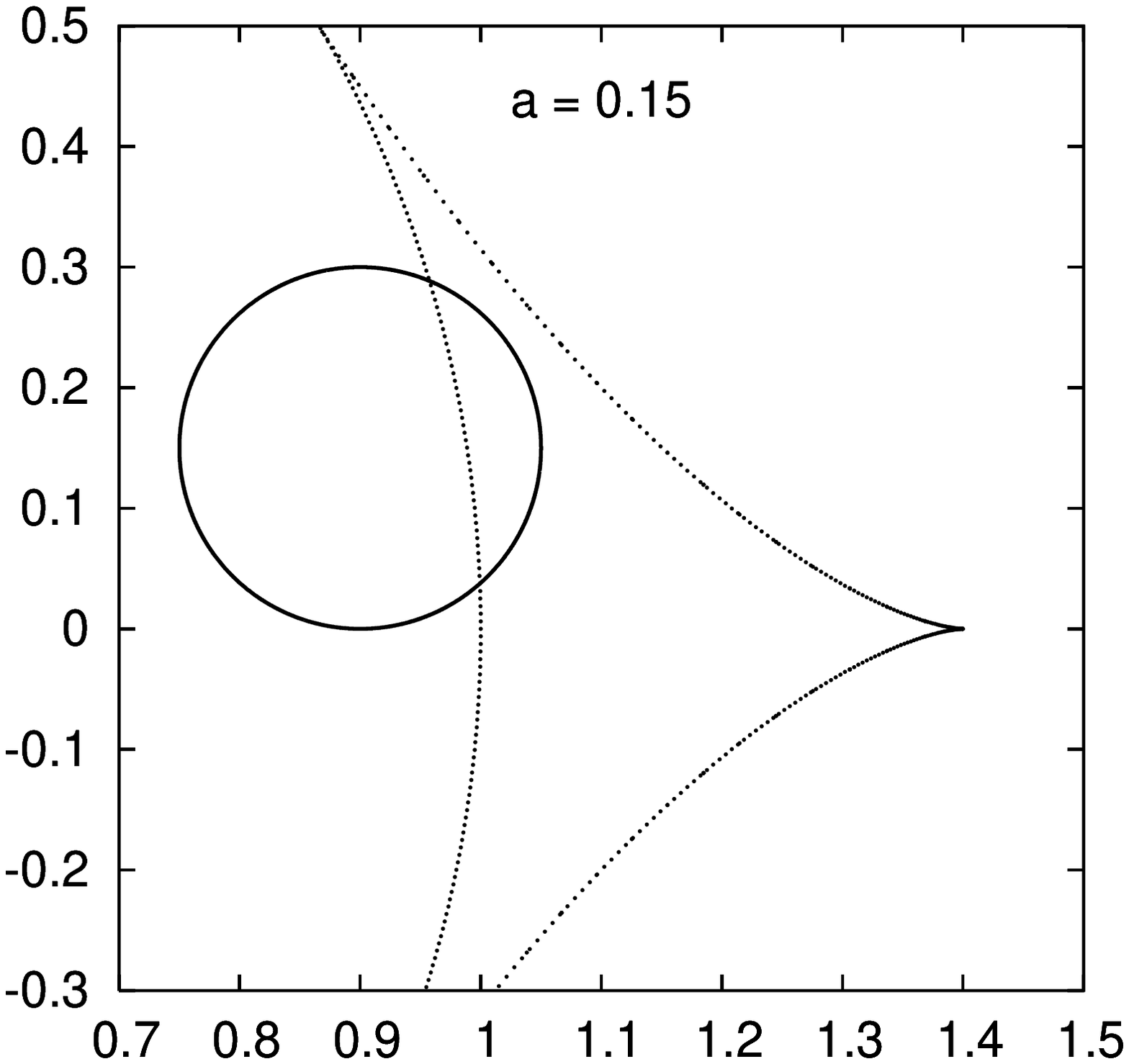}
\includegraphics[width=0.2\textwidth]{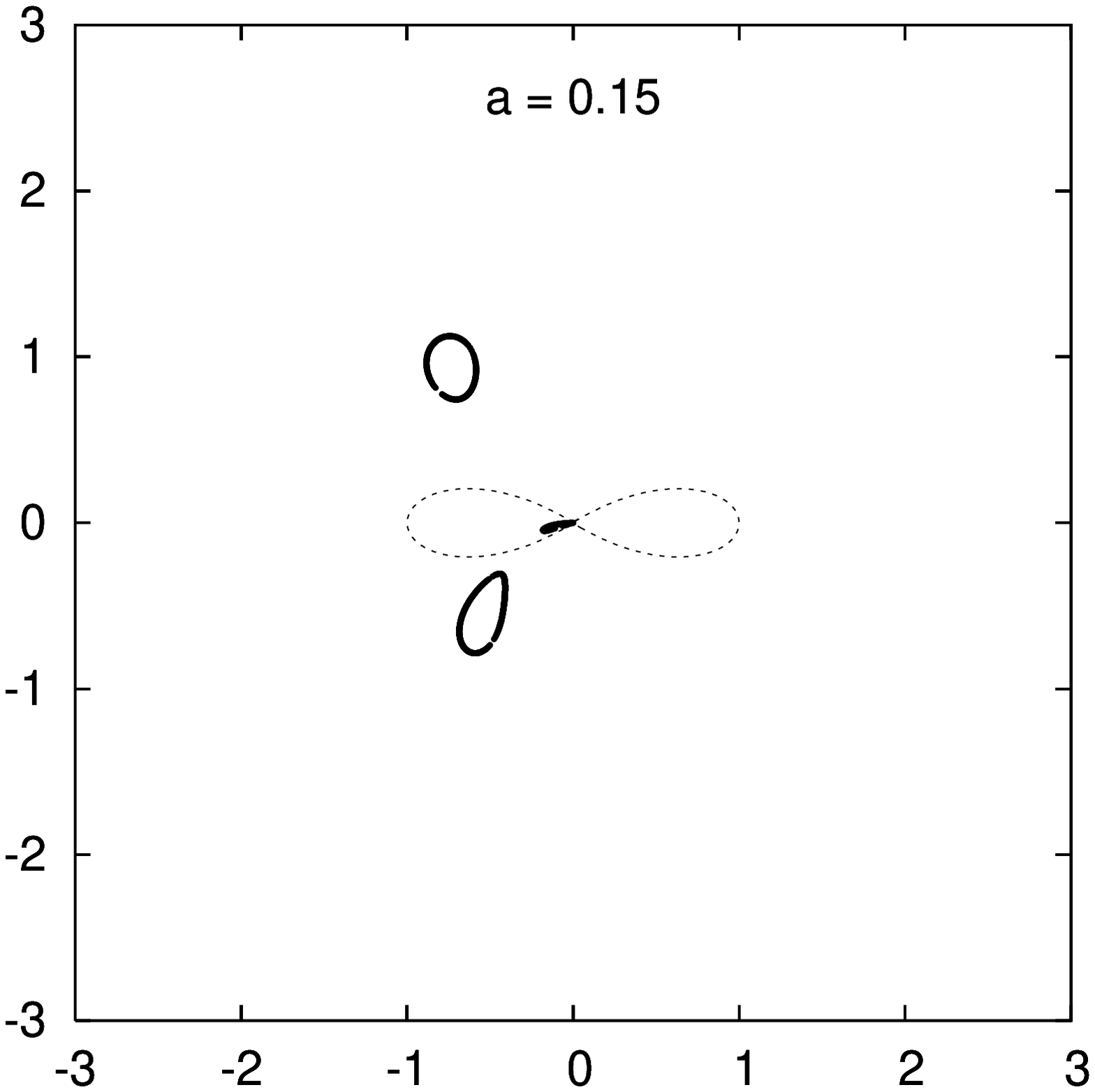}

\includegraphics[width=0.2\textwidth]{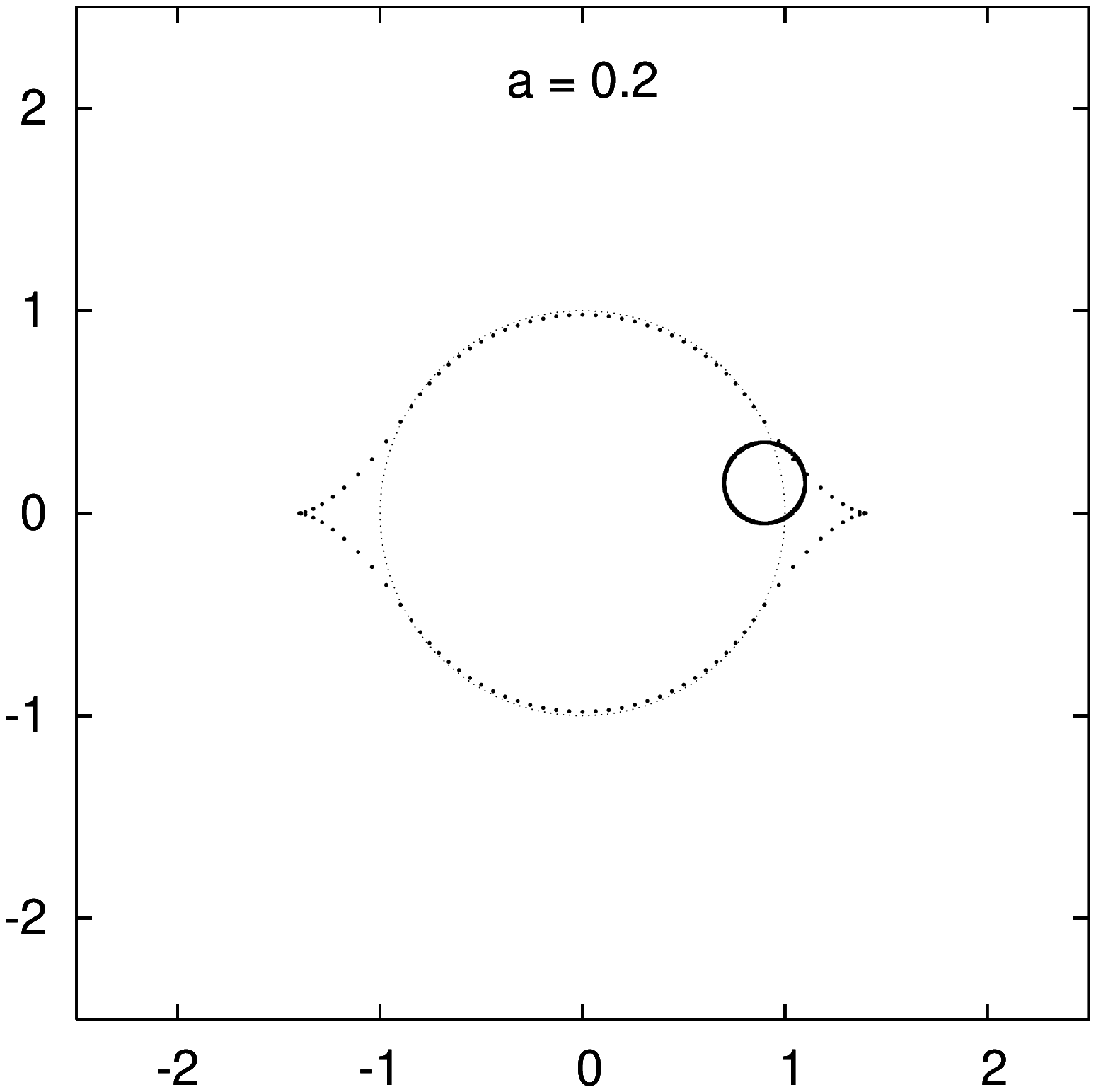}
\includegraphics[width=0.2\textwidth]{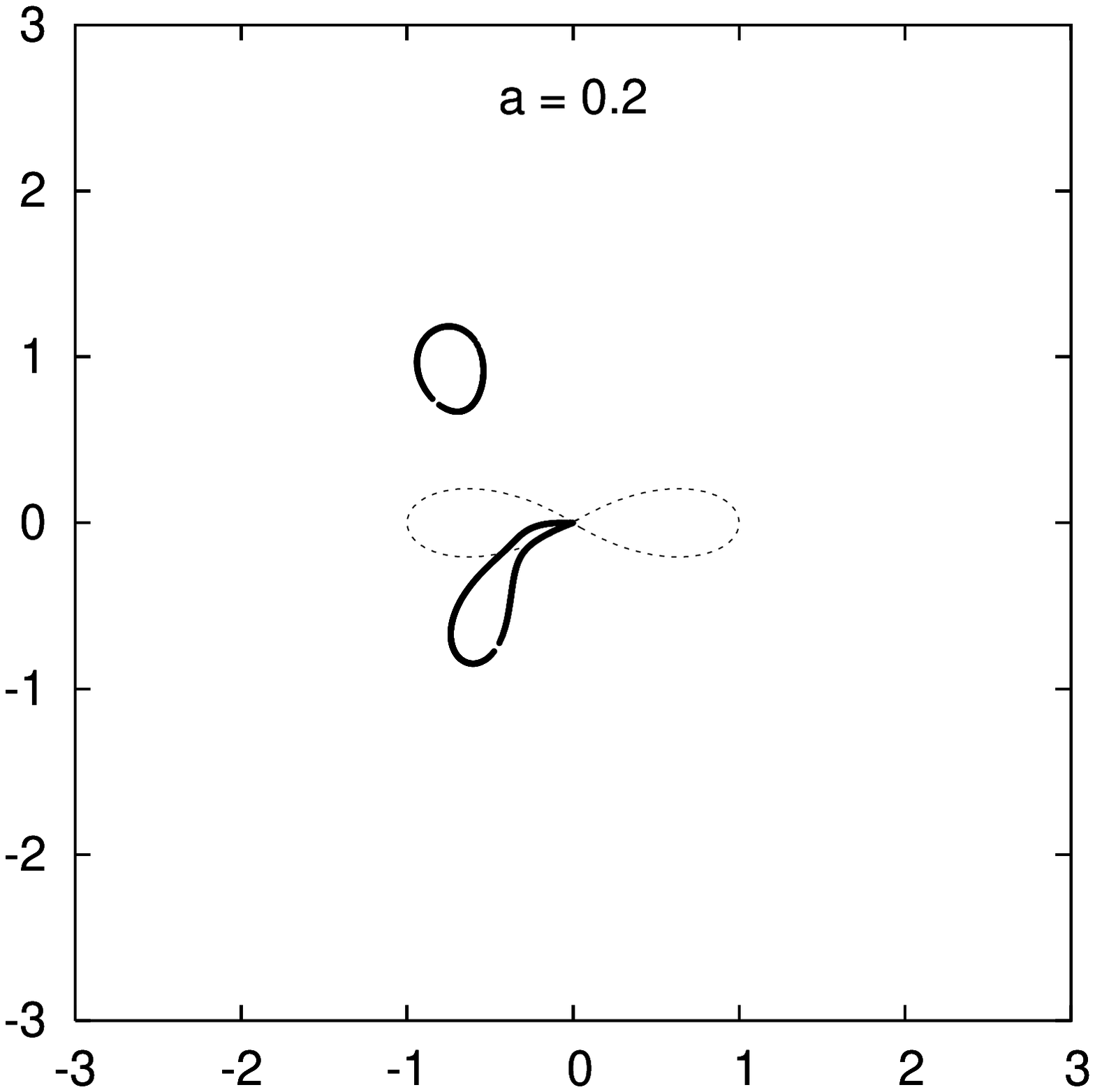}

\includegraphics[width=0.2\textwidth]{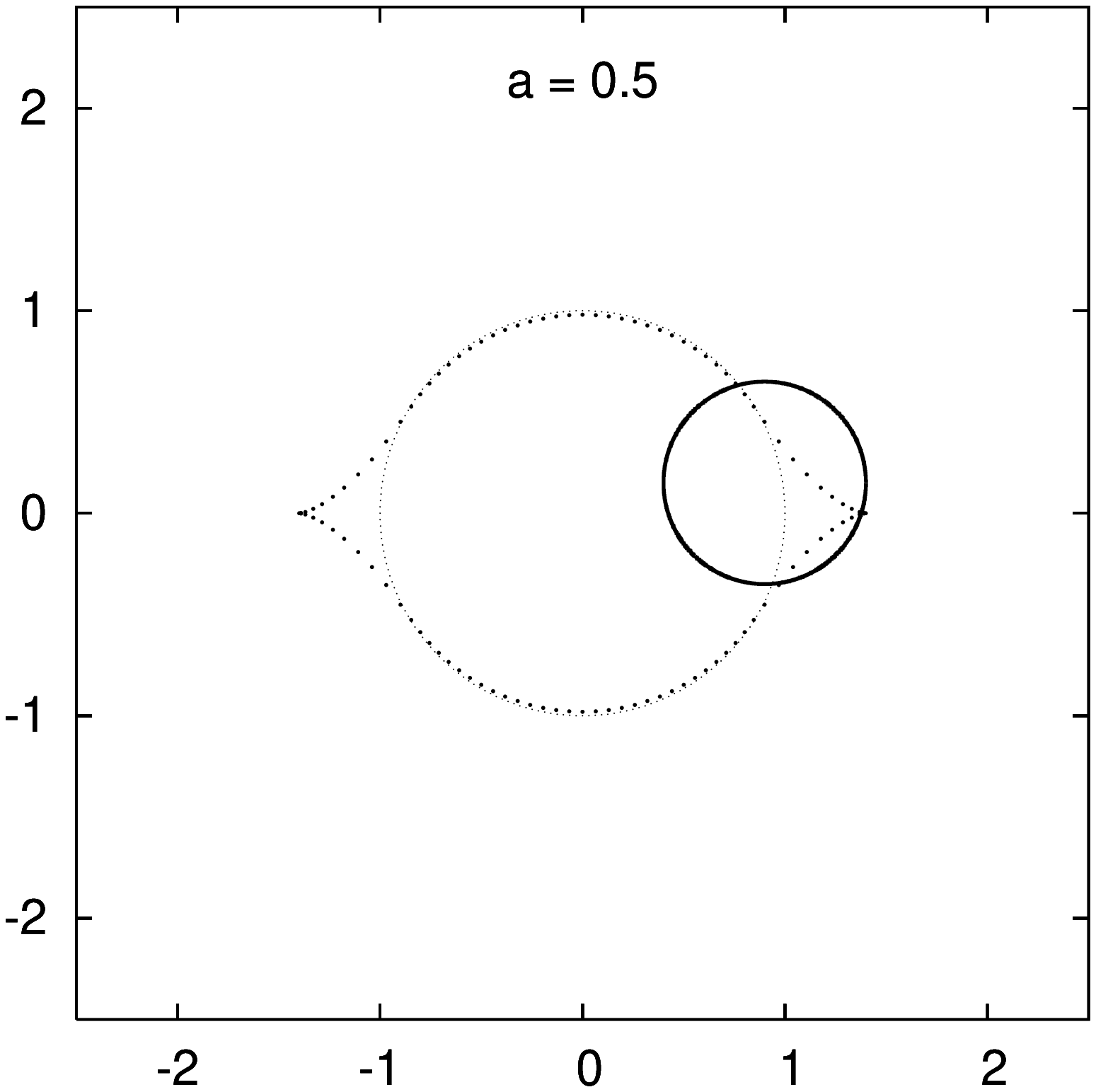}
\includegraphics[width=0.2\textwidth]{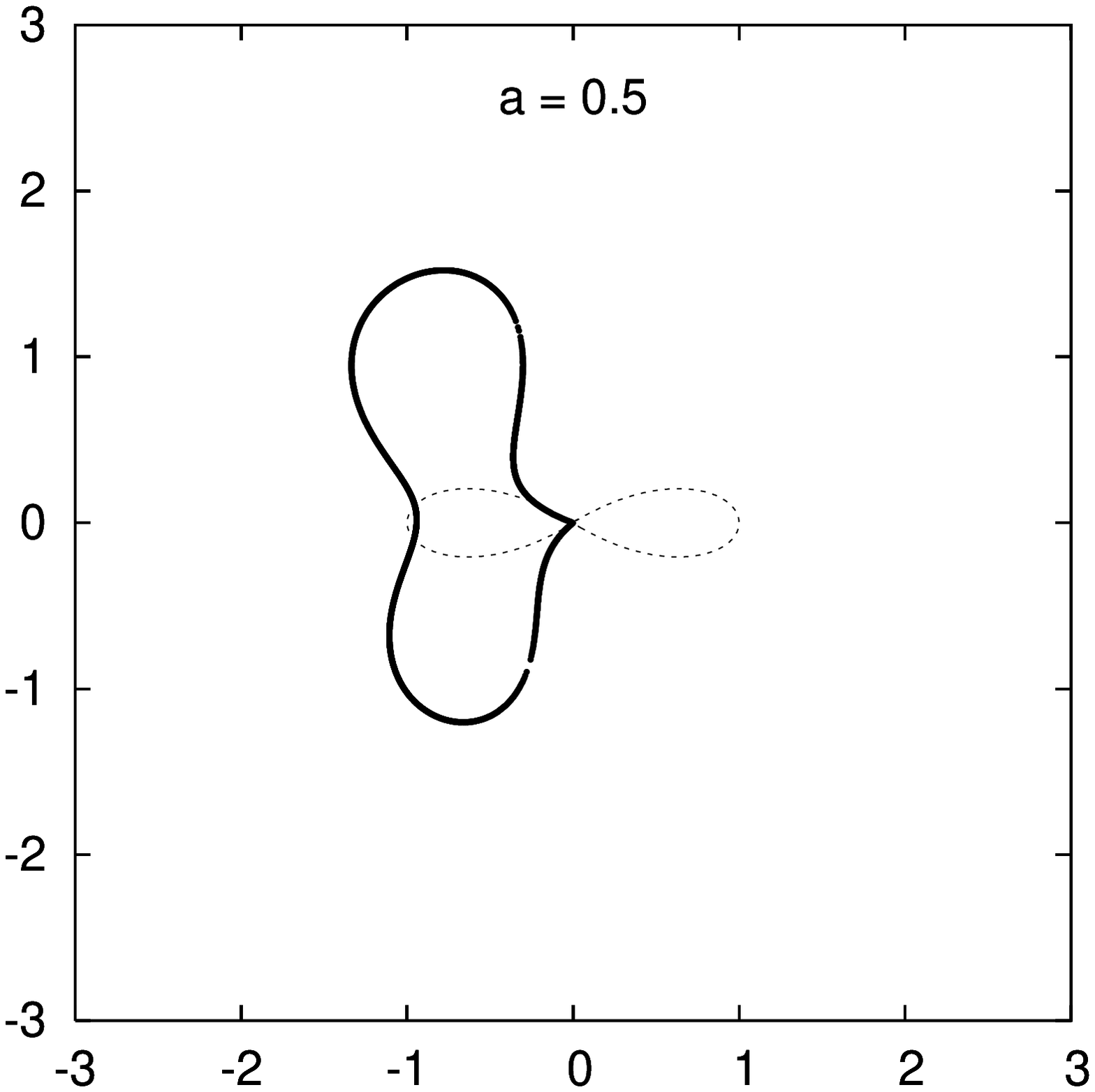}

\includegraphics[width=0.2\textwidth]{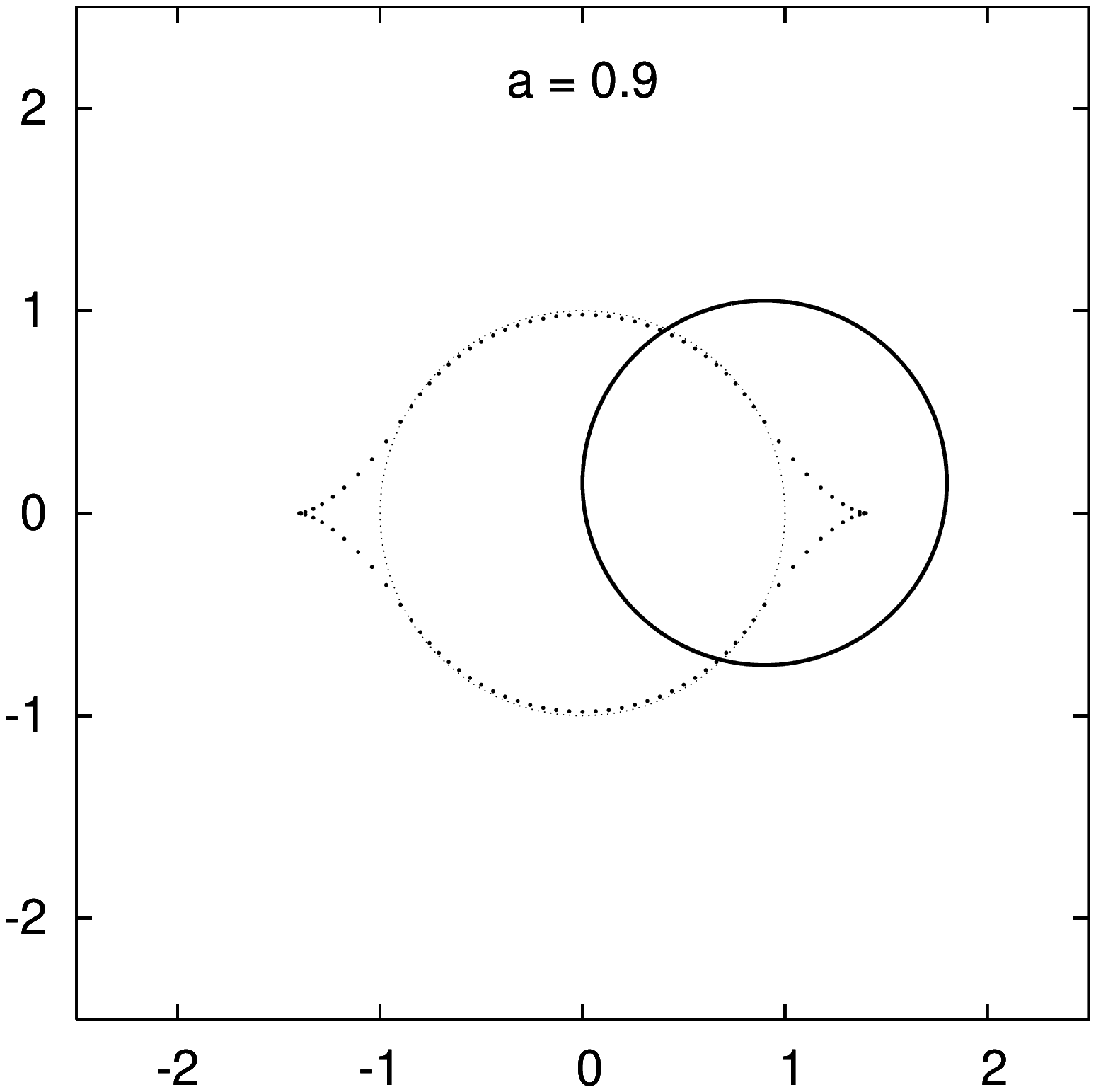}
\includegraphics[width=0.2\textwidth]{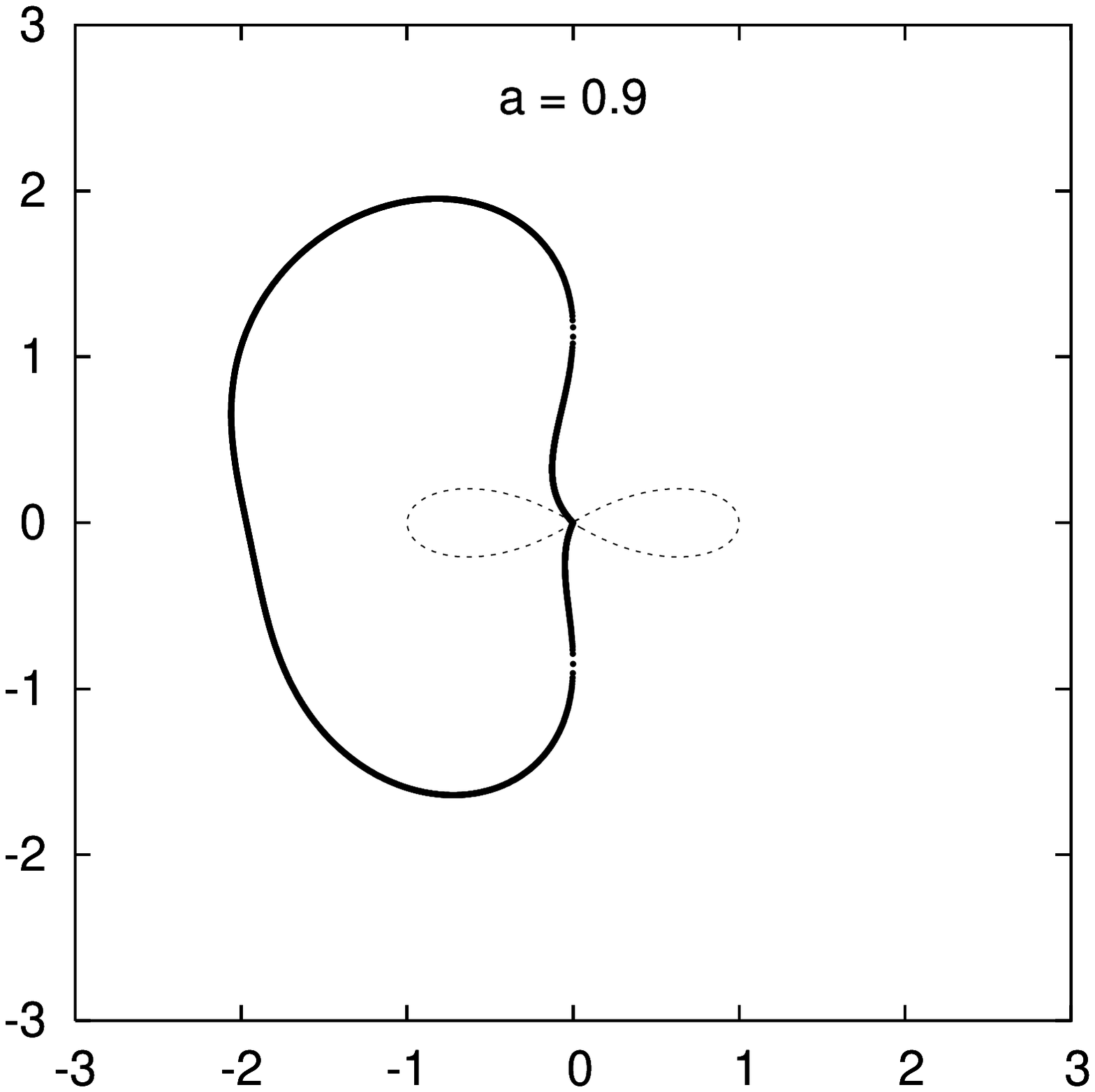}

\includegraphics[width=0.2\textwidth]{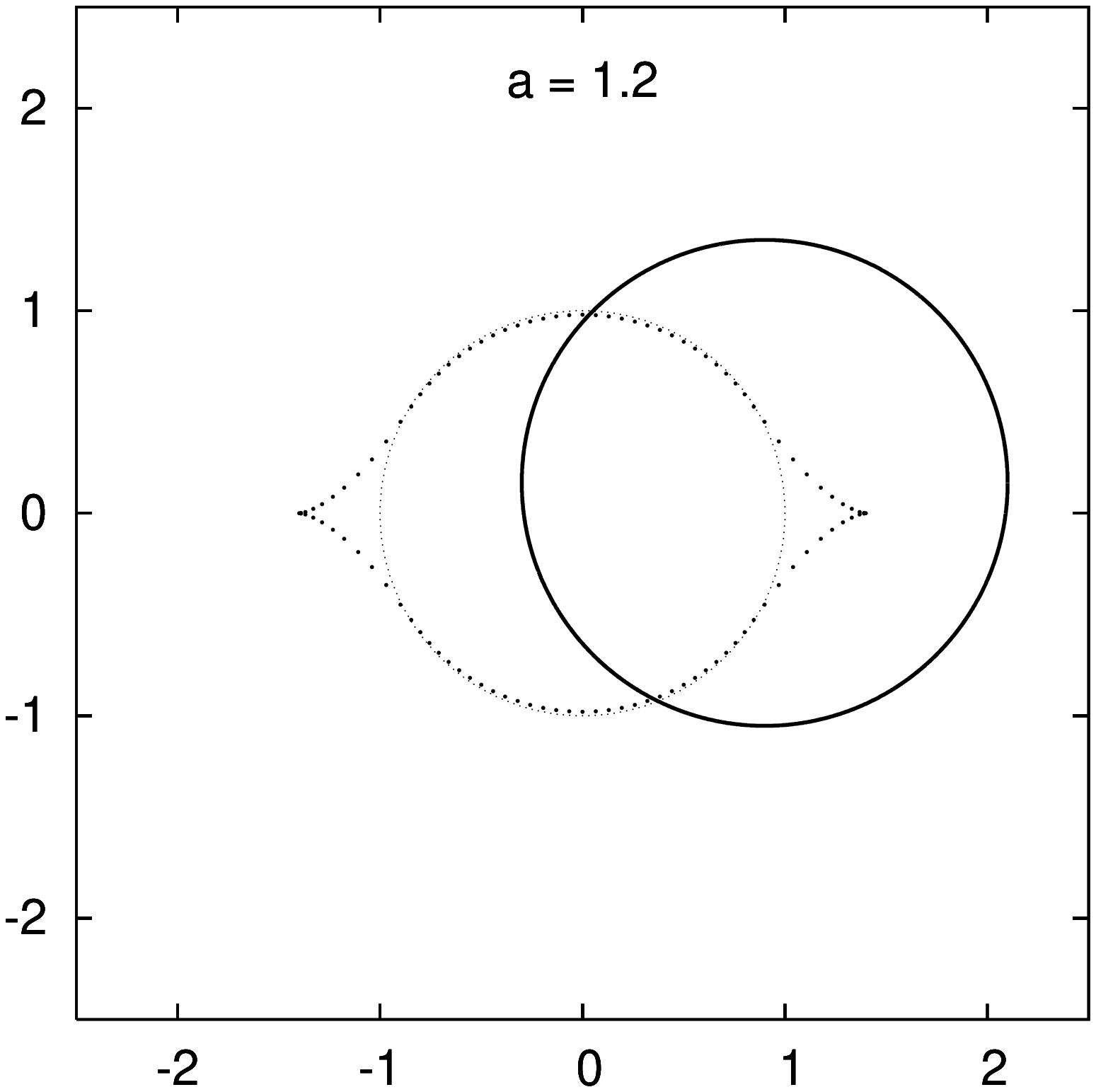}
\includegraphics[width=0.2\textwidth]{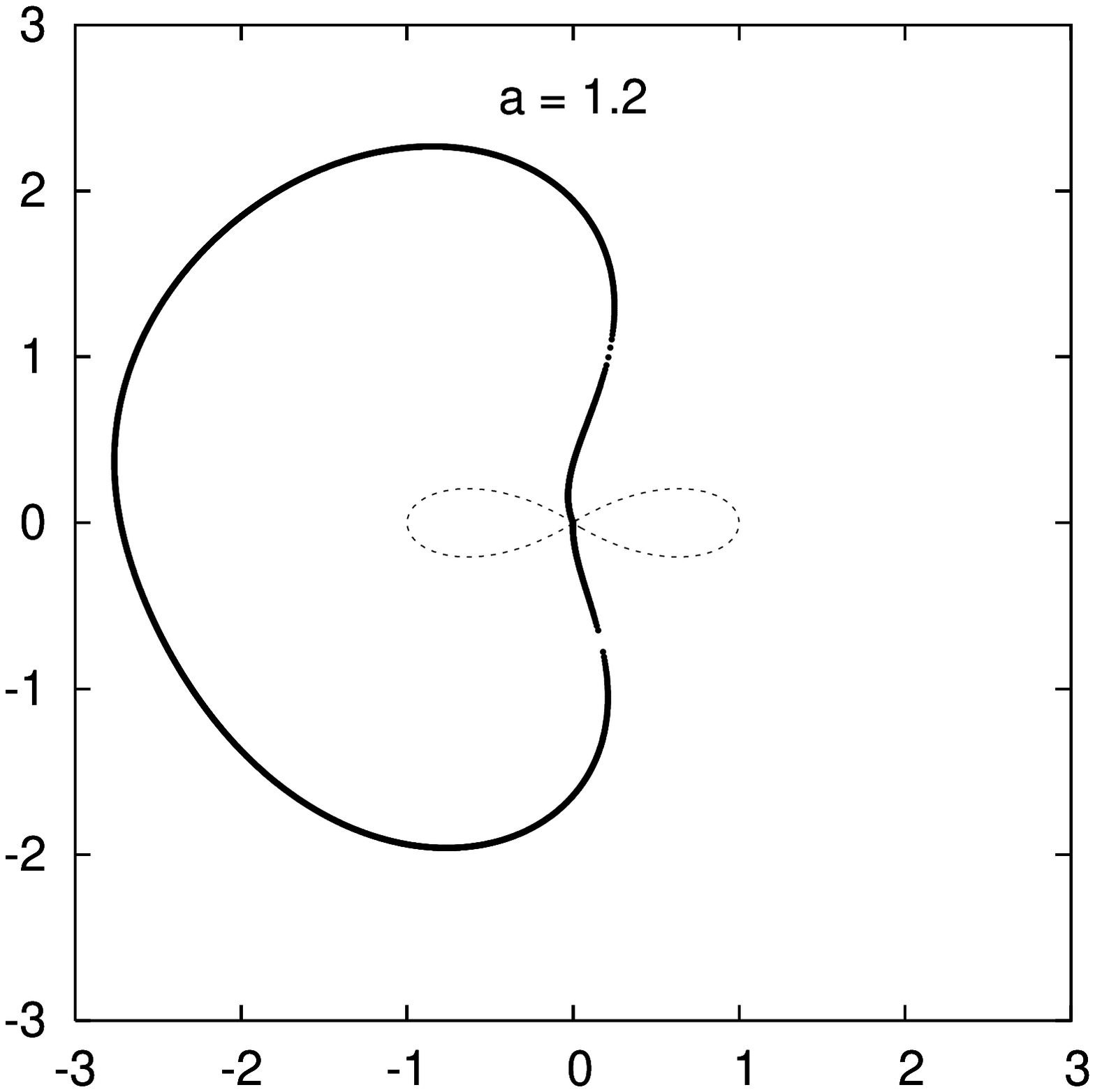}

\caption{
Similar to Fig.~\ref{fig:positiveimages}, but for negative parity.
In these examples, $\kappa=\gamma=0.7$, $(u_0,v_0)=(0.9,0.15)$, $b=1$,
and $a$ is increased from $0.04$ to $1.2$.
}\label{fig:negativeimages}
 \end{center}
\end{figure}

Fig.~\ref{fig:negativeimages} is similar to Fig.~\ref{fig:positiveimages}
except that the unperturbed image has negative parity with
$\kappa=\gamma=0.7$ ($\mu_0 = -2.5$), and we use a different source
position.  Between $a=0.04$ and $a=0.15$, the source crosses the caustic
separating the two and three-image regions, and the image configurations
show the appearance of a small third image near the origin.  By $a=0.2$,
the source has intersected the caustic again, which corresponds to the
merging of two of the images.  Increasing the source further results in
the growth and merger of the images, and by $a=1.2$ we are again beginning
to see the ellipse that would be expected from only the convergence and
shear field.

\subsection{Magnification of a finite source}
\label{sec:mag}

Having found the image configurations for finite sources, we now seek the
magnifications.  Since gravitational lensing conserves surface brightness,
the change of flux is due solely to the change in size of the source when
it is lensed.  If the source has a uniform surface brightness, then the
magnification is the ratio of the area of the image(s) to the area of the
source.

\begin{figure}
 \begin{center}
\includegraphics[width=0.45\textwidth]{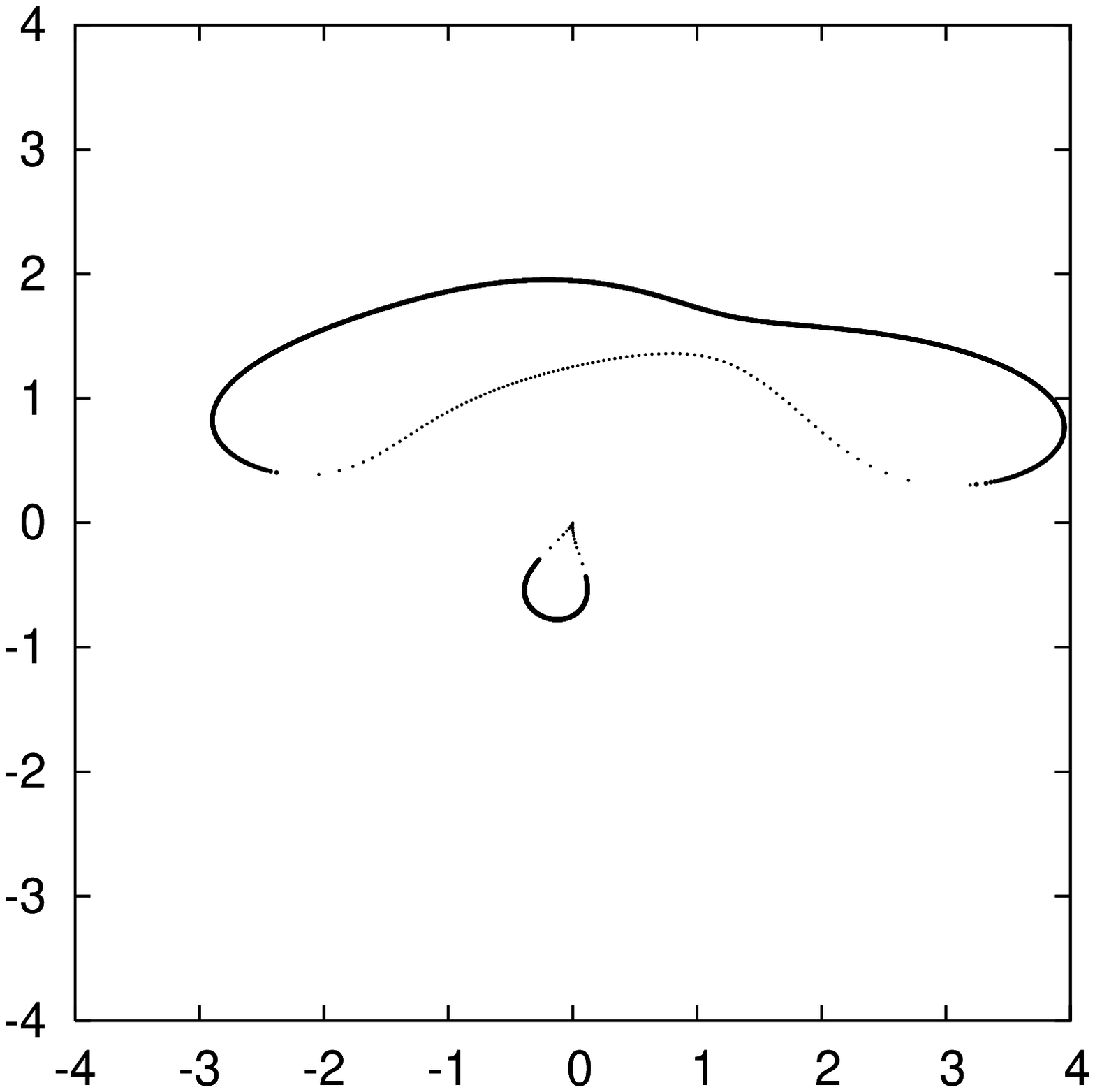}
\caption{
Image configuration for $\kappa=\gamma=0.3$, $(u_0,v_0)=(0.2,0.6)$,
$(a,b)=(0.1,1.0)$.  The solid line shows $r_{+}(\theta)$, which forms
the outer boundary of the images.  The dotted line shows $r_{-}(\theta)$,
which forms the inner boundary.
}\label{fig:outerinner}
 \end{center}
\end{figure}

Our parametric solution for the image boundaries allows us to compute
the image area, if we take care to understand the different solution
regimes.  Where $r_{+}(\theta)$ and $r_{-}(\theta)$ are real and
positive, they form the outer and inner boundaries (respectively) of
the images (see Fig.~\ref{fig:outerinner}).  The image area is then
\begin{equation}
  \frac{1}{2} \int_I \left[ r_{+}^2(\theta) - r_{-}^2(\theta) \right]\,
    d\theta\,,
\end{equation}
where $I$ is the range of $\theta$ over which the solution is defined
(i.e., where $B \ge 0$).  If only $r_{+}(\theta)$ is real and positive,
it forms the complete boundary of the image and the image area is
$\frac{1}{2} \int_I r_{+}^2(\theta)\,d\theta$.  Finally, where
$r_{+}(\theta)$ and $r_{-}(\theta)$ are both negative, there is no
contribution to the image area.  (Note that $r_{-} < r_{+}$ for all
parameter values and all $\theta$, so there is never an area contribution
due to $r_{-}(\theta)$ alone.)

Thus, the total magnification (ratio of image size to source size) can be
written as,
\begin{equation}
\mm \equiv \frac{\mu}{\mu_0}
= \frac{1}{2\pi \mu_0 a^2} \int d\theta \cases{
  r_{+}^2(\theta) - r_{-}^2(\theta) & if $B>0$ and $A > \sqrt{B}$ \cr
  r_{+}^2(\theta)                   & if $B>0$ and $A < \sqrt{B}$ \cr
  0                                 & if $B<0$ or  $A + \sqrt{B} < 0$
} \label{eq:mageqn}
\end{equation}
Since we are primarily interested in how much the flux \emph{changes}
due to the presence of the SIS perturber, we have normalized
eq.~(\ref{eq:mageqn}) with respect to the magnification $\mu_0$
produced by the convergence and shear alone.  We refer to this as
the `normalized magnification.' Unfortunately, the integral in
eq.~(\ref{eq:mageqn}) cannot be evaluated analytically.  Still, it
requires only a one-dimensional numerical integral, which means that
the analytic solution of the lens equation yields a much faster
calculation than a conventional two-dimensional numerical integral.
\citet{inoue} give analytic approximations for $\mm$ in the limit of
a large source (also see Appendix A), but we are interested in the
exact result for a wide range of source sizes.

We can now understand the effects that the mass sheet degeneracy
in the macromodel have on the substructure analysis.  Adding a
mass sheet rescales the macromodel as shown in eq.~(\ref{eq:sheet}).
From eqs.~(\ref{eq:macromagnification}) and (\ref{eq:boundary}), we
then see that $\mu_0$ and $r_{\pm}^2$ are both rescaled as
$(1-\kappa_{\rm sheet})^{-2}$.  As a result, the normalized
magnification $\mm$ is \emph{unchanged} by the addition of the mass
sheet.  We conjecture that this result is special to the SIS clump
model, and it would be interesting to consider other clump models.
However, for our purposes the remarkable implication is that our
substructure analysis is completely unaffected by the mass sheet
degeneracy.

\begin{figure}
 \begin{center}
\includegraphics[width=0.55\textwidth]{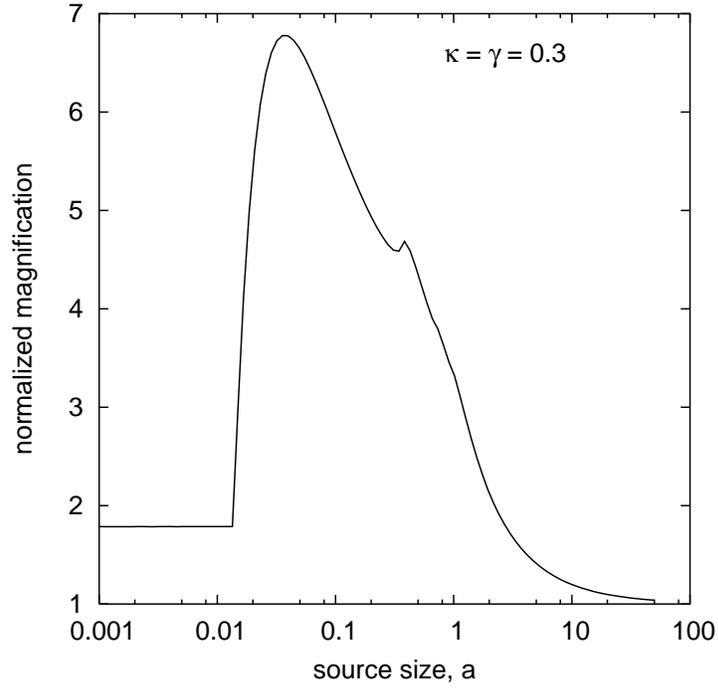}
\caption{
Normalized magnification as a function of source size for
$\kappa=\gamma=0.3$, $(u_0,v_0)=(0.2,0.6)$, $b=1$.  The sharp increase
at $a \approx 0.015$ corresponds to a transition from two images to
three (see text).
}\label{fig:magnificationcurve}
 \end{center}
\end{figure}

Fig.~\ref{fig:magnificationcurve} shows the normalized magnification
as a function of source size for the source from
Fig.~\ref{fig:positiveimages}.  The curve has several notable features
that can be understood in terms of the image configurations in
Fig.~\ref{fig:positiveimages}.  First, for $a \la 0.01$ the source
does not come into contact with the caustics and the magnification is
basically independent of source size.  The sharp increase in magnification
between $a=0.01$ to $0.02$ corresponds to the appearance of a third image
as the source begins to cross the fold caustic.   The magnification then
comes to a large peak, followed by a smaller peak near $a = 0.8$.
Comparing to the fourth row of Fig.~\ref{fig:positiveimages}, we see
that this secondary peak occurs when the source begins to come into
contact with the caustic cusps.  Finally, as $a$ becomes large, the
magnification approaches unity; that is, for large sources the effect
of the SIS perturber becomes negligible, as expected.  Nevertheless, it
should be noted that even at $a/b=50$, the image flux is still perturbed
by 3.7 per cent.

\begin{figure}
 \begin{center}
\includegraphics[width=0.6\textwidth]{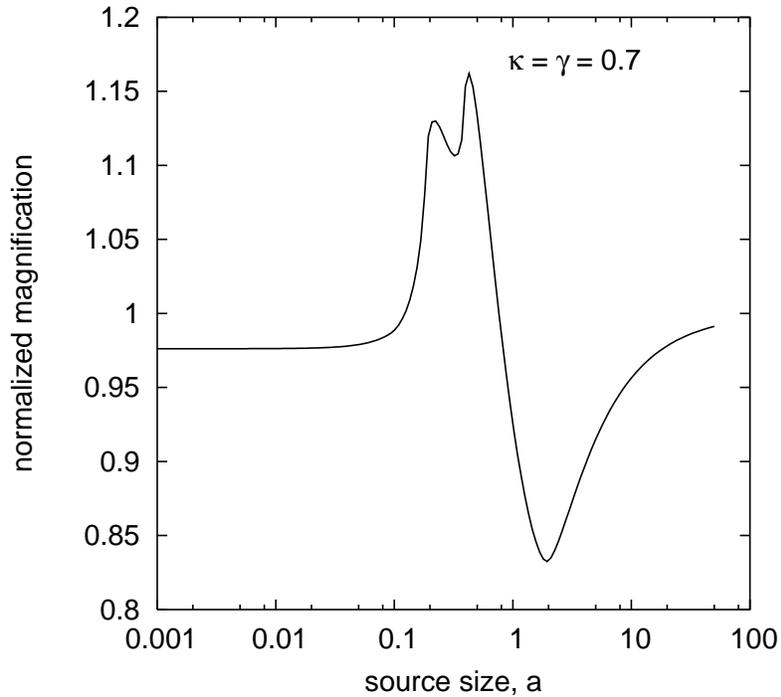}
\caption{
Similar to Fig.~\ref{fig:magnificationcurve}, but for $\kappa=\gamma=0.7$,
$(u_0,v_0)=(0.9,0.15)$, $b=1$.  The features of the curve are due to
the source coming into contact with the caustics, as shown in
Fig.~\ref{fig:negativeimages}.
}\label{fig:negmagcurve}
 \end{center}
\end{figure}

Fig.~\ref{fig:negmagcurve} shows the normalized magnification as a
function of $a$ for the source from Fig.~\ref{fig:negativeimages}.  
The parameters here are the same as the previous figure except that
this case has negative parity ($\kappa=\gamma=0.7$).  For an SIS clump
in front of a negative parity image, most source positions yield
\emph{de}magnification relative to the background convergence and shear
field \citep{analytics}.  Indeed, for this position the normalized
magnification is less than 1 for most source sizes.  However, there is
a region of magnification (relative to the convergence and shear field)
for $0.1 \la a \la 1.0$.  We can again match some of the features of
this plot to the image configurations from Fig.~\ref{fig:negativeimages}.
There is very little change in the normalized magnification until
$a \approx 0.1$, which corresponds to the source coming into contact
with the caustic.  There is a peak at $a \approx 0.2$ as the source
starts to come into contact with the caustic cusp, followed by a
shallow dip at $a \approx 0.3$ as the source begins to occupy more of
the demagnification region of the source plane.  As the source size
is increased further, there is then another maximum followed by a
minimum at $a \approx 2$.  The normalized magnification for this
negative parity case also approaches unity for large $a$, with the
flux at $a/b = 50$ differing from unity by 1 per cent.  Comparing the
positive and negative parity cases gives the interesting result that,
at large $a$, an SIS perturber has less effect on the magnification of
a negative parity image than on an equivalent positive parity image.

\begin{figure}
 \begin{center}
\includegraphics[width=0.2\textwidth]{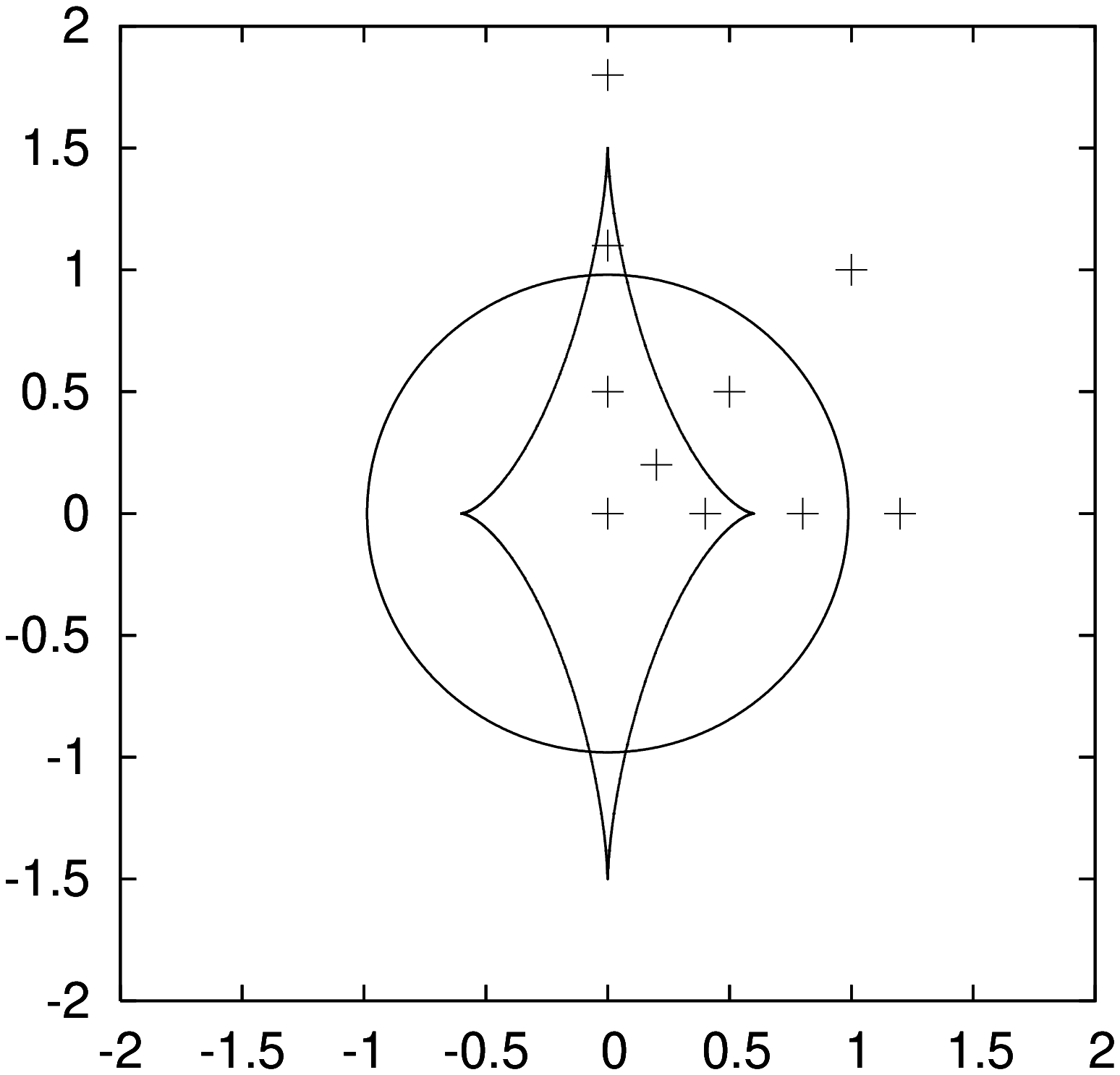}
\includegraphics[width=0.2\textwidth]{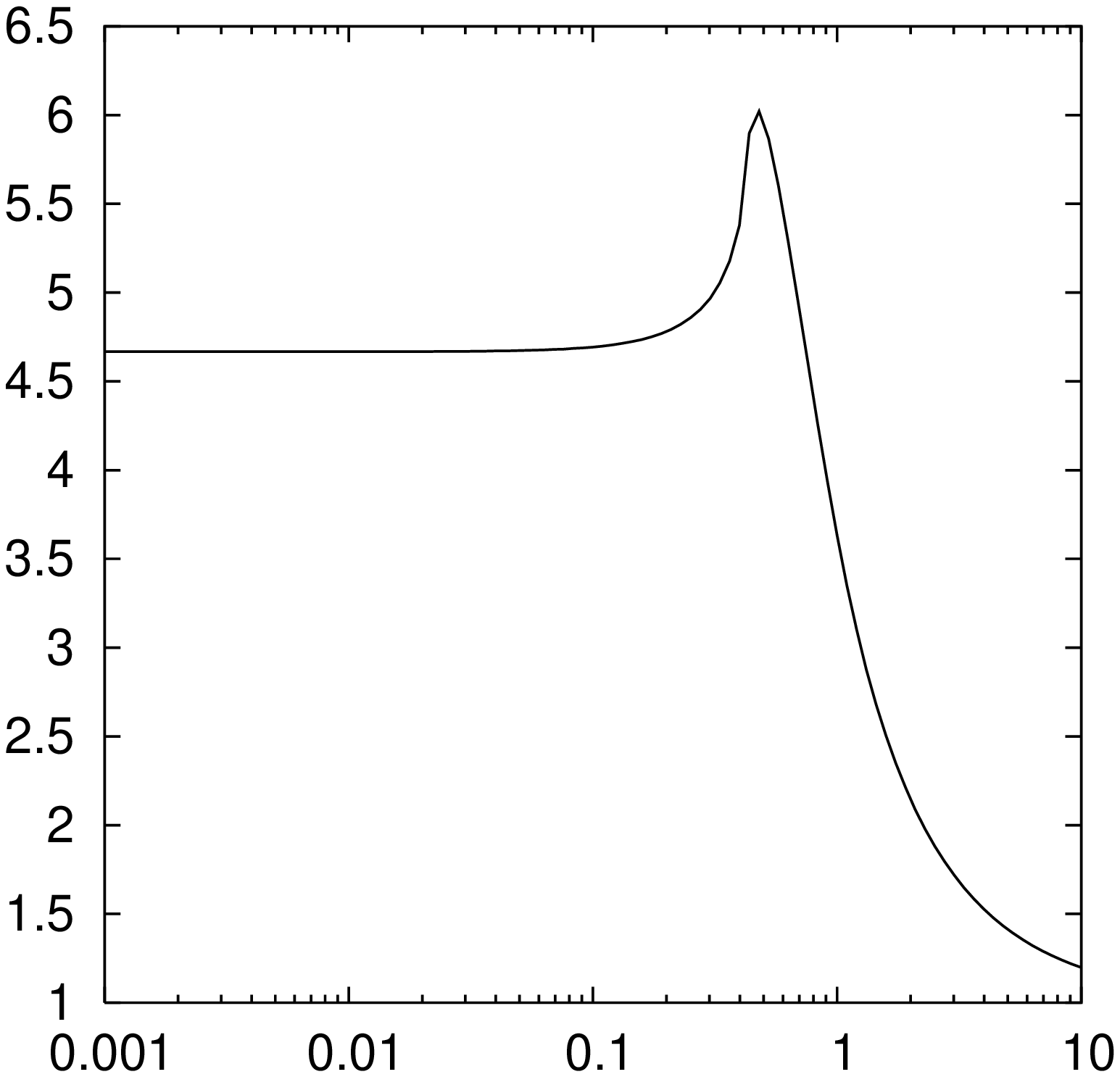}
\linebreak
\linebreak
\includegraphics[width=0.2\textwidth]{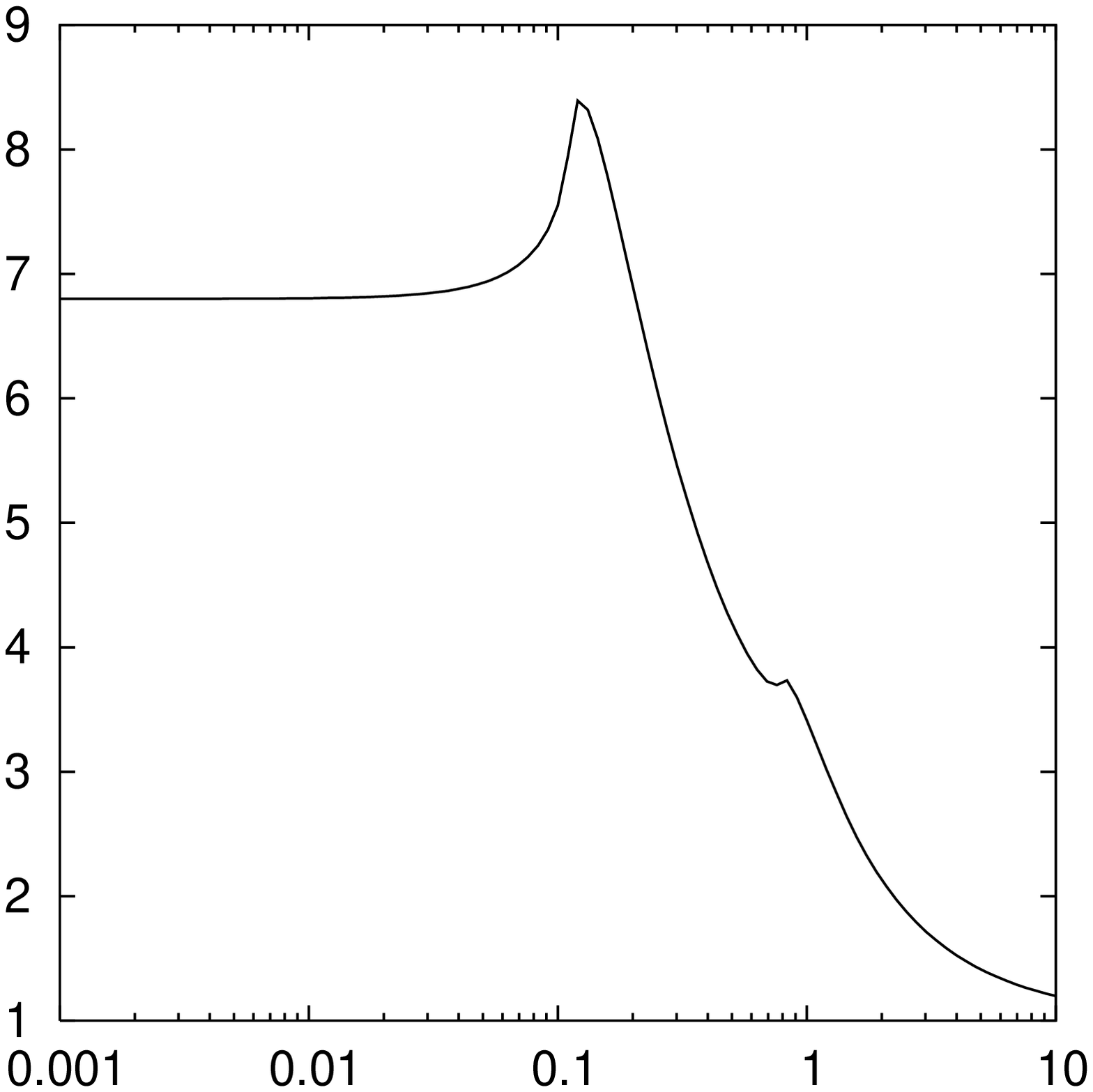}
\includegraphics[width=0.2\textwidth]{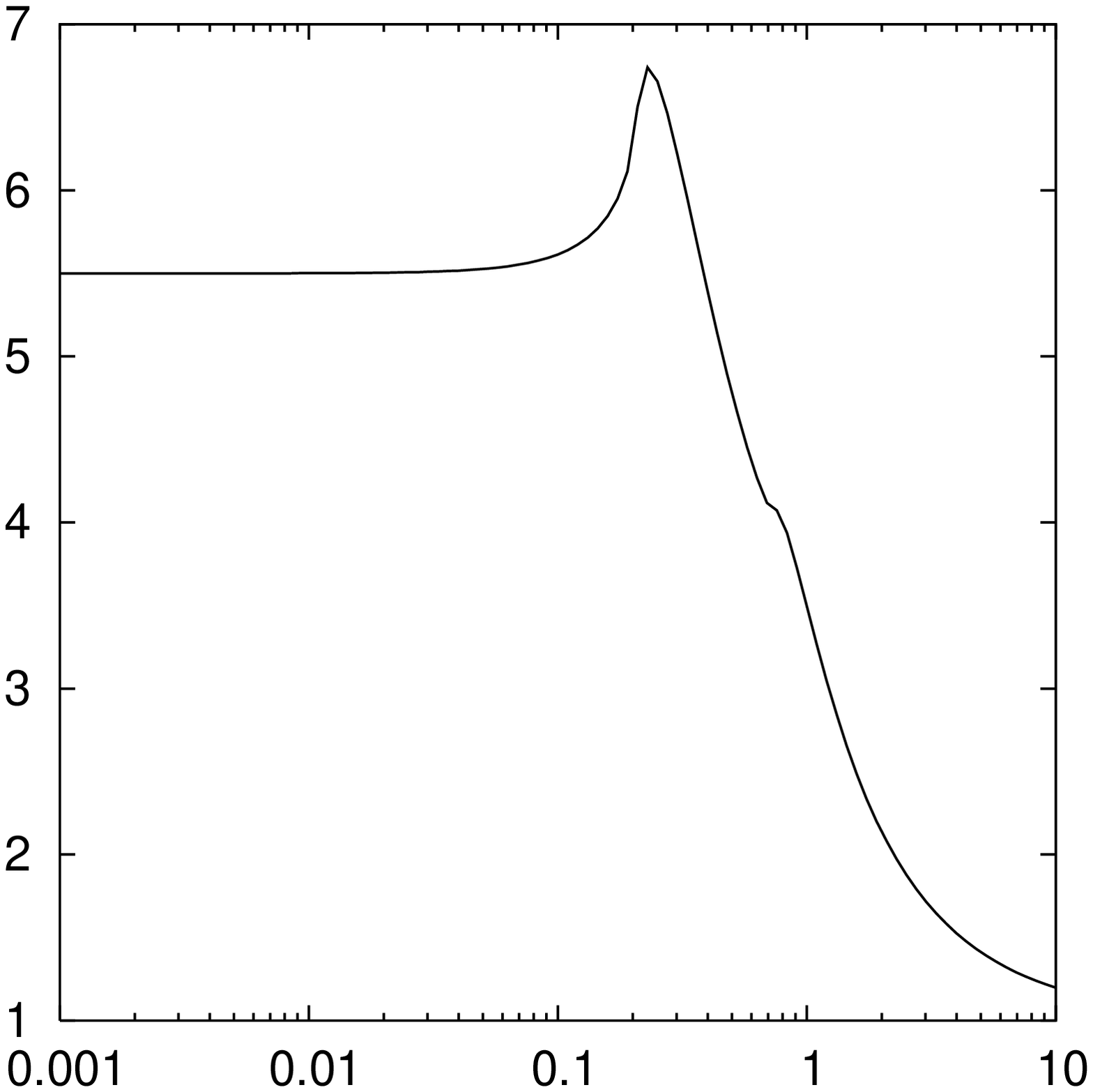}
\includegraphics[width=0.2\textwidth]{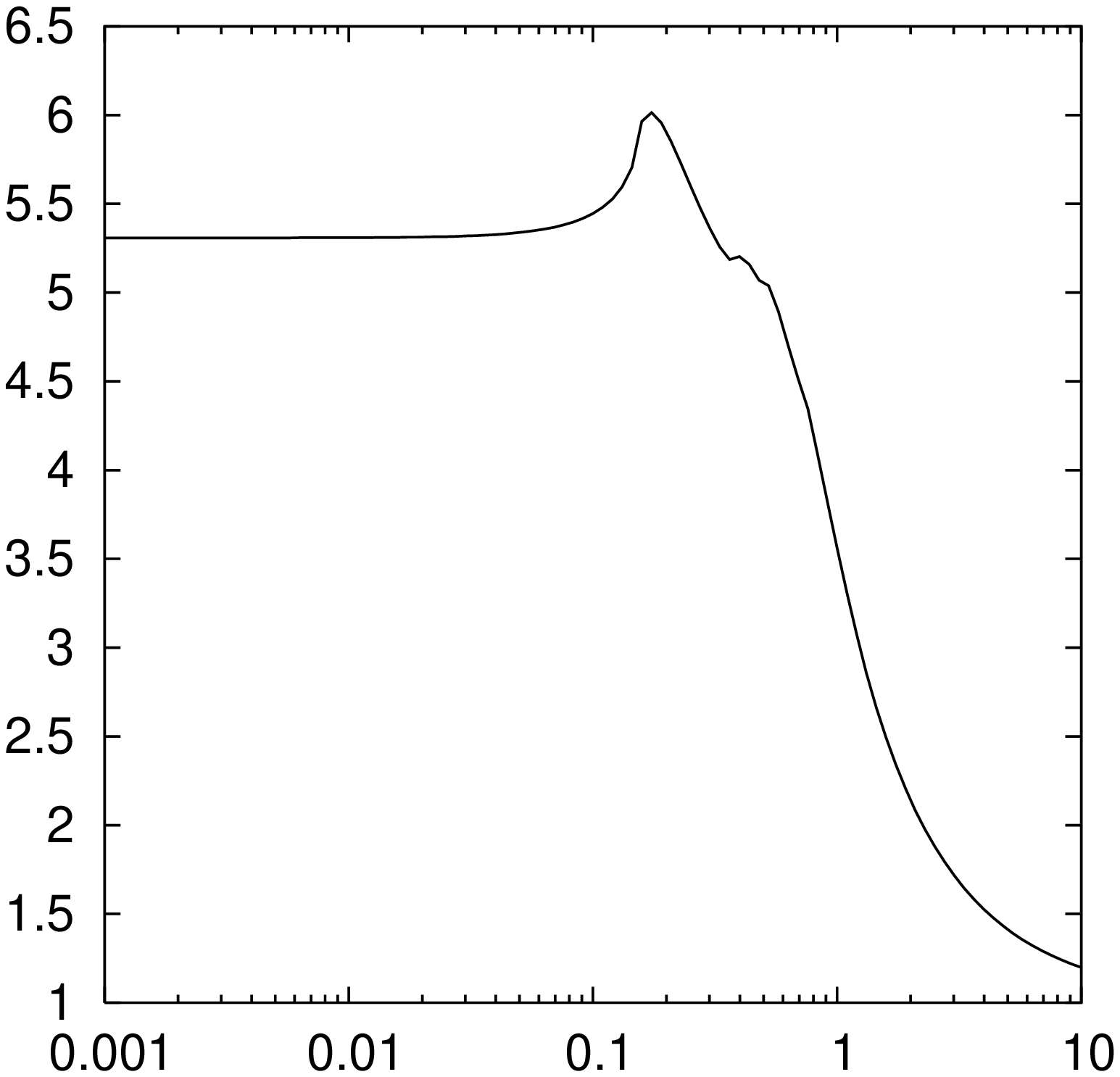}
\linebreak
\includegraphics[width=0.2\textwidth]{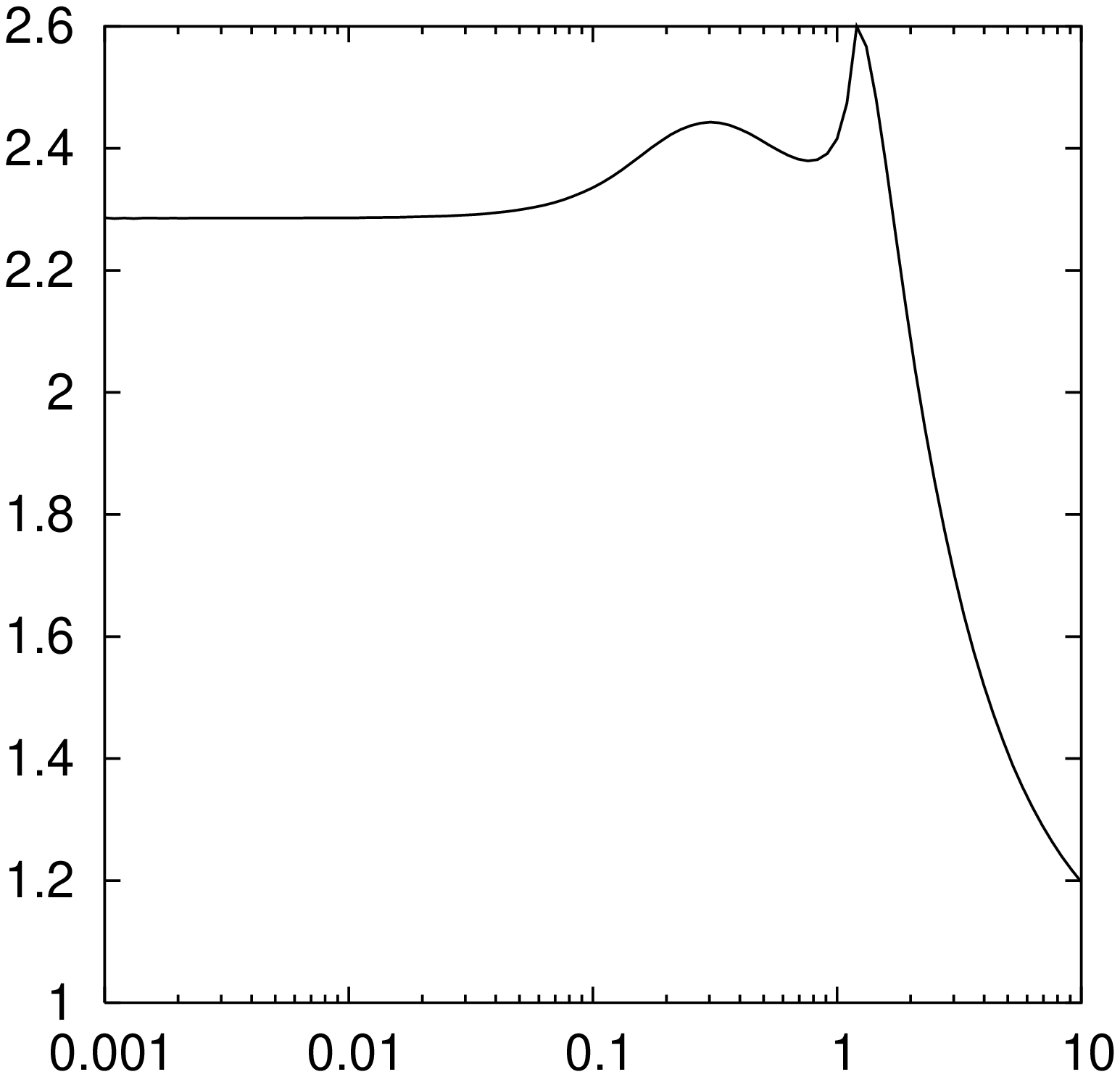}
\includegraphics[width=0.2\textwidth]{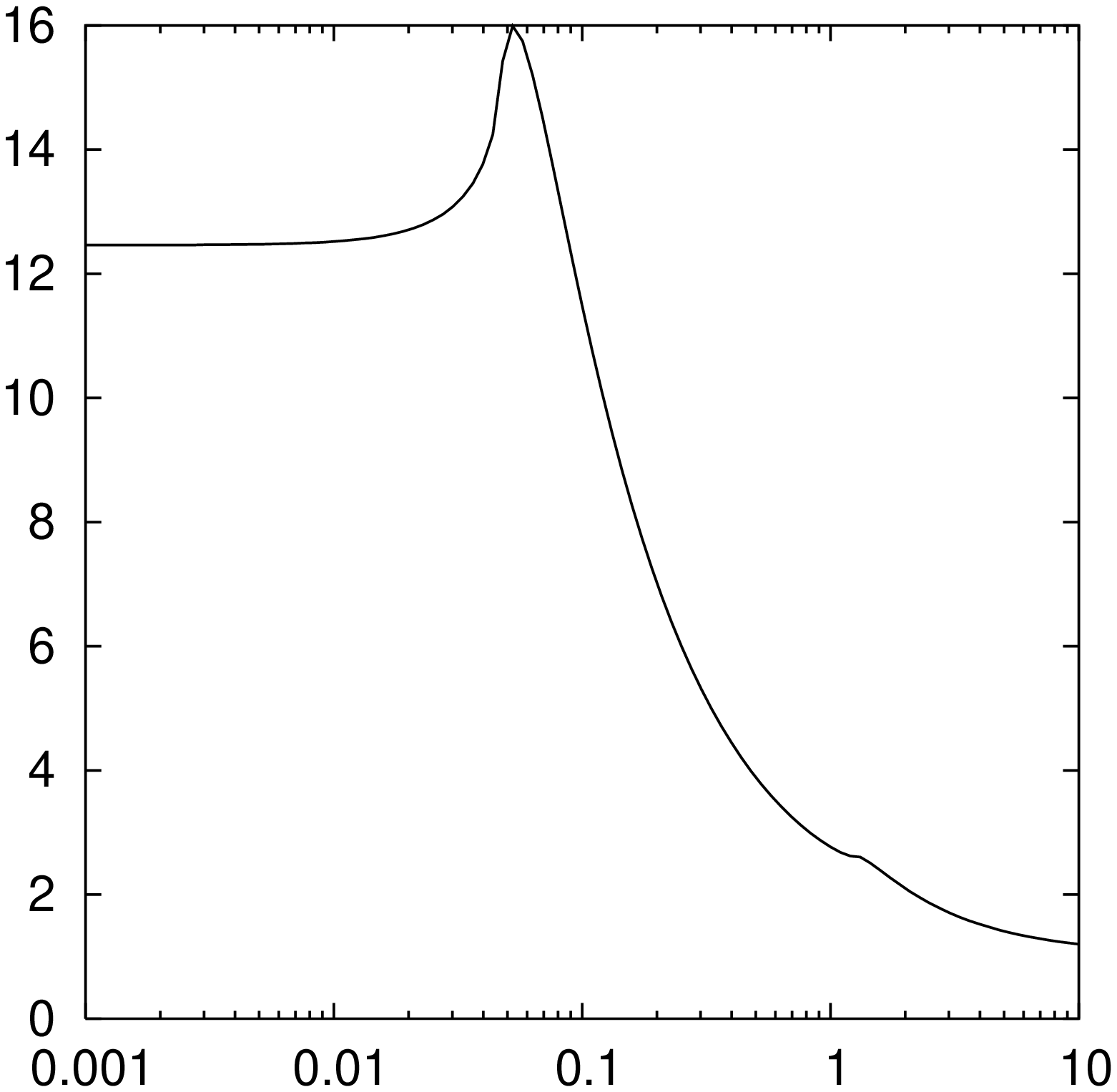}
\includegraphics[width=0.2\textwidth]{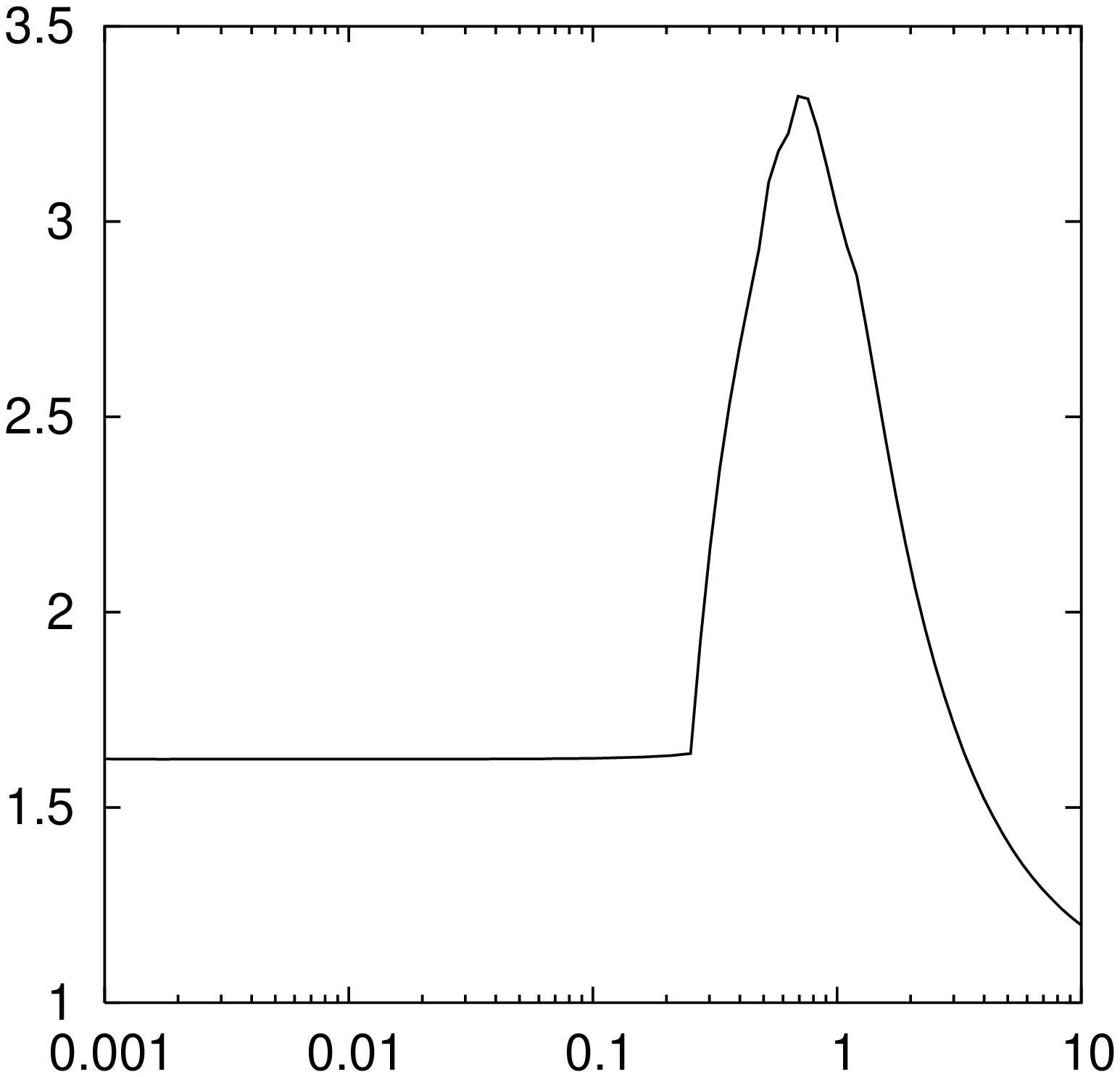}
\linebreak
\includegraphics[width=0.2\textwidth]{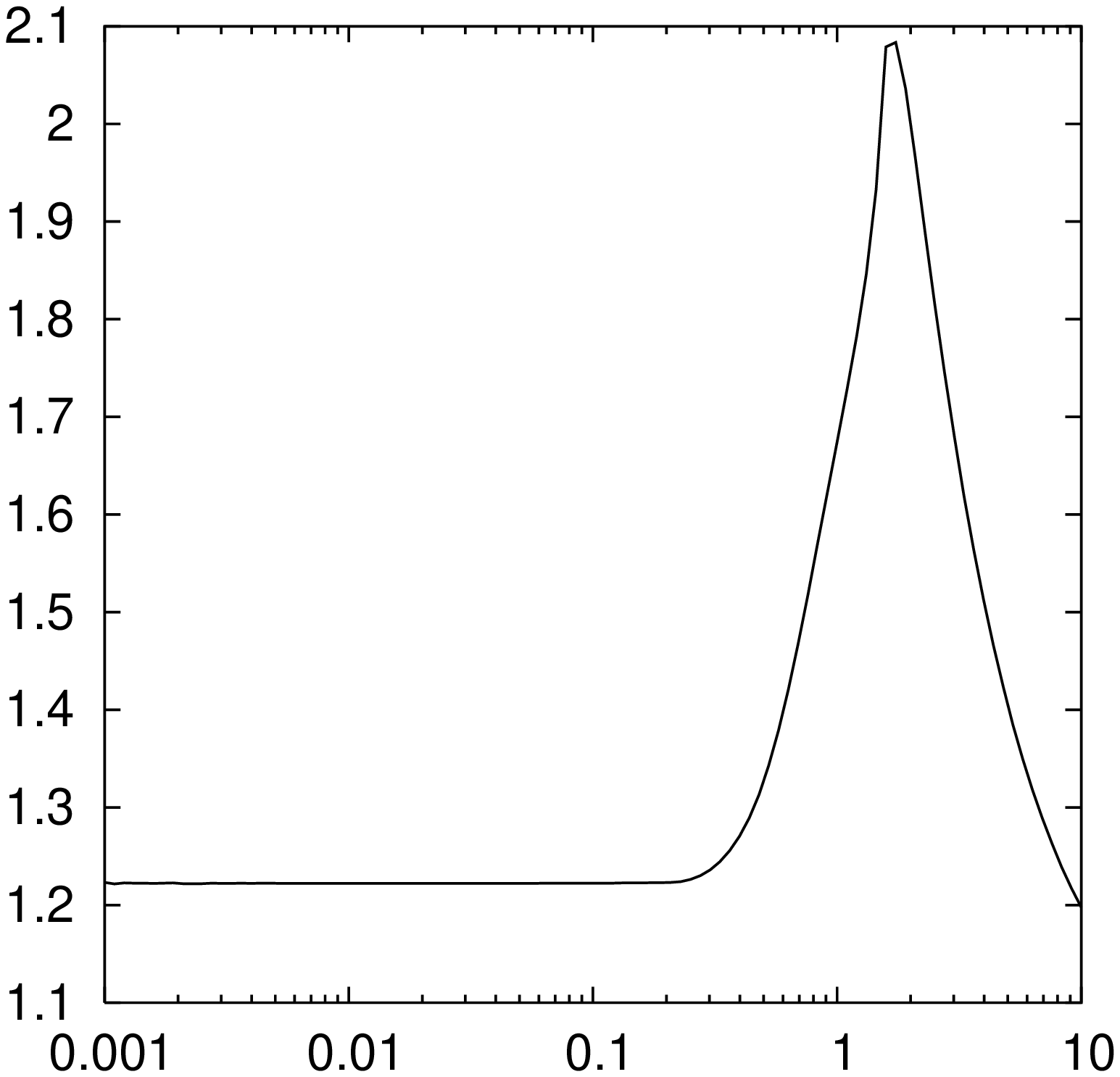}
\includegraphics[width=0.2\textwidth]{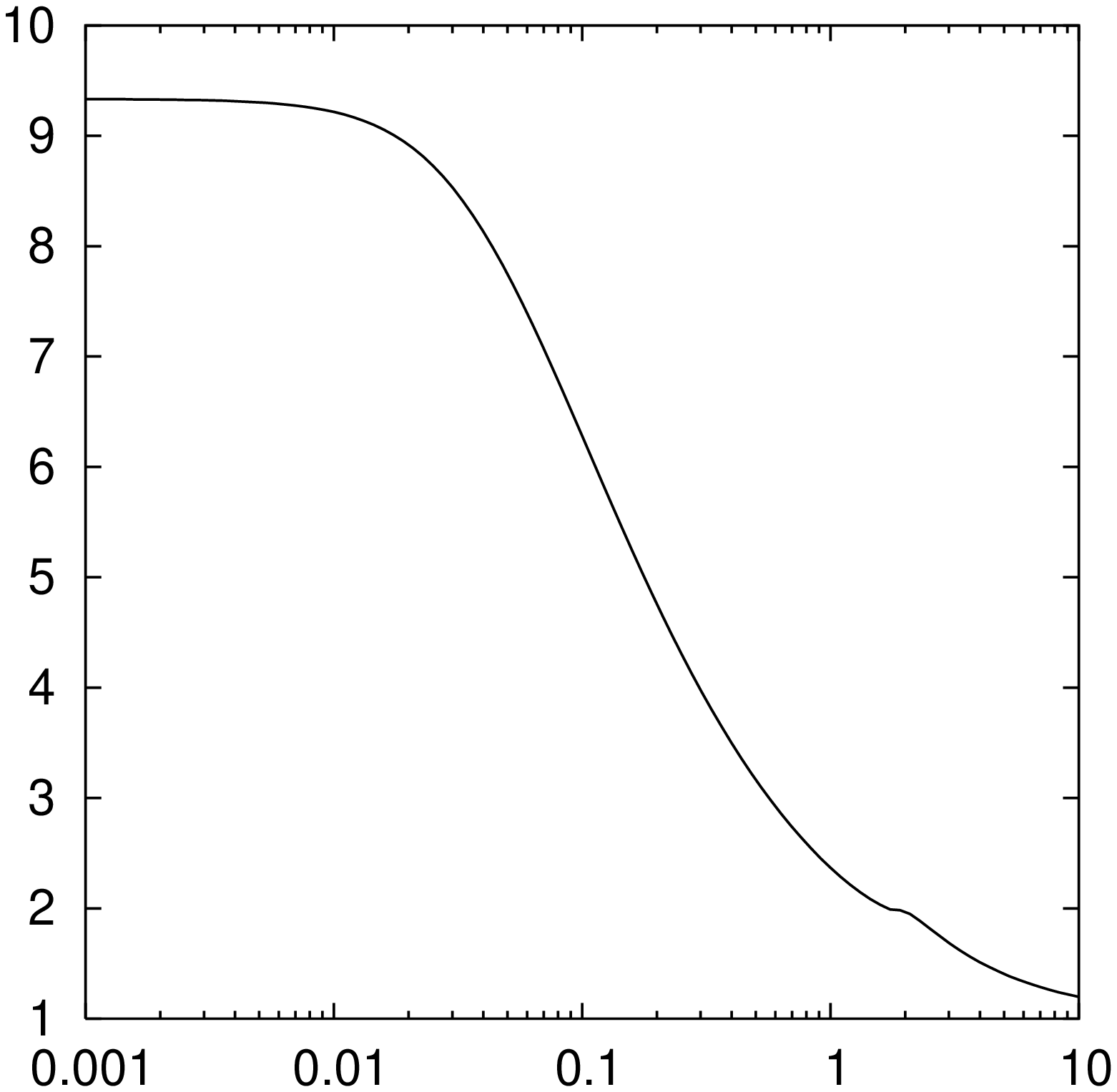}
\includegraphics[width=0.2\textwidth]{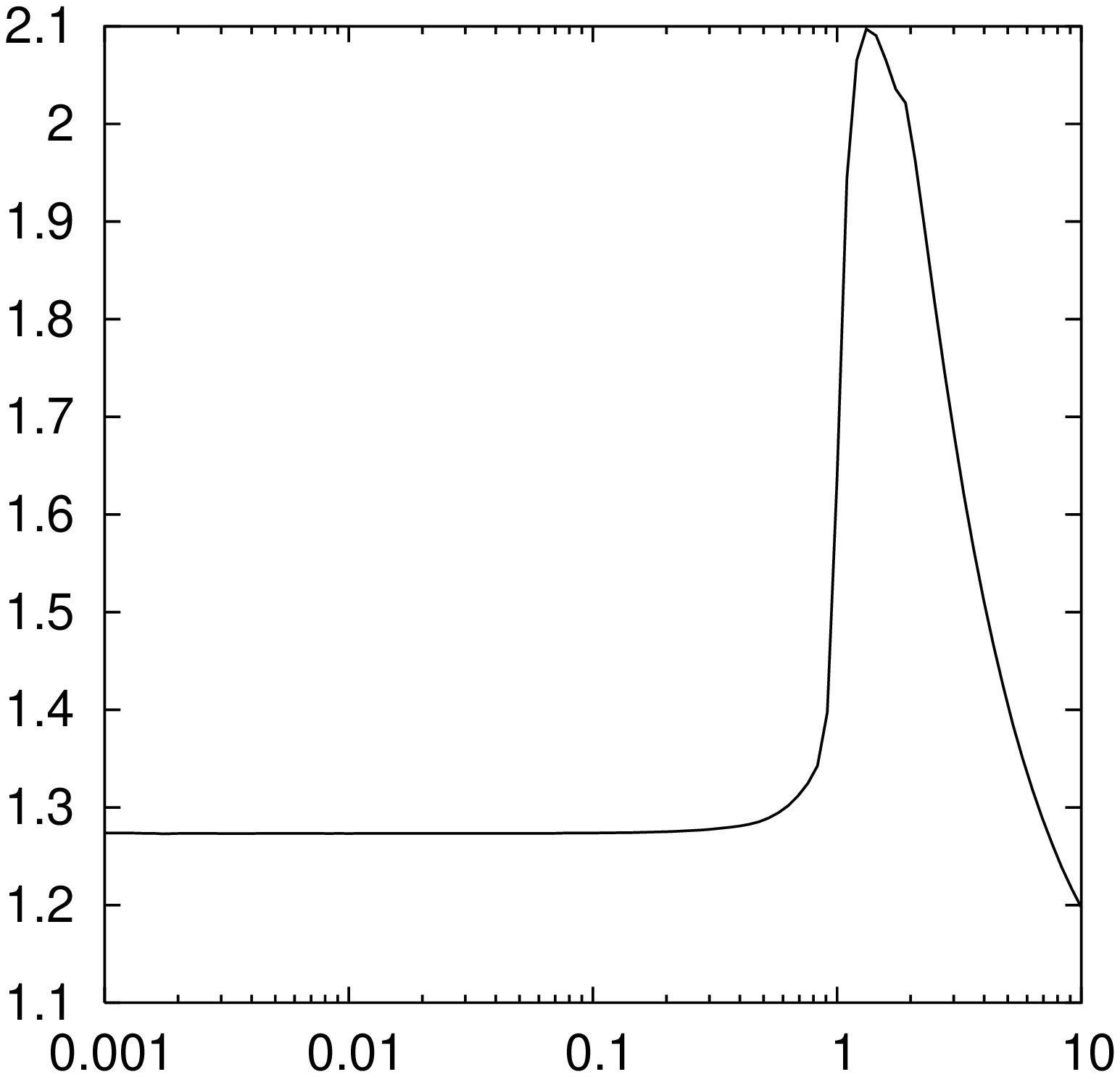}
\caption{
Normalized magnification as a function of source size for a positive
parity case ($\kappa = \gamma = 0.3$), with $b=1$.  The top left panel
shows the caustics and source positions.  The top right panel shows the
magnification curve for a source at the origin.  The left column represents
moving the source along u-axis, $u_0=0.4,0.8,1.2$ (top to bottom); the 
middle column represents moving along the v-axis, $v_0=0.5,1.1,1.8$; and
the right column represents moving along the line $v_0=u_0$, with
$u_0=0.2,0.5,1.0$.
}\label{fig:poscurves}
 \end{center}
\end{figure}

Since the magnification is largely determined by encounters with
caustics, we now study how the magnification versus source size curve
changes as the source \emph{position} is varied.  Fig.~\ref{fig:poscurves}
shows the normalized magnification curves for different source positions,
for a positive parity case ($\kappa=\gamma=0.3$).  The upper left panel
shows the caustics and source positions for the plot's other panels,
and the upper right panel shows the normalized magnification curve for
the source located at the origin.

Most of these curves have at least one, and in some cases two, peaks
where the normalized magnification increases sharply.  These peaks are,
in general, associated with the source boundary crossing a caustic.
It should also be noted that, while some peak heights are relatively
low with a normalized magnification of around 2, the normalized
magnification can become as high as about 16 (second row middle column).
This case is particularly illuminating as the main peak turns out to
coincide with the source crossing the upper cusp caustic while the
secondary peak around $a \sim 1.0$ corresponds to the source coming
into contact with the left and right cusps.  The next panel (row three
column two) is also notable for the plateau at low values of $a$, and
again there is a small secondary peak around $a \sim 1$--$2$ that
corresponds to the source coming into contact with the left and right
cusps.  Finally, two general features of the plots are striking.  First,
all of the curves remain fairly constant at small $a$ where the source
does not intersect the caustics of the SIS, implying that the source
does not `feel' the structure of the perturber before it comes into
contact with these caustics.  Second, although all of the plots tend
towards unity as expected for a large source, they do seem to deviate
from unity fairly uniformly at $a \ga 10$; in particular, all of the
magnifications are $\mm \simeq 1.2$ at $a=10$.  In Appendix A we
formalize this result by showing that the normalized magnification is
independent of source position, to first order in $1/a$.  Since it is
possible to measure flux ratios with percent-level precision (after
correcting for time delays; see \citealt{fas1608}), an important
implication is that even large sources relatively far from the mass
clump can be perturbed at a detectable level.

\begin{figure}
 \begin{center}
\includegraphics[width=0.2\textwidth]{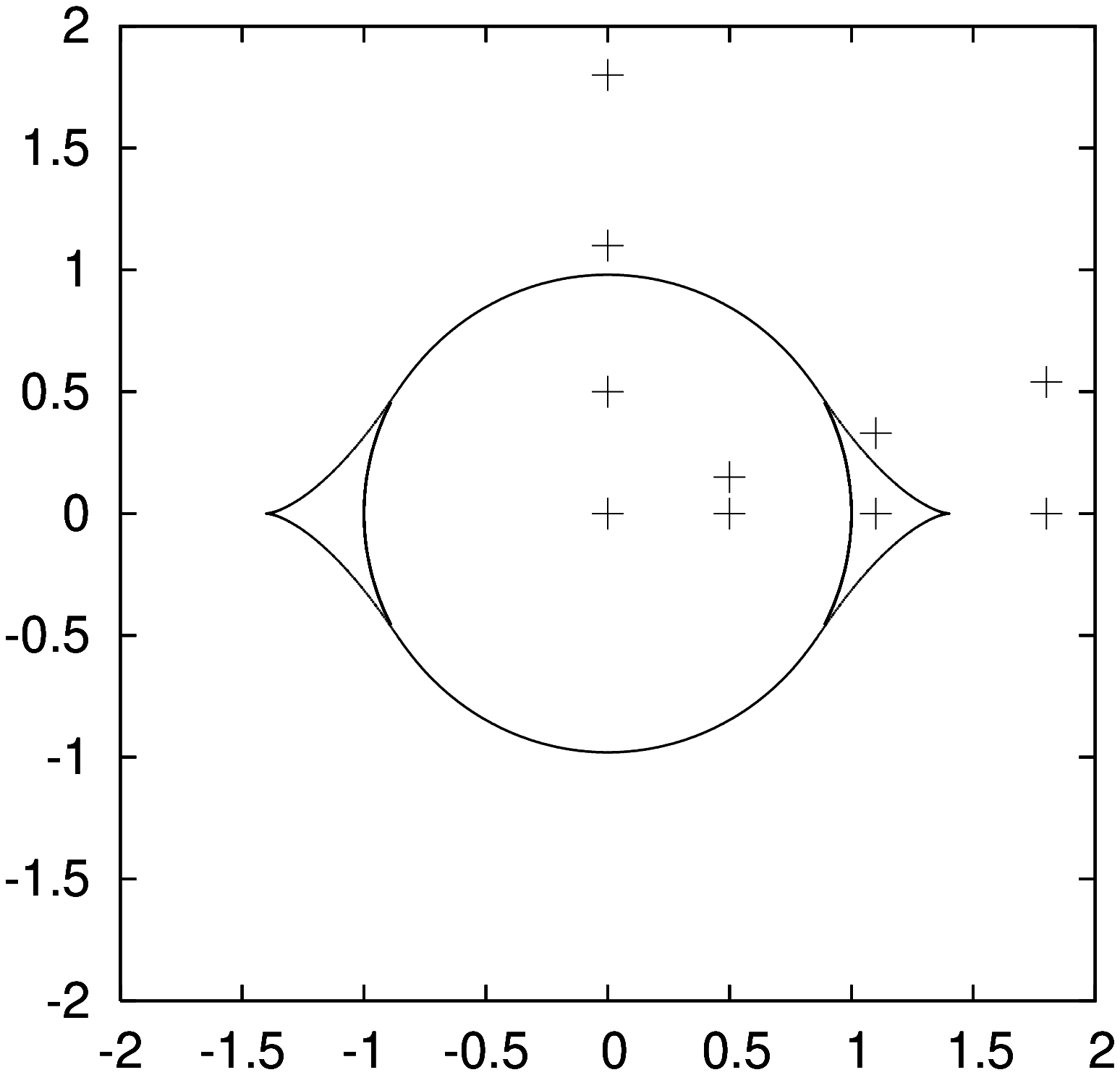}
\includegraphics[width=0.2\textwidth]{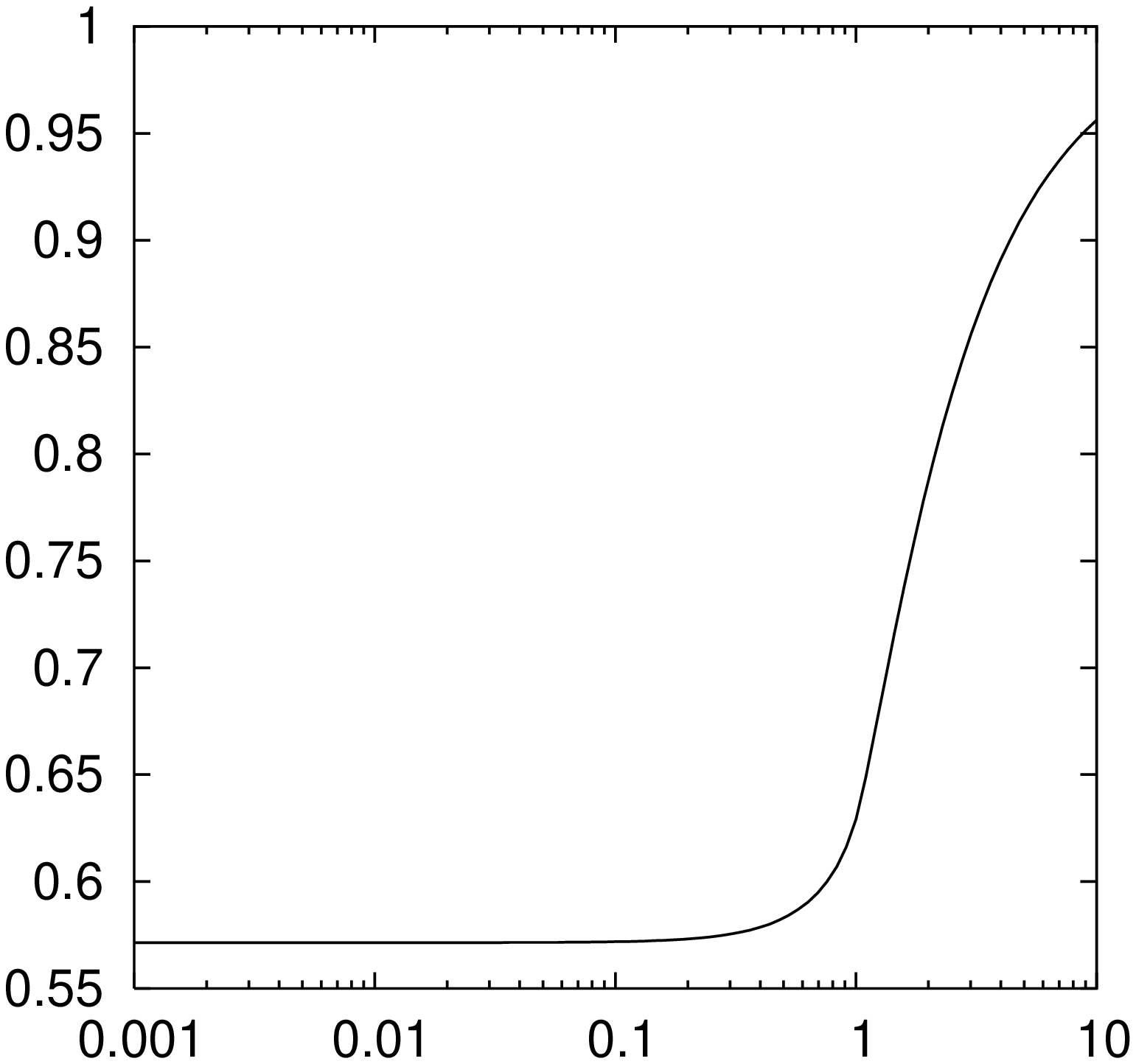}
\linebreak
\linebreak
\includegraphics[width=0.2\textwidth]{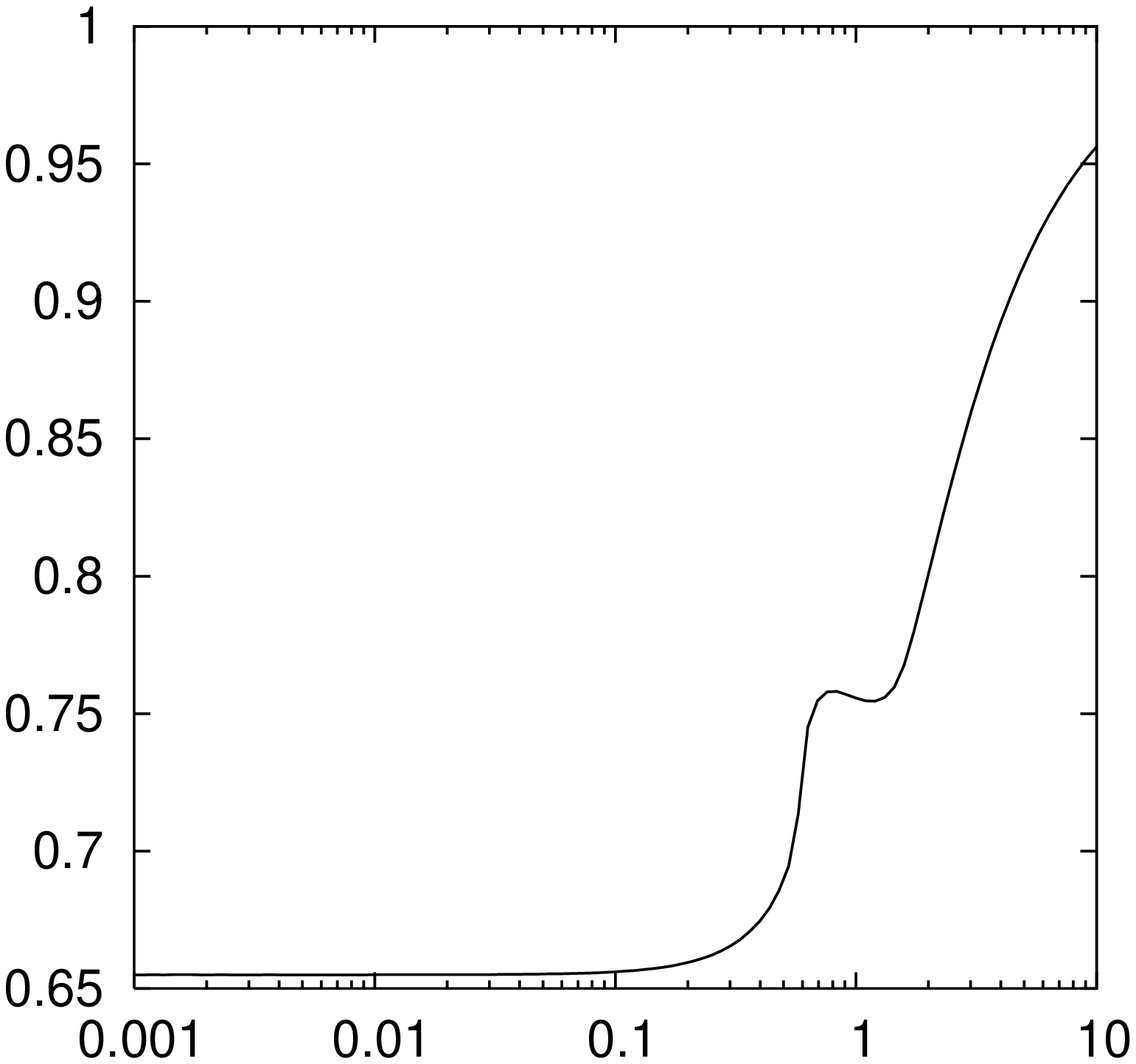}
\includegraphics[width=0.2\textwidth]{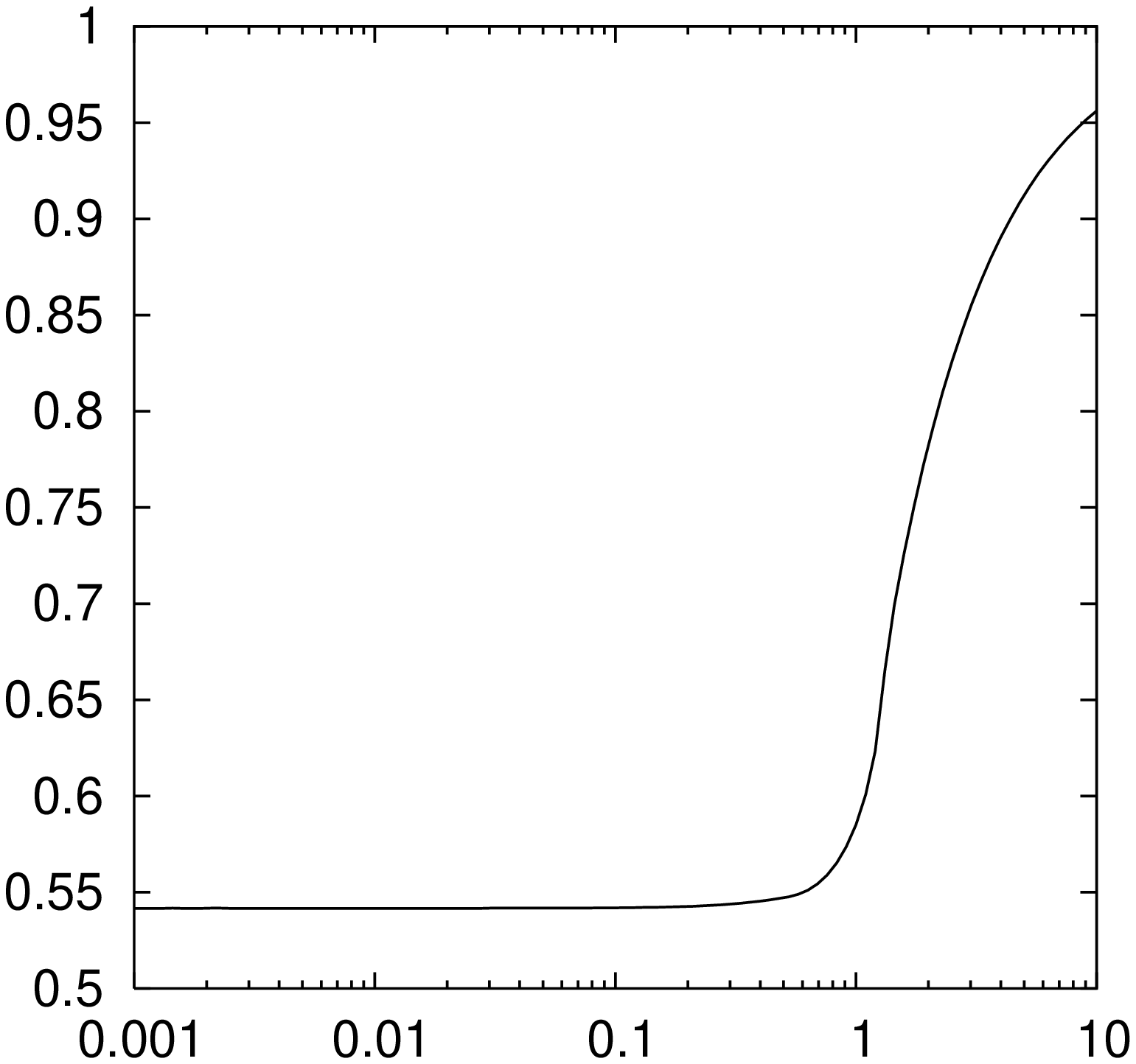}
\includegraphics[width=0.2\textwidth]{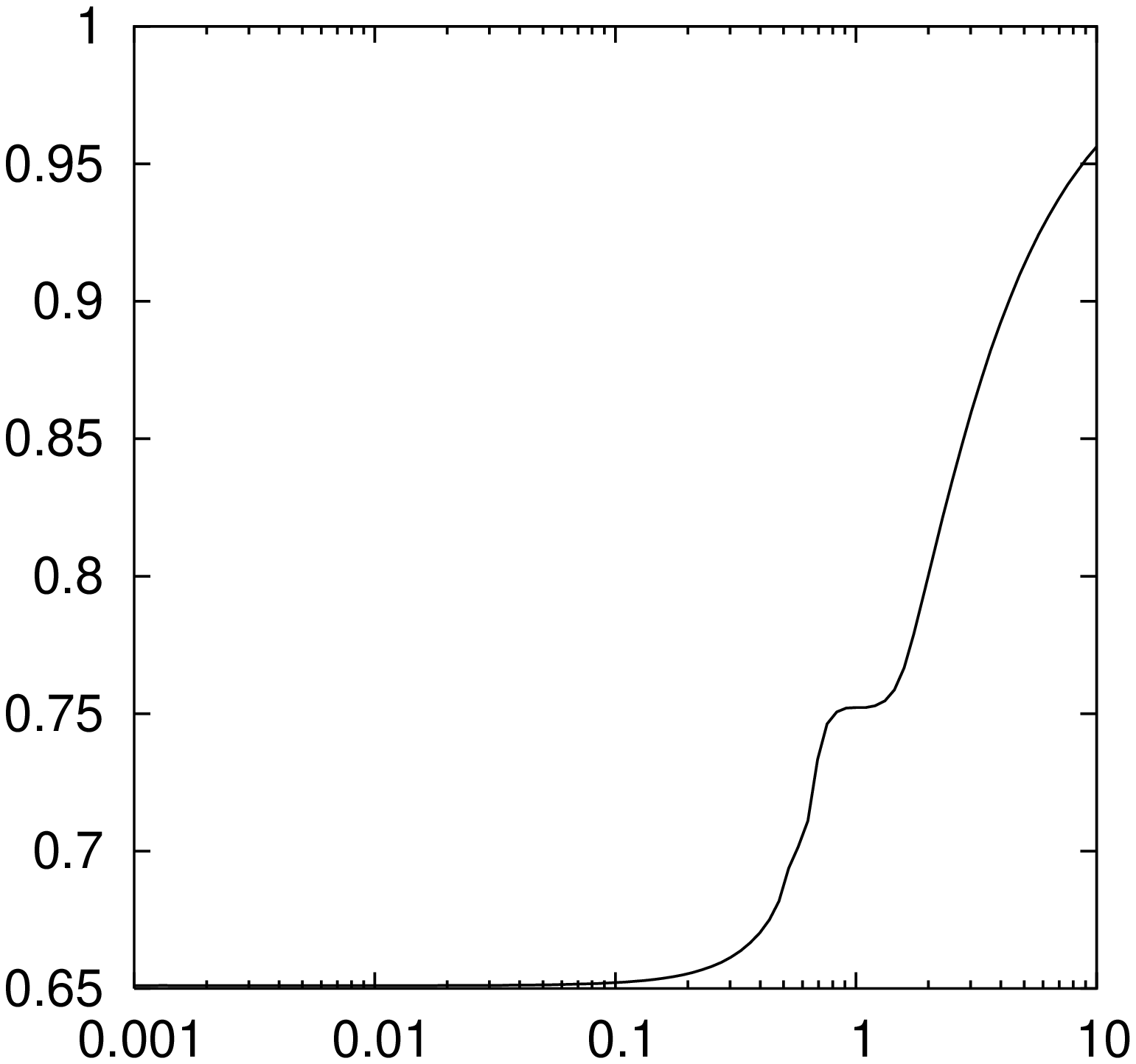}
\linebreak
\includegraphics[width=0.2\textwidth]{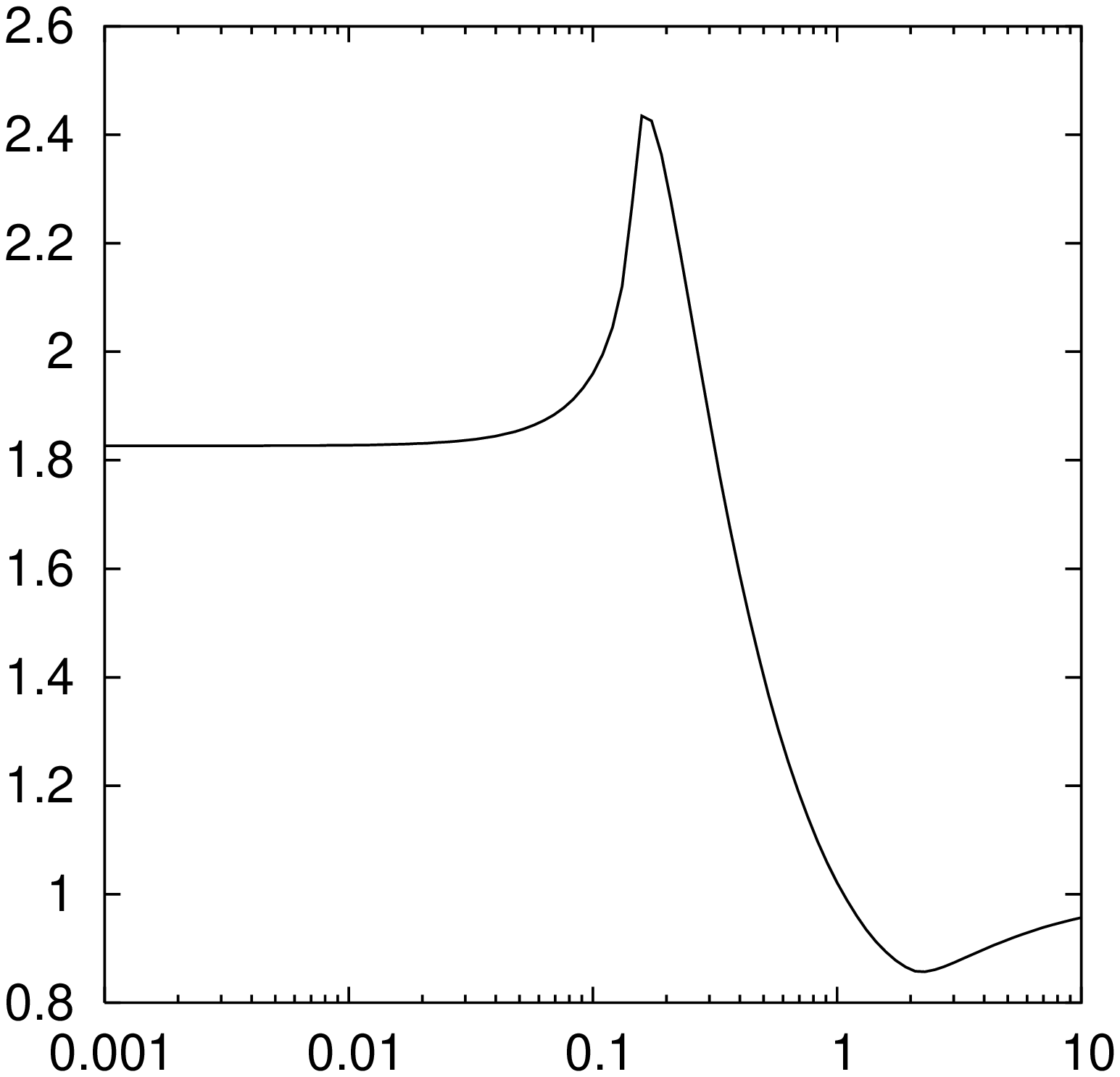}
\includegraphics[width=0.2\textwidth]{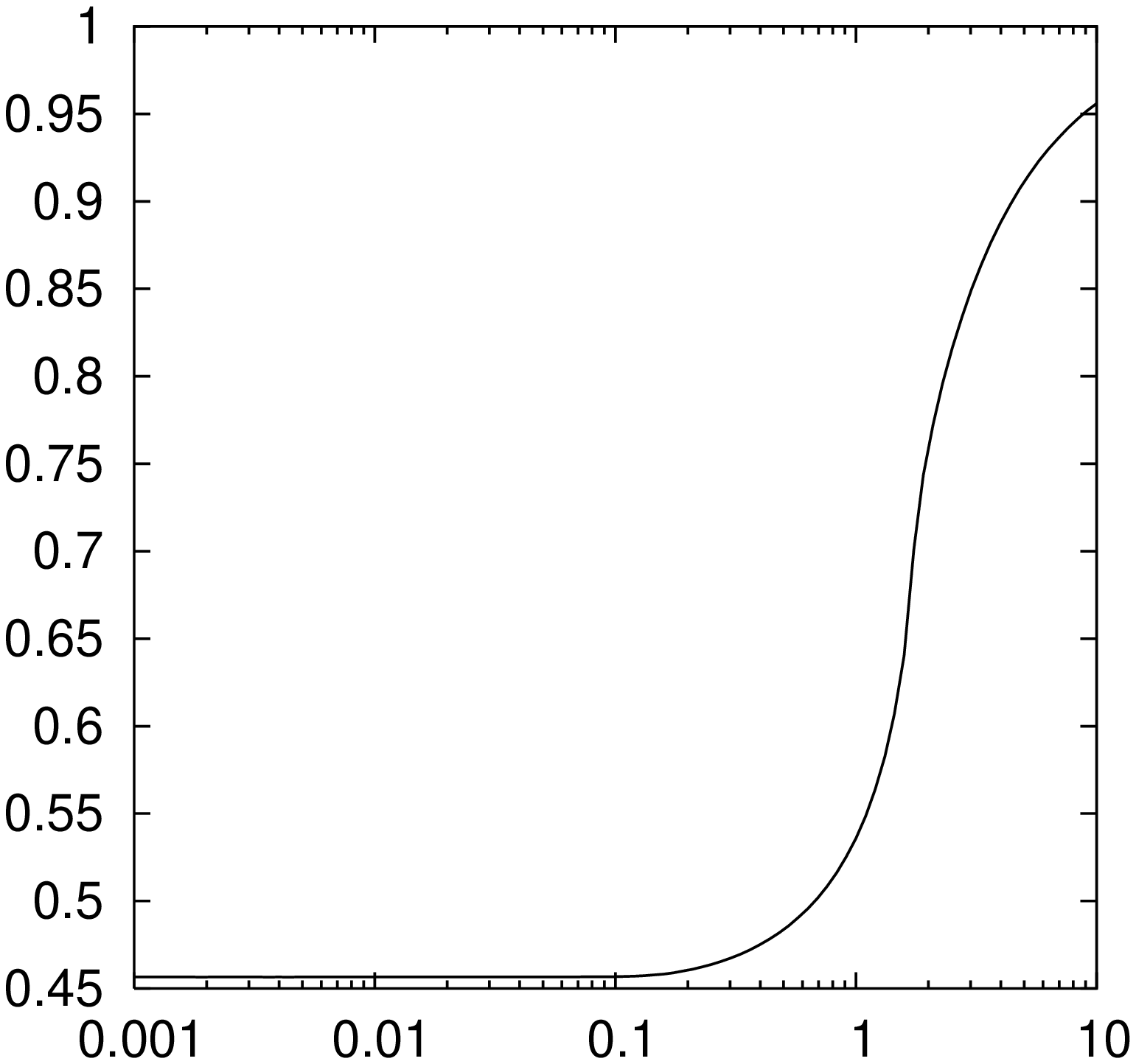}
\includegraphics[width=0.2\textwidth]{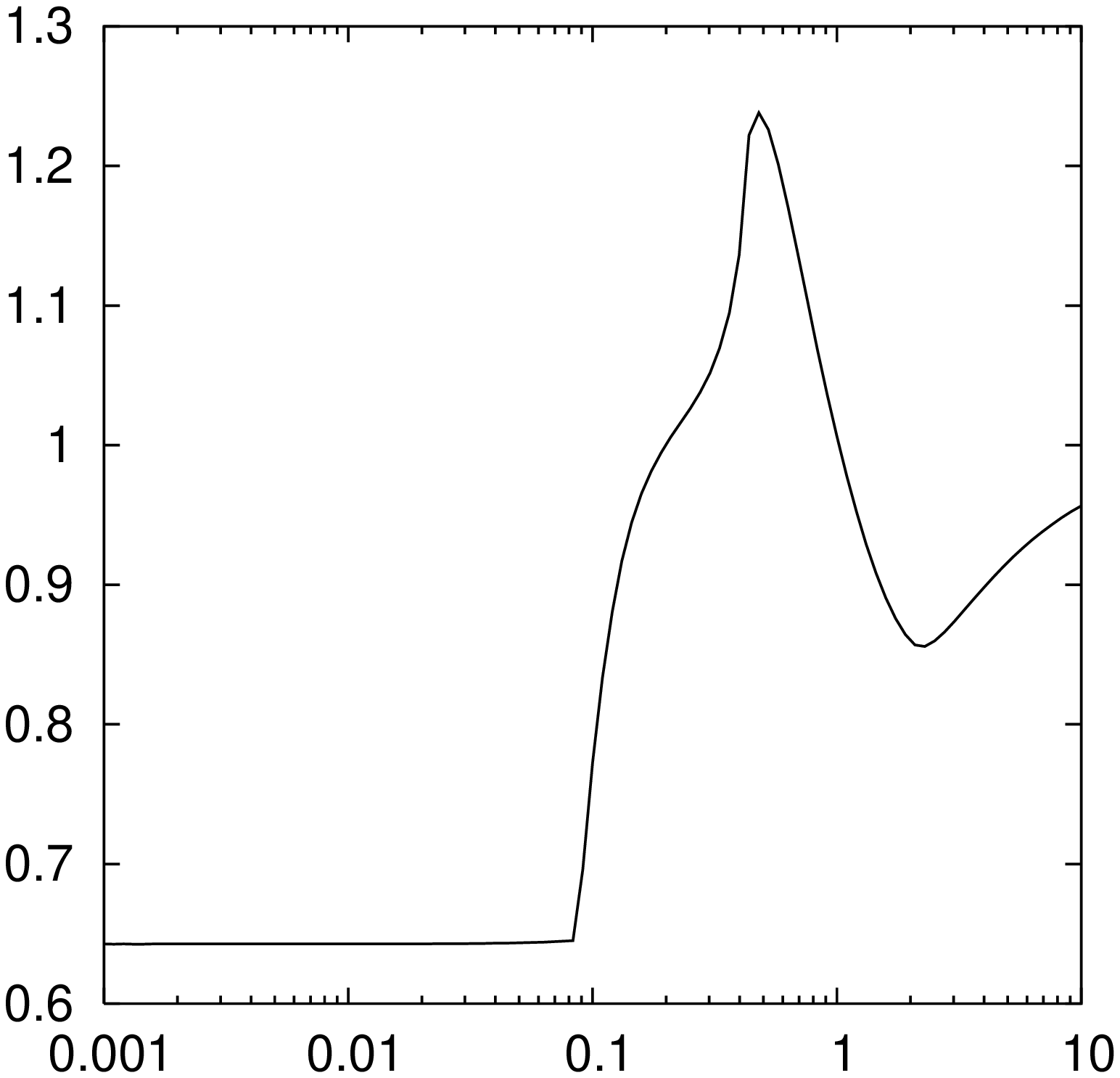}
\linebreak
\includegraphics[width=0.2\textwidth]{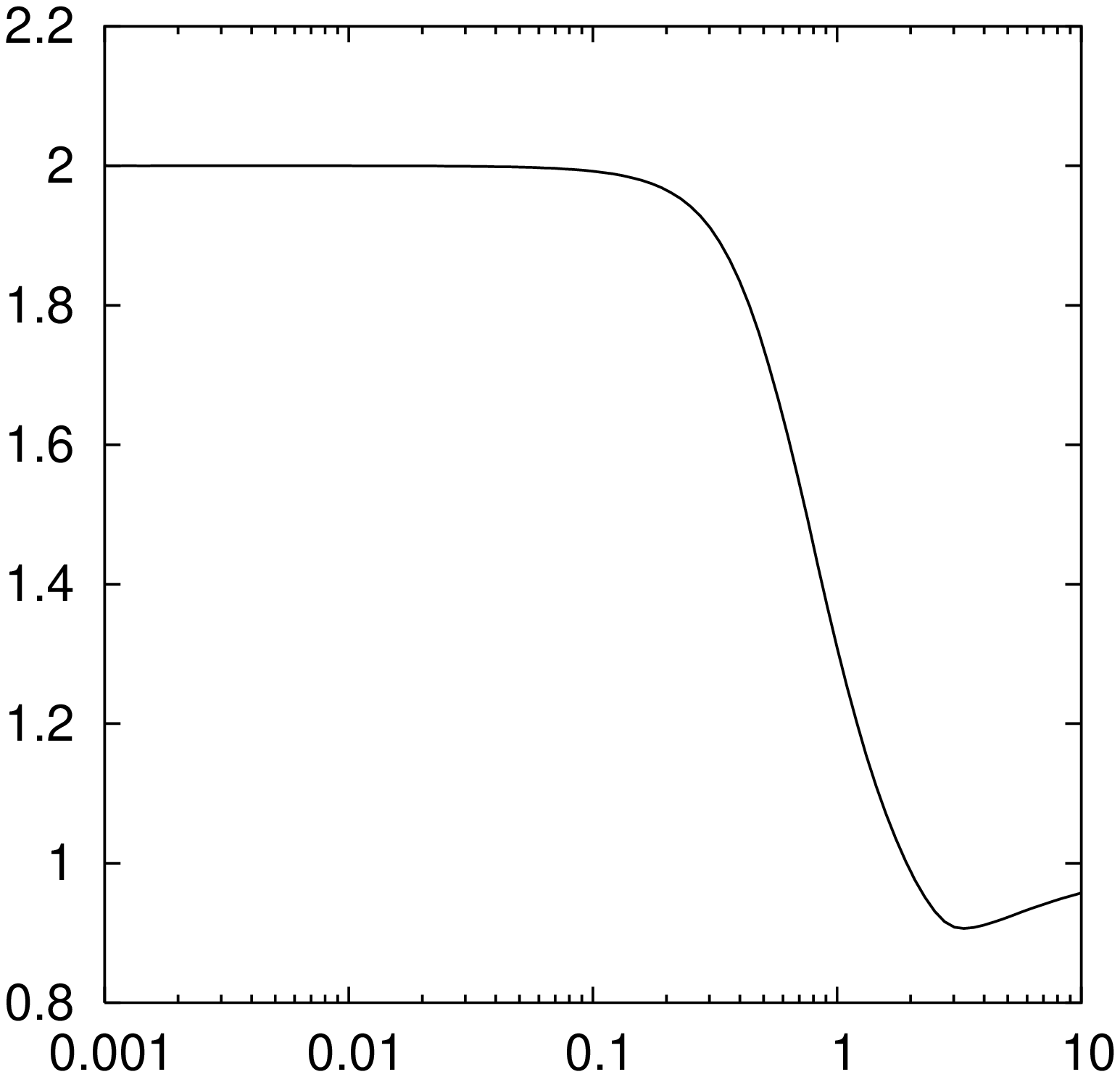}
\includegraphics[width=0.2\textwidth]{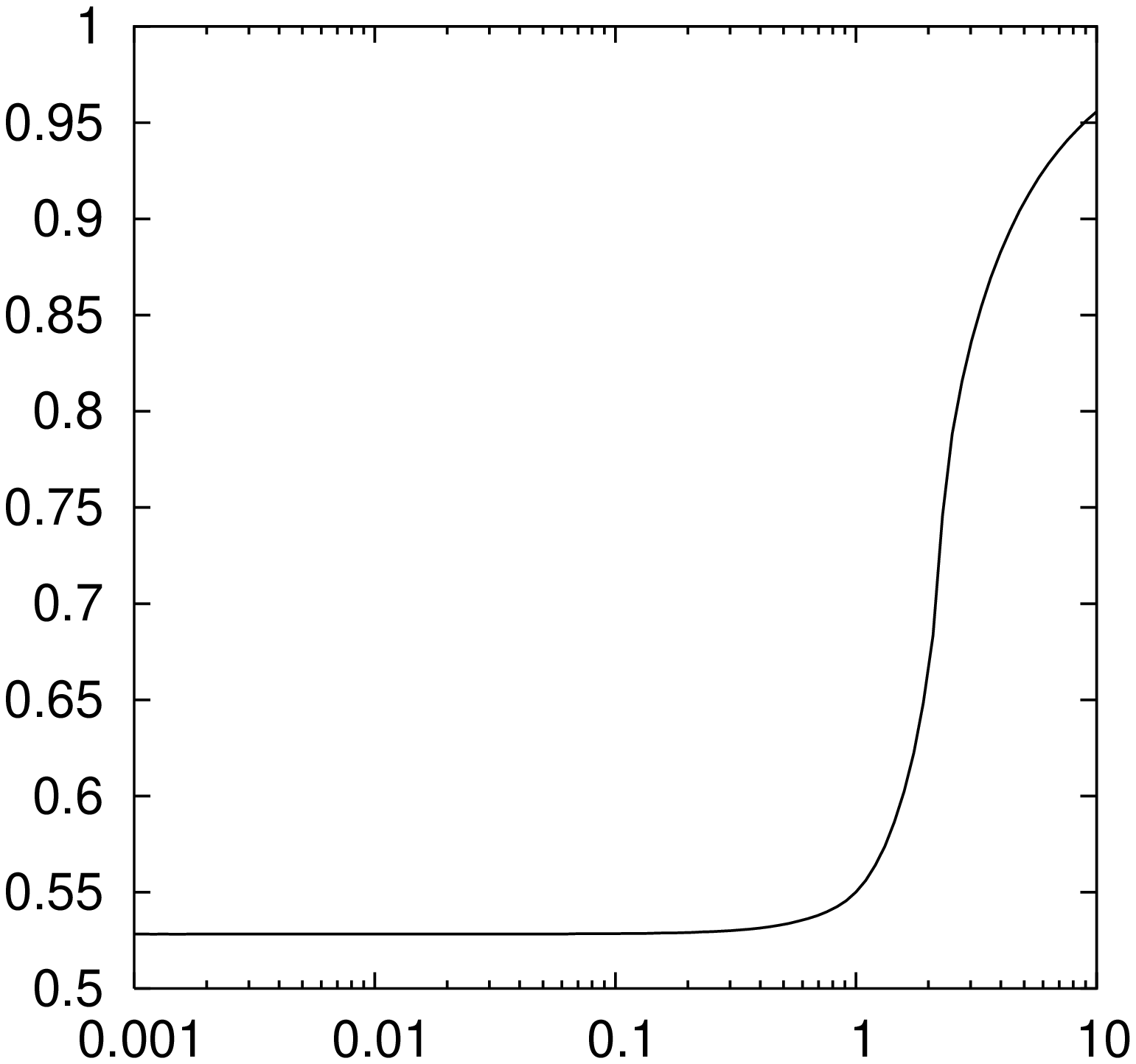}
\includegraphics[width=0.2\textwidth]{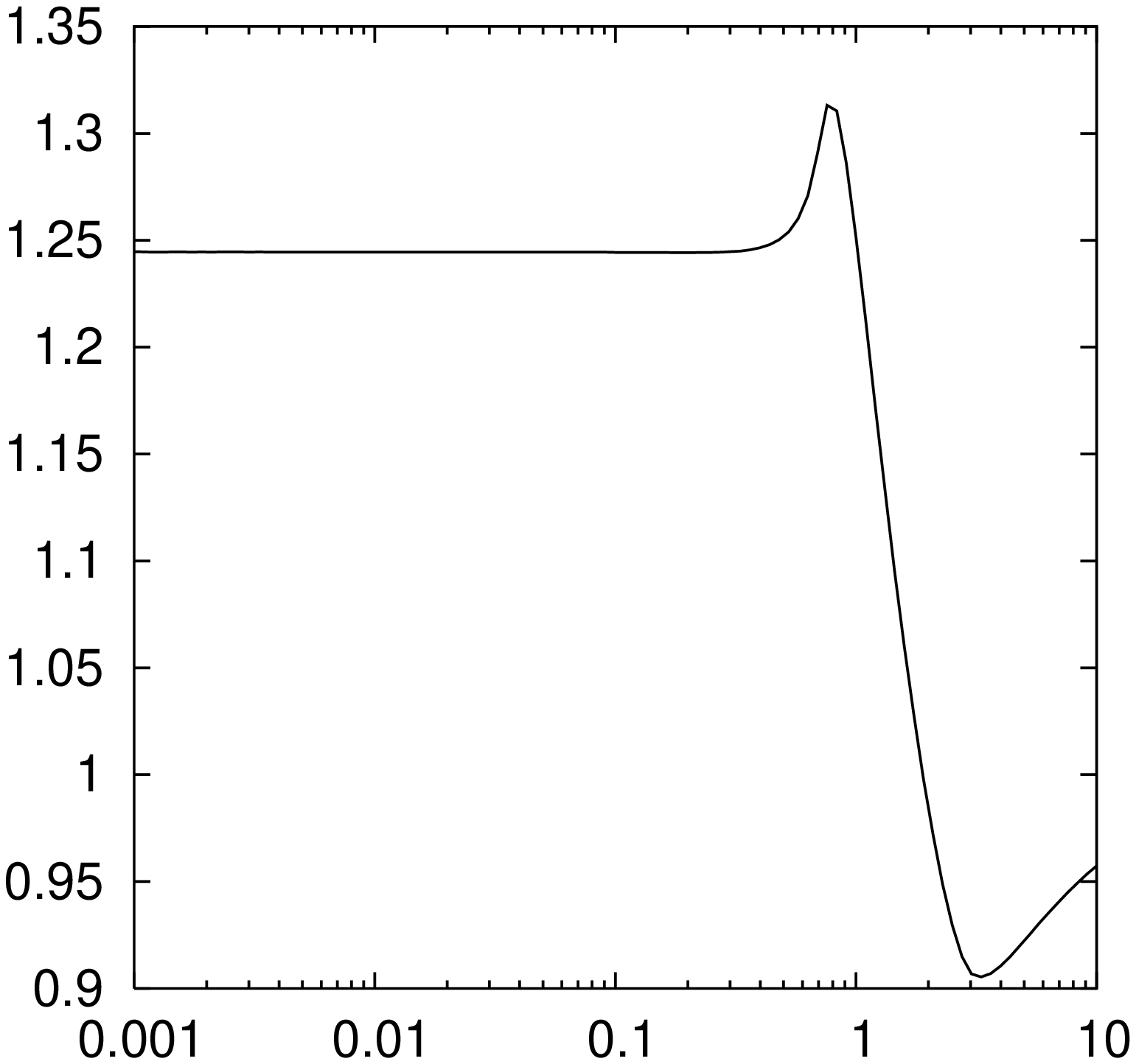}
\caption{
Similar to Fig.~\ref{fig:poscurves}, but for a negative parity case
($\kappa=\gamma=0.7$).
}\label{fig:negcurves}
 \end{center}
\end{figure}

Fig.~\ref{fig:negcurves} shows the normalized magnification curves at
various source positions for a negative parity case ($\kappa=\gamma=0.7$).
The behavior for the negative parity case is a bit more complex in that 
both magnification and demagnification (relative to the convergence
and shear field) are seen.  For example, at a source position of
$u_0,v_0=1.1,0.0$ (second row first column), the source is magnified at
low $a$, rises to a peak at $a \simeq 0.2$, falls to a demagnified valley
at $a \simeq 2$, and then rises toward unity for larger $a$.  As in the
positive parity case, we see that for small $a$ there is relatively little
structure in the curves.  Also, the magnifications are again fairly uniform
for $a \ga 10$ (see Appendix A), with $\mm \simeq 0.95$ at
$a=10$.

\subsection{Center of flux position}
\label{sec:pos}

Although we have determined the image boundary and magnification of
a finite source, we have not yet quantified the image position.  If
the separate component images are not resolved, what matters is the
`centre of flux' of the image configuration, which can be computed
as
\begin{equation}
  \vect X = \frac{\int {\vect x}\,f({\vect x})\,d{\vect x}}
    {\int f({\vect x})\,d{\vect x}},
\end{equation}
where $f({\vect x})$ is 1 inside the image and 0 outside.  (This is
the optical analog of the centre of mass.) The difference between
this `centre of flux' and the original image position in the
absence of the clump, $\vect X_0$, is the astrometric shift due to
the perturber,
\begin{equation}
  \delta \vect X = \vect X - \vect X_0.
  \label{eq:astroperturbeq}
\end{equation}
Our solution for $r_{\pm}(\theta)$ yields a straightforward calculation
for the astrometric perturbation, provided that we account for the
different solution regimes as in eq.~(\ref{eq:mageqn}).

\begin{figure}
 \begin{center}
\includegraphics[width=0.3\textwidth]{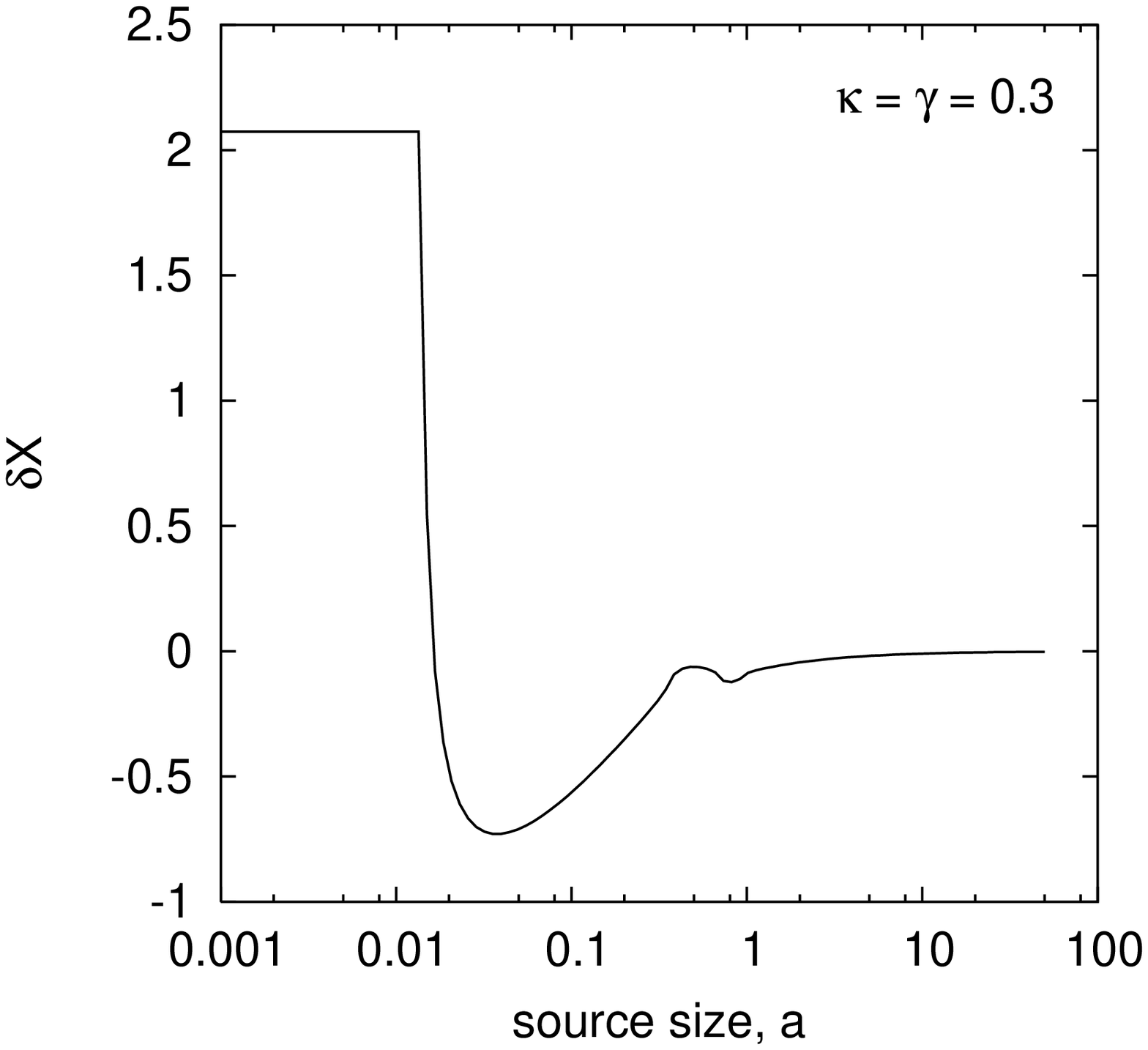}
\includegraphics[width=0.3\textwidth]{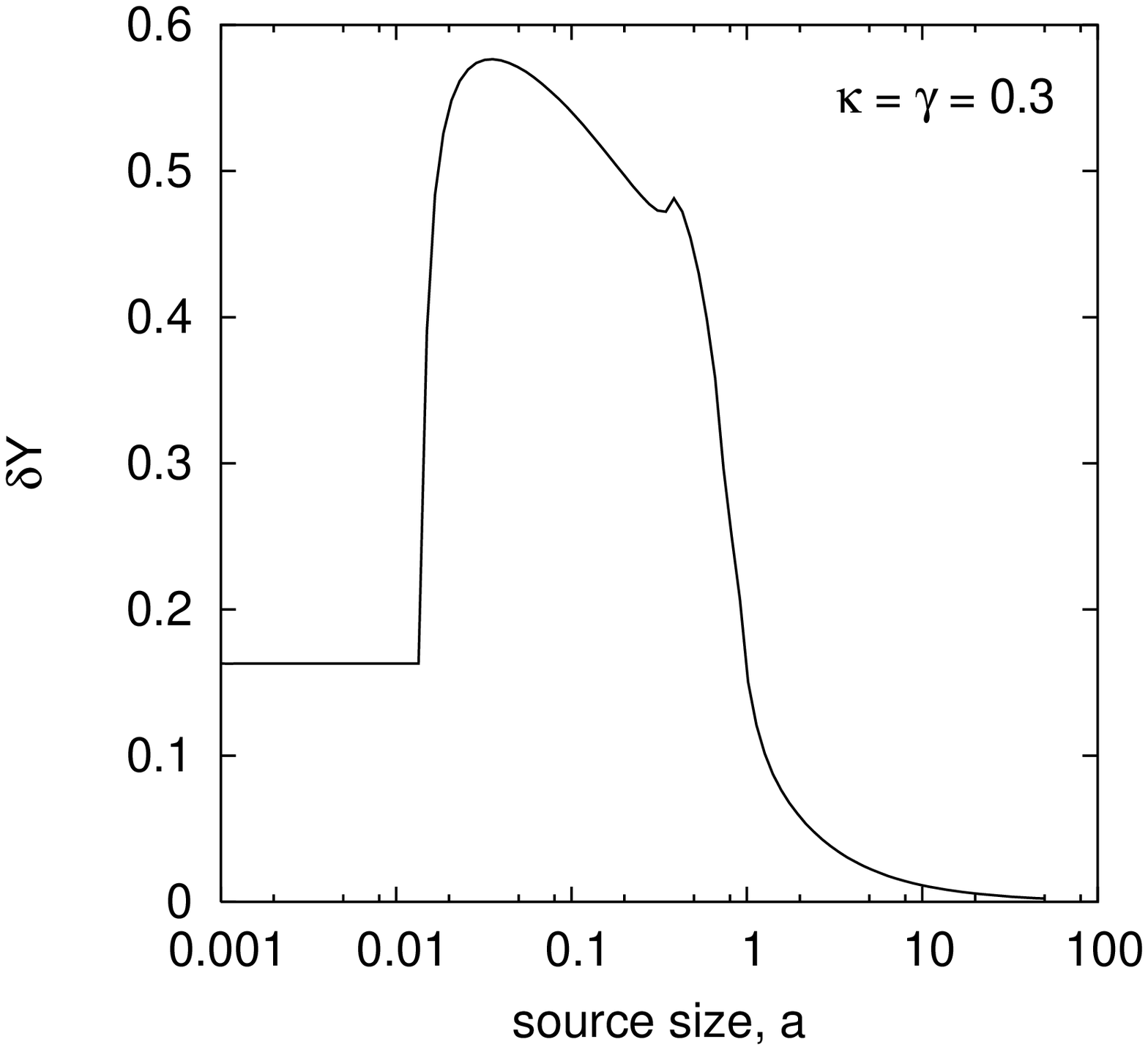}
\includegraphics[width=0.3\textwidth]{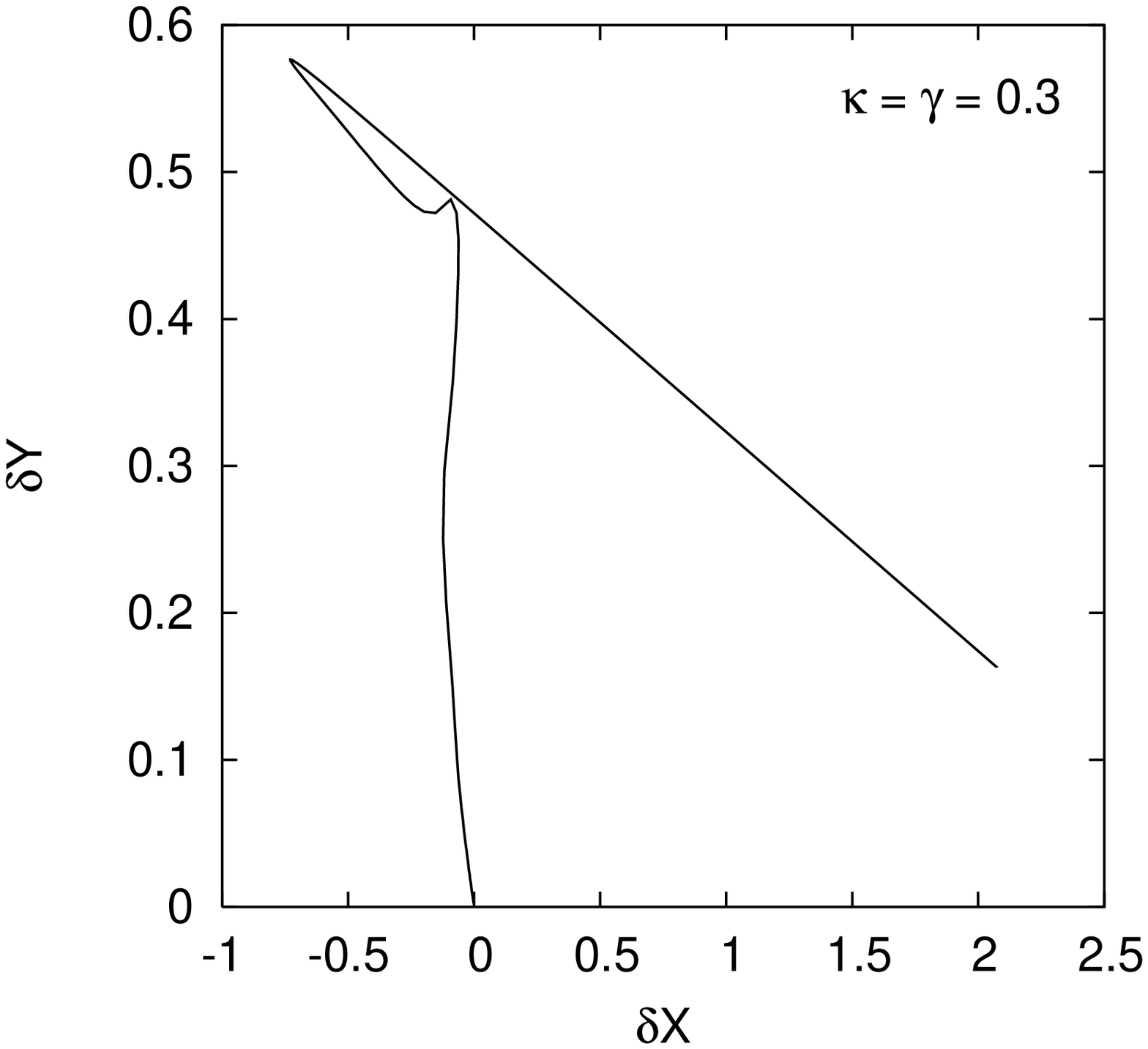}
\caption{
Astrometric perturbation as a function of source size for $b=1$,
$(u_0,v_0)=(0.2,0.6)$, and $\kappa=\gamma=0.3$.  The large change
between $a=0.01$--$0.02$ corresponds a change from two to three images
(see Fig.~\ref{fig:positiveimages}).  For a large source, the perturber
has little effect on the centre of flux position as expected.
}\label{fig:positiveastropert}
 \end{center}
\end{figure}

Fig.~\ref{fig:positiveastropert} shows the astrometric perturbation
as a function of source size for the source from
Fig.~\ref{fig:positiveimages}.  As with the magnification calculation,
we see that for a source size below about $a=0.015$ the astrometric
perturbation remains fairly constant.  Comparison with the image
configuration from Fig.~\ref{fig:positiveimages} shows that the
centre of flux position lies on the line joining the two images that
are seen at $a=0.01$.  Between $a=0.01$ and $a=0.02$ there is a sudden
change as the source intersects the caustic.  The emergence of a bright
third image rapidly pulls the centre of flux away from the line joining
the initial two images.  For a large source size, we would expect that
the perturbing SIS would have little effect on the centre of flux of
the image configuration and indeed as $a$ gets large, the astrometric
perturbation tends towards zero.

\begin{figure}
 \begin{center}
\includegraphics[width=0.3\textwidth]{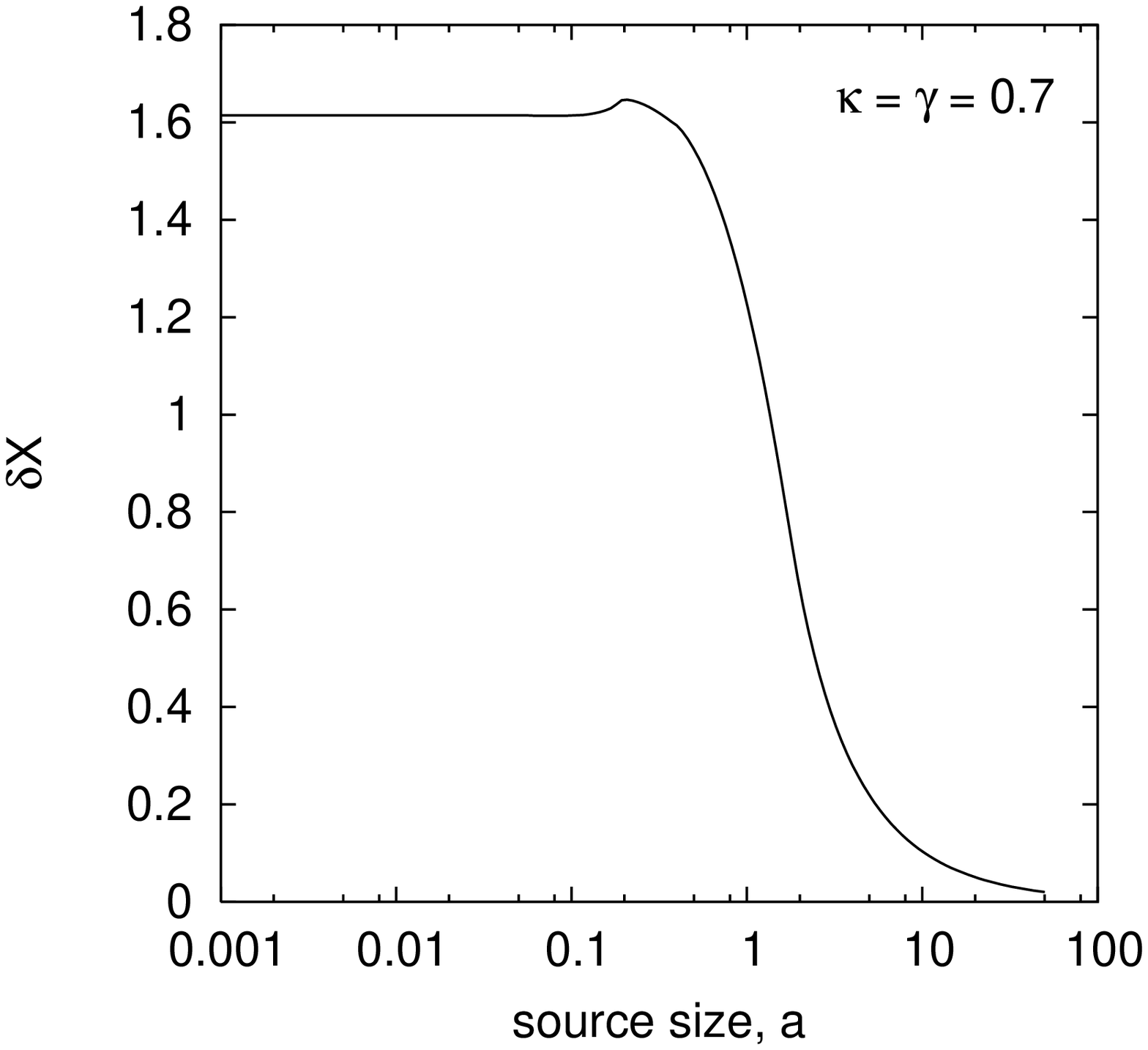}
\includegraphics[width=0.3\textwidth]{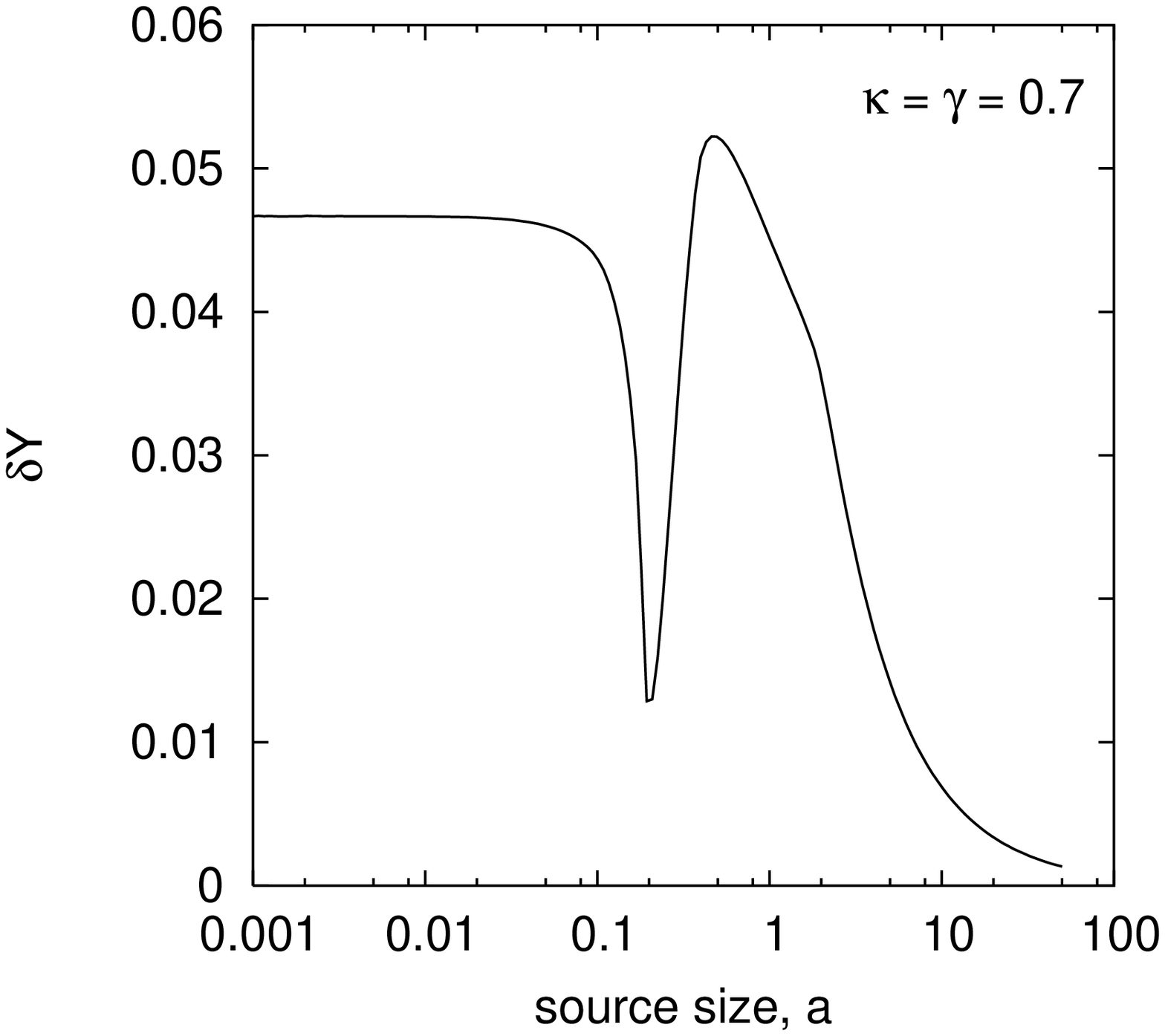}
\includegraphics[width=0.3\textwidth]{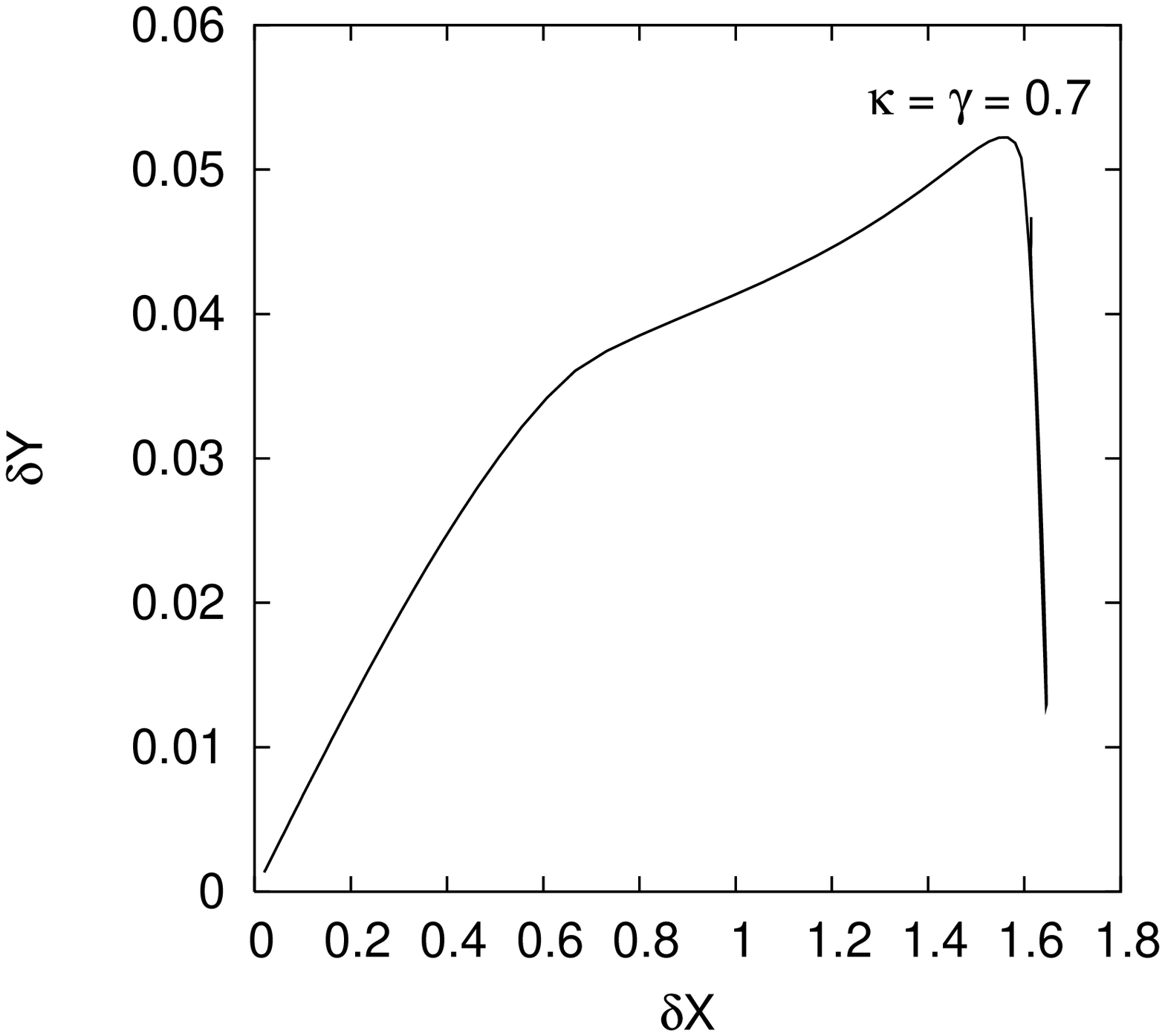}
\caption{
Astrometric perturbation as a function of source size for $b=1$,
$(u_0,v_0)=(0.9,0.15)$, and $\kappa=\gamma=0.7$.  The large change at
around $a=0.2$ corresponds to the source intersecting a cusp caustic
(see Fig.~\ref{fig:negativeimages}).
}\label{fig:negativeastropert}
 \end{center}
\end{figure}

Fig.~\ref{fig:negativeastropert} shows the analogous results for a
negative parity case with $\kappa=\gamma=0.7$.  There is not much change
in the astrometric perturbation when $a$ is less than about 0.1, but
there is then a dramatic dip in both curves between $a=0.01$ and $0.02$.
Returning to Fig.~\ref{fig:negativeimages}, we see that this large
change in the centre of flux position occurs when the source begins to
intersect the caustic.

Like stellar microlensing \citep[e.g.,][]{tr}, substructure lensing
can produce astrometric shifts of order several Einstein radii.  The
difference is in the scale: the Einstein radius for stars is of order
micro-arcseconds, while for substructure it is milli-arcseconds.  Thus,
astrometric perturbations due to substructure should be detectable with
radio interferometry, and perhaps even with space-based optical or
infrared imaging.  While small position shifts might be degenerate with
small changes in the macromodel, the dependence of the shift on source
size (and hence wavelength) would provide a clear signature of
substructure lensing.  Astrometric shifts are related to shape
perturbations \citep[see][]{me2}, but do not require resolved image
shapes.  A full analysis of prospects for observing astrometric
perturbations and using them to constrain substructure is beyond the
scope of this paper, but warrants further study.

\subsection{Comments}
\label{sec:comments}

To review, we have studied how the magnification depends on the size
of the source and its position relative to the caustics.  The specific
details depend on our assumption of an SIS clump, which is a toy model,
but the basic principles should be more general.  One point is that the
image properties are basically independent of the source size until the
source is large enough to encounter the caustics; that threshold of
course depends on the source position.  In the other extreme, sources
more than an order of magnitude larger than the clump Einstein radius
can still be perturbed at the percent level, and that precision can be
obtained with careful observations \citep[e.g.,][]{fas1608}.  In other
words, the conventional wisdom that a source does not `feel' lensing
structure on scales smaller than itself is not really accurate beyond
order of magnitude estimates.

In between these extremes, there is significant structure in the
magnification versus source size curves.  Therefore, in principle,
comparing the flux of an image at different wavelengths corresponding
to different source sizes could reveal a wealth of information about
the size and location of substructure.  Prospects for doing that are
good: recent observations have demonstrated the ability to measure
flux ratios not only for radio and optical continua, but also for
optical emission lines and mid-IR emission \citep{agol,0435b,met}.
We would advocate concerted effort to obtain and analyse such
panchromatic observations of lenses with flux ratio anomalies (see
\citealt{met} for an example).  Even more exciting is the possibility
of measuring astrometric shifts along with flux perturbations; further
study is needed to determine the feasibility and value of such
measurements.

\section{Placing Limits on Substructure Size}

While a few lens systems have been observed at many wavelengths, the
ones that are most interesting for millilensing are still limited to
radio continuum observations (plus perhaps broad-band optical data).
Nevertheless, it is still possible to place important \emph{lower}
bounds on the substructure responsible for observed flux ratio
anomalies.  In this section we customize our general analysis of
substructure lensing to this application.

\subsection{Maximally affected images}

\begin{figure}
 \begin{center}
\includegraphics[width=0.42\textwidth]{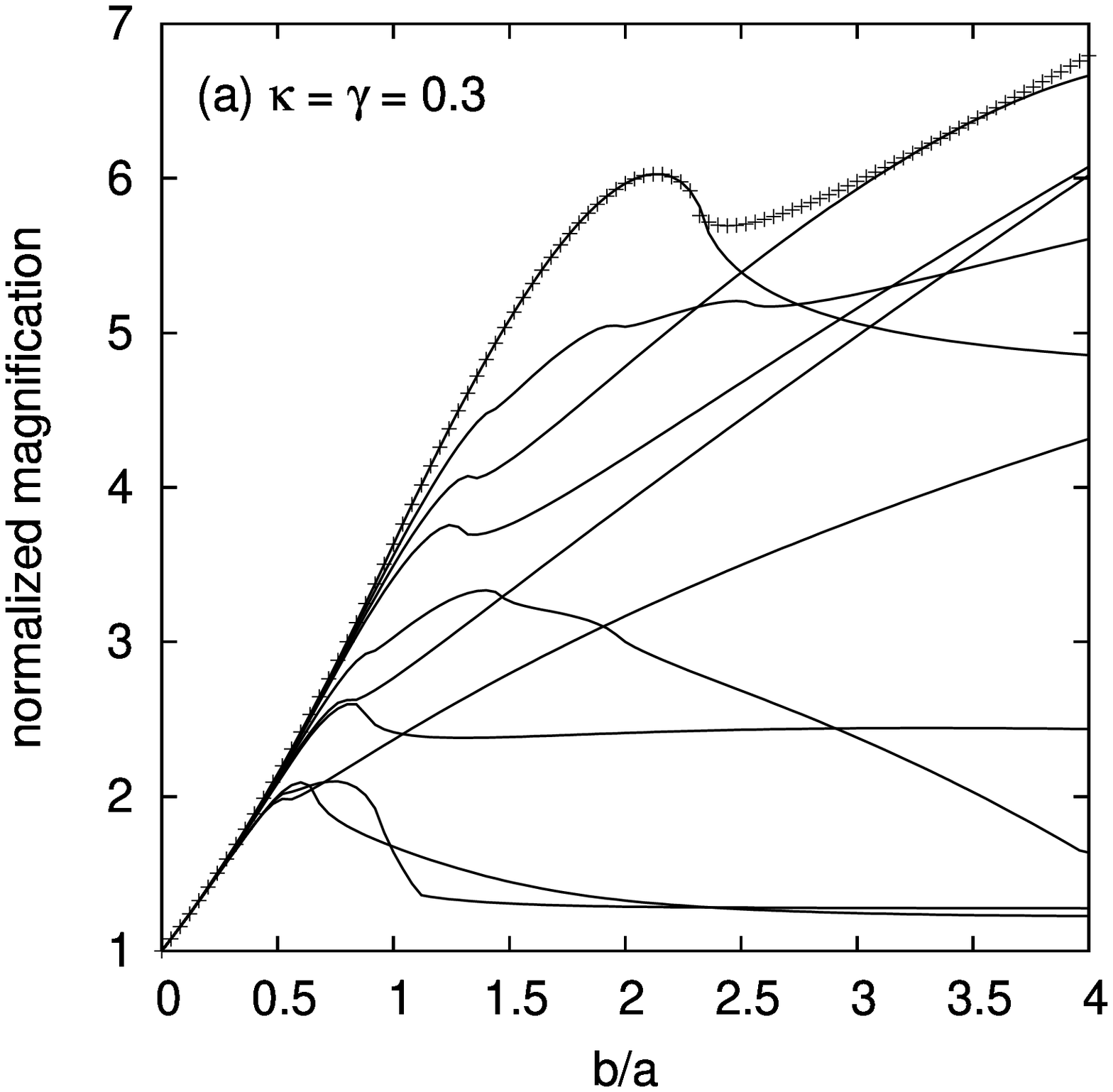}
\includegraphics[width=0.43\textwidth]{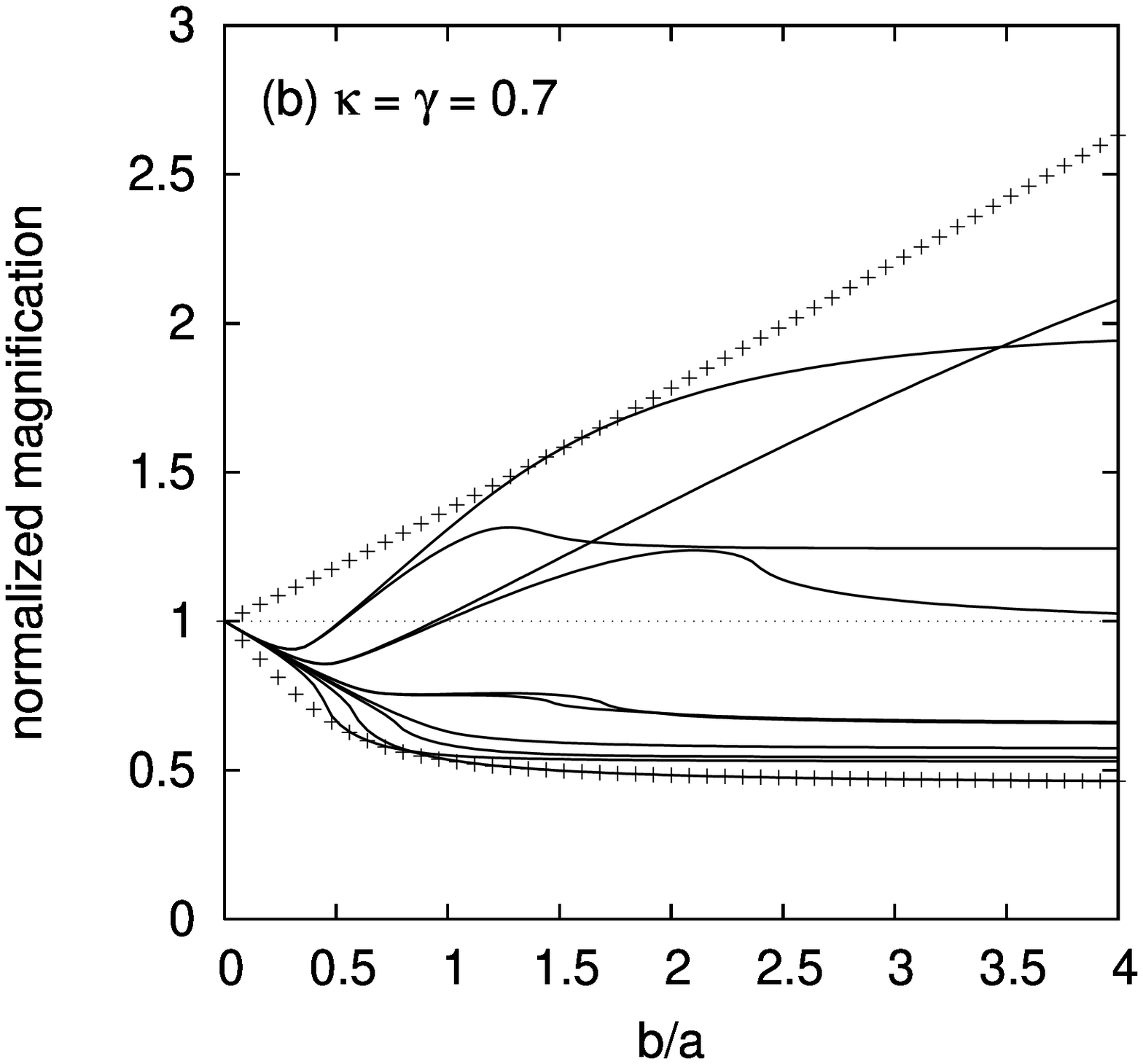}
\caption{
Normalized magnification as a function of $b/a$ for a positive parity
image (a) and a negative parity image (b).  The crosses represent the
maximum (maximized over source position) or minimum magnification for
a given $b/a$.  The solid lines show $\mm$ versus $b/a$ at various
fixed source positions, from Figs.~\ref{fig:poscurves} and
\ref{fig:negcurves}.
}\label{fig:fmax}
 \end{center}
\end{figure}

We have seen that changing the position of the source relative to the
perturber (or vice versa) has a dramatic effect on the magnification
versus source size curves.  However, in general we do not know the
relative position, so it is useful to determine the bounds on the
magnification that can be produced by a given perturber for a given
source size.  In practice, this amounts to setting the ratio $b/a$ of
perturber and source sizes (as well as the background field $\kappa$
and $\gamma$), and then maximizing or minimizing the normalized
magnification over $u_0$ and $v_0$.  Fig.~\ref{fig:fmax}a shows
the bounds as a function of $b/a$ for a positive parity image with
$\kappa=\gamma=0.3$.  For comparison, the solid lines show curves
for fixed source positions (from Fig.~\ref{fig:poscurves}).  (The
lower bound is trivial, $\mm = 1$, since an SIS perturber in front
of a positive parity image never produces demagnification.)  At small
$b/a$, all of the curves have roughly the same behavior, again
illustrating that at large source size the change in magnification
is independent of position.  At large $b/a$, $\mm_{\rm max}$ grows
to infinity since, in the limit of an infinitesimal source, placing
the source on the caustic yields infinite magnification.

\begin{figure}
 \begin{center}
\includegraphics[width=0.45\textwidth]{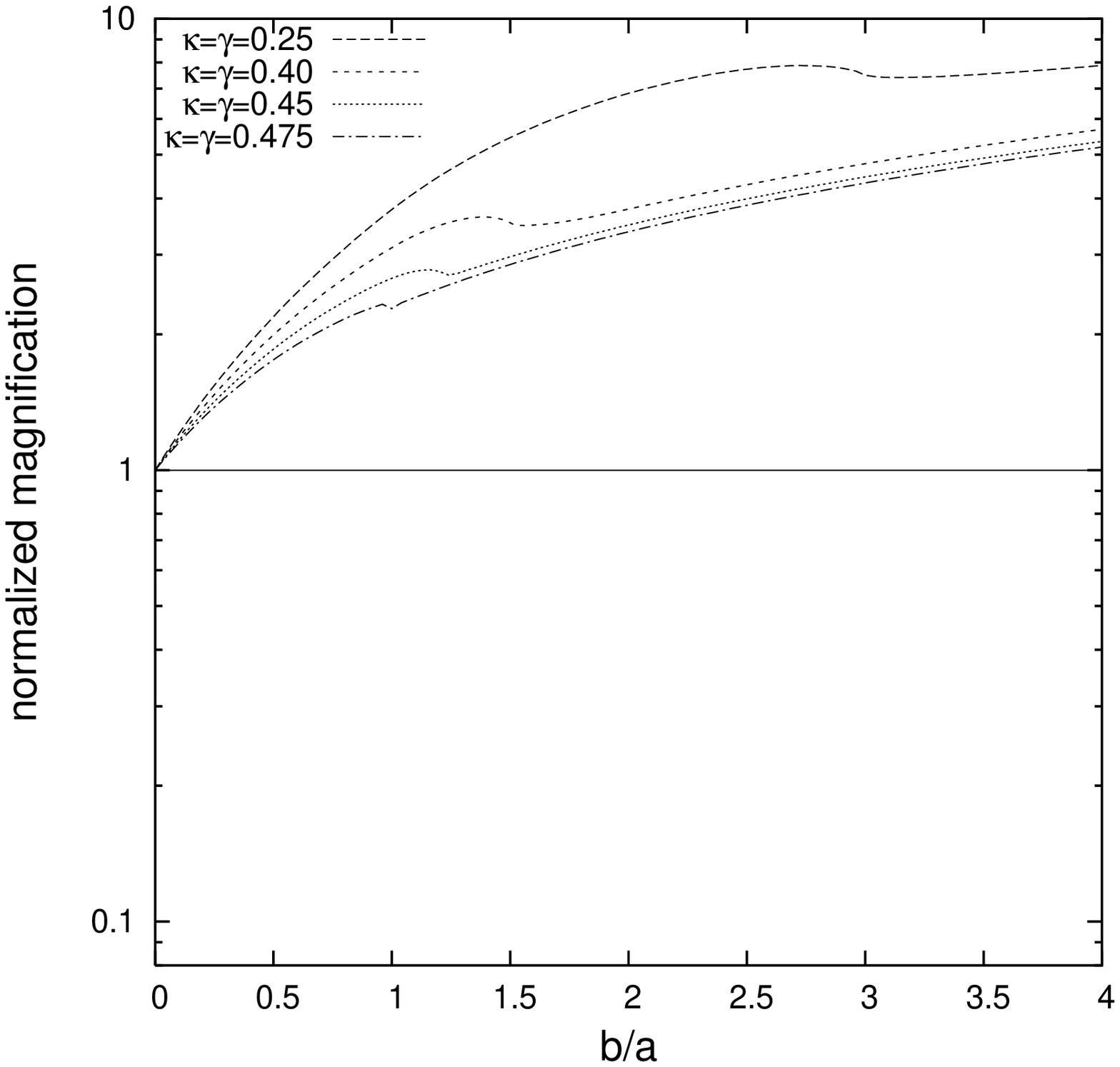}
\includegraphics[width=0.45\textwidth]{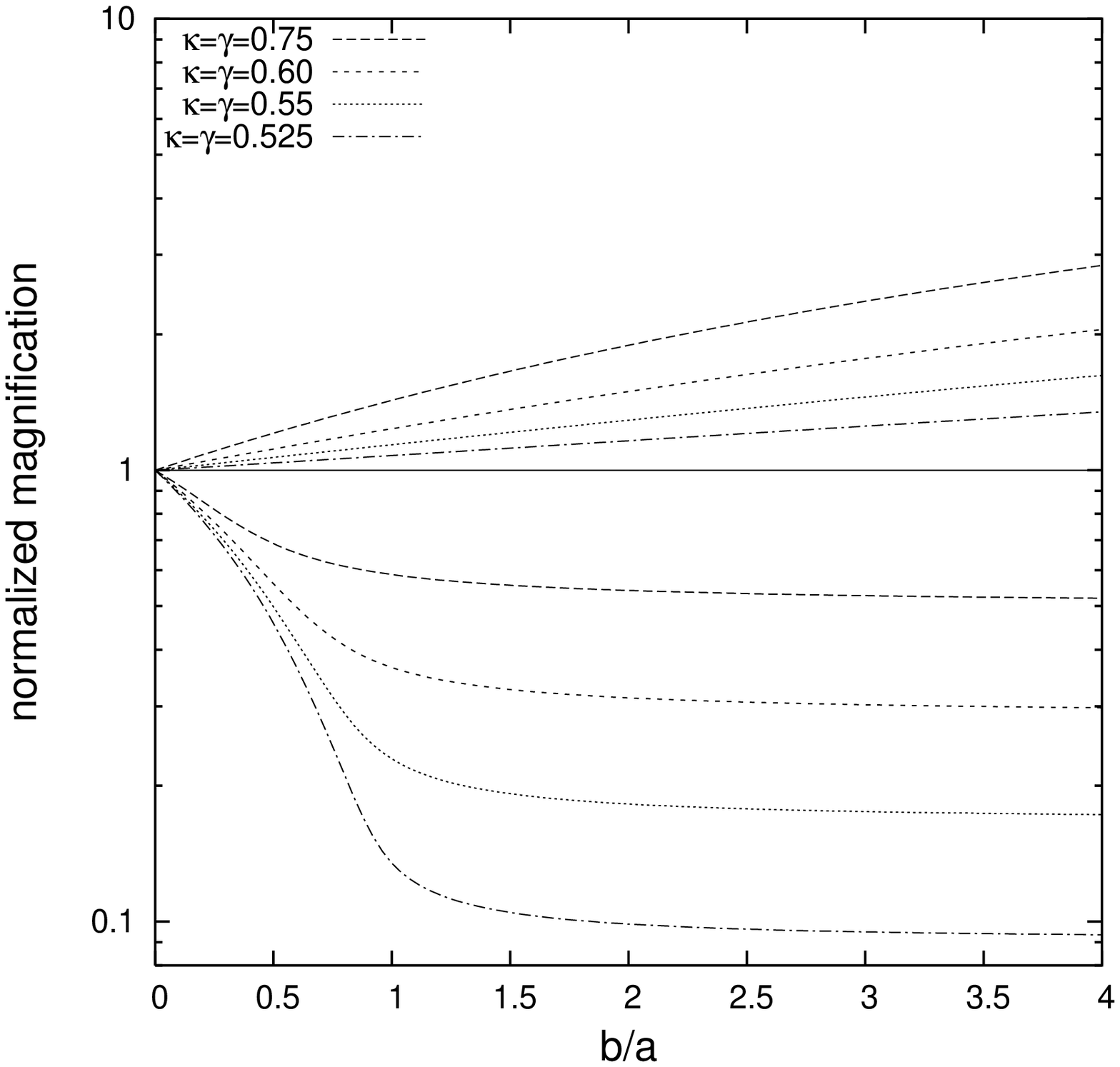}
\caption{
Maximum and minimum normalized magnification versus $b/a$, for different
values of $\kappa$ and $\gamma$.
}\label{fig:fmaxkapgam}
 \end{center}
\end{figure}

The analogous results for a negative parity case are shown in
Fig.~\ref{fig:fmax}b.  Here we have both $\mm_{\rm max}$ and
$\mm_{\rm min}$ curves since an SIS in front of a negative parity
image can produce both magnification and demagnification.\footnote{Again,
by magnification or demagnification we mean images brighter or
fainter than produced by the convergence and shear field alone.}
The curves of $\mm$ versus $b/a$ at various source position from
Fig.~\ref{fig:negcurves} are again shown for comparison.  An important
feature is that the $\mm_{\rm min}$ curve approaches a constant value
at large $b/a$.  In the limit of an infinitely small source, a negative
parity image allows infinite magnification but \emph{not} infinite
\emph{de}magnification.  The shapes of the $\mm_{\rm max}$ and
$\mm_{\rm min}$ curves depend on the macromodel through $\kappa$ and
$\gamma$, as shown in Fig.~\ref{fig:fmaxkapgam}.

\subsection{Limits from a single flux measurement}

A key result from Fig.s~\ref{fig:fmax} and \ref{fig:fmaxkapgam}
is that we can use any measured $\mm \ne 1$ to place a lower bound
on $b/a$, \emph{without needing to know the relative positions of
the perturber and source}.  This bound comes from the fact that,
although the region below the $\mm_{\rm max}$ curve is completely
accessible by choosing appropriate values for $u_0$ and $v_0$, the
region \emph{above} the $\mm_{\rm max}$ curve is excluded by definition.
For any observed $\mm_{\rm obs} > 1$, we simply find the value of
$b/a$ where $\mm_{\rm max}(b/a) = \mm_{\rm obs}$, and that gives us
the lower limit on the size of the perturber (relative to the size
of the source).  The bound can be understood physically with the idea
that a source cannot `feel' a perturber that is much smaller than
itself.  Conversely, there is no upper bound because a source that
is small relative to the perturber can be placed as far from or as
close to the caustics as necessary to reproduce any observed
magnification.  (Similar reasoning applies to both the magnification
and demagnification regimes in the negative parity case.)

For the positive parity case, increasing $\kappa$ and $\gamma$
(increasing $\mu_0$) lowers the $\mm_{\rm max}$ curve, or equivalently,
increases the minimum value of $b/a$ required to produce a given
normalized magnification (see Fig.~\ref{fig:fmaxkapgam}).  For the
negative parity case, decreasing $\kappa$ and $\gamma$ (increasing
$|\mu_0|$) lowers both the $\mm_{\rm max}$ and $\mm_{\rm min}$ curves.
This is equivalent to increasing the minimum value of $b/a$ required to
produce a given $\mm > 1$, or decreasing the minimum value of $b/a$
required to produce a given $\mm < 1$.  Although the lower bound on
$b/a$ does depend on $\kappa$ and $\gamma$, these parameters are well
constrained by the macromodel (see Fig.~\ref{fig:1422contkapgam} below).

\begin{figure}
 \begin{center}
\includegraphics[width=0.45\textwidth]{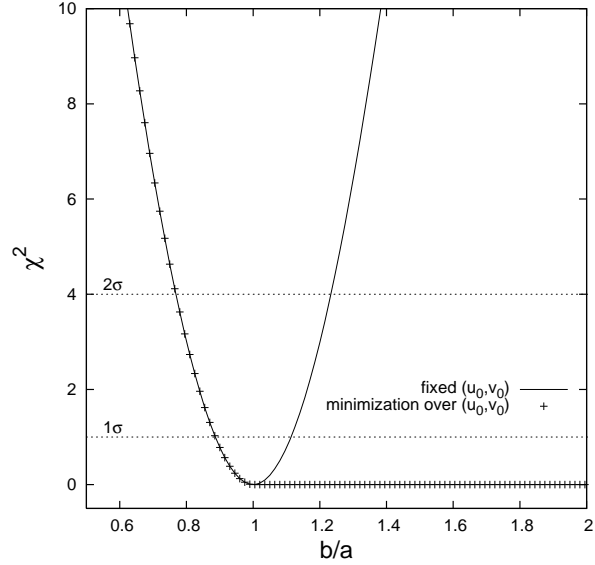}
\caption{
$\chi^2$ versus $b/a$ for a mock measurement of the normalized
magnification.  The solid line shows the result if we fix the source
at its input value, while the crosses show the result if we optimize
over the source position.
}\label{fig:chisqwithmin}
 \end{center}
\end{figure}

If the observed image flux $f_{\rm obs}$ were known precisely, then
$\mm_{\rm obs} \equiv f_{\rm obs}/f_0$ could be used to place a strict
lower bound on $b/a$.  Of course, flux measurement uncertainties smear
the bound, and the simplest way to incorporate the uncertainties is
to define a goodness of fit,
\begin{equation}
  \chi^2_{\rm sub}\left( \frac{b}{a}; u_0, v_0; \kappa, \gamma \right)
  = \left[ \frac{\mm_{\rm mod}(b/a;u_0,v_0;\kappa,\gamma)
    - \mm_{\rm obs}}{\sigma_{\rm obs}} \right]^2 .
  \label{eq:chisub}
\end{equation}
Fig.~\ref{fig:chisqwithmin} shows a sample $\chi^2$ analysis, where
we generated a mock measurement of $\mm_{\rm obs} = 3.63$ assuming 
$\kappa = \gamma = 0.3$, $a=b=5$, $u_0=v_0=0$, and
$\sigma_{\rm obs} = 0.1 \times \mm_{\rm obs}$.  In the figure, the solid line
shows $\chi^2$ versus $b/a$ if we fix the source at the origin, while
the crosses show the result if we optimize over the source position.
(We always fix $\kappa$ and $\gamma$, because they are determined well
enough by the macromodel; see \refsec{1422} below.)  If the source
position were known, we would get both upper and lower limits on $b/a$.
If the source position is unknown, we lose the upper limit but still
get the lower limit as discussed above.  This is the more interesting
limit anyway, since for a given flux ratio anomaly it is useful to know
the \emph{smallest possible} perturbing mass that could produce the
anomaly.

\section{Application to Observed Lens Systems}

Before we can apply our millilensing theory to derive substructure
mass bounds, we must first figure out which of the images are
perturbed.  We must also determine the convergence and shear that
create the background in which the clump lives.  To do this, the
idea is to fit a smooth macromodel to an observed lens, identify
any images that cannot be fit, and attribute the discrepancy to
substructure.  We emphasize that this process is independent of any
assumptions about the nature of the substructure.  It does depend
on our choice of macromodel, but the models we use are standard in
millilensing analyses \citep[e.g.,][]{DK,KD,met,metcalfLOS}.
Dependence on the substructure model enters only when we bring in
the method from \S 3 to derive constraints on the masses of the
substructures.

\subsection{Methodology}

For the macromodel, we consider two related models.  In the first
case, we treat the lens galaxy as a singular isothermal ellipsoid
(SIE), which is a simple but useful model that is consistent with
many lensing, dynamical, and X-ray observations \citep[e.g.,][]
{fabbiano,rix97,gerhard,tk2016,kt1608,rkk}.  The model has surface
mass density
\begin{equation} \label{eq:SIE}
  \kappa(r,\theta) = \frac{R_{\rm ein}}
    {2 r \sqrt{1-\epsilon\cos2(\theta-\theta_\epsilon)}}\ ,
\end{equation}
where $R_{\rm ein}$ is the macromodel Einstein radius, $\epsilon$ is an
ellipticity parameter related to the axis ratio $q$ of the ellipse
by $q^2 = (1-\epsilon)/(1+\epsilon)$, and $\theta_\epsilon$ is the
orientation angle of the ellipse major axis.  A simple SIE model is
insufficient to fit most 4-image lenses, so we add an external shear
term to represent the effects of other mass in the environment of
the lens galaxy \citep{KKS}.  The $N_p=10$ model parameters are then:
the position, Einstein radius, ellipticity, and orientation of the
lens galaxy; the amplitude and direction of the external shear; and
the position and flux of the source.

In the alternate macromodel, we keep the same monopole but allow a
more general angular structure.  Expanding the lens potential in
multipoles, we write \citep[see][]{SaasFee}
\begin{equation}
  \phi(r,\theta) = R_{\rm ein}\,r
  + \frac{R_{\rm ein}^4}{2r^2}\,\gamma_{\rm int}\cos2(\theta-\theta_{\rm int})
  + \frac{2^2}{2}\,\gamma_{\rm ext}\cos2(\theta-\theta_{\rm ext}) + \ldots
\end{equation}
The first term is the potential for a singular isothermal sphere
(a mass model given by eq.~\ref{eq:SIE} with $\epsilon=0$).  The
second term represents the shear due to mass within the Einstein
radius, where $\gamma_{\rm int}$ and $\theta_{\rm int}$ are the internal
shear amplitude and direction.  The third term represents the shear
due to mass outside the Einstein radius, which can now include a
contribution from the outer parts of the lens galaxy halo in
addition to a contribution from the larger lens environment.
Compared to the ellipsoid+shear model, the internal+external shear
model allows a more general structure for the lens galaxy, but at
the expense of not representing true elliptical symmetry very well
(since the multipole series is truncated).  Both macromodels have
the same number of parameters.

We define the macromodel goodness of fit to include contributions
from the image positions, the image fluxes, and the galaxy position
(if known):
\begin{eqnarray}
  \chi^2_{\rm tot} &=& \chi^2_{\rm pos} + \chi^2_{\rm flux} + \chi^2_{\rm gal} , \\
  \chi^2_{\rm pos}  &=& \sum_{A,B,C,D} \frac{ \left(\vect x_{\rm mod} - \vect x_{\rm obs} \right)^2}{\Delta \vect x_{\rm obs}^2} , \\
  \chi^2_{\rm flux} &=& \sum_{A,B,C,D} \frac{ \left(f_{\rm mod} - f_{\rm obs} \right)^2}{\Delta f_{\rm obs}^2} , \\
  \chi^2_{\rm gal}  &=& \frac{ \left(\vect X_{\rm mod} - \vect X_{\rm obs} \right)^2}{\Delta \vect X_{\rm obs}^2} ,
\end{eqnarray}
where ${\vect x}$ values are the positions of the images, $f$ values
are the fluxes of the images, ${\vect X}$ is the position of the
galaxy, and $\Delta$ indicates measured uncertainties in the
respective quantities.

Evidence for substructure is revealed when the macromodel fails to
fit the observed fluxes,\footnote{Failure to fit the image positions
could also provide evidence for substructure (see \refsec{pos}),
but we expect that in most cases existing data are not good enough
to detect this effect.  In B1422+231, the position uncertainties
from VLBA maps are very precise \citep{pat}, but as our formalism
is not currently equipped to use astrometric perturbations to
constrain substructure, we inflate the errorbars.  The ability of
astrometry to probe substructure certainly deserves further study.}
but to understand the substructure we need to identify \emph{which}
of the images are perturbed.  To do that, we systematically relax
the flux constraints and refit the macromodel.  (We always fit the
positions of all four images.)  For example, if fitting all four
fluxes fails, then we try to fit the fluxes of images A/B/C, then
images A/B/D, then A/C/D, and finally B/C/D.  If one of those cases,
say A/C/D, does provide an acceptable fit, then we settle on the
hypothesis that image B is the one most likely to be perturbed by
substructure.  Should relaxing the flux constraints on one image at
a time fail to produce an acceptable fit, we consider the six
different possibilities for relaxing two of the flux constraints.
(There is no point in relaxing three flux constraints, because the flux
of one image can always be fit trivially.)  When we fit all four fluxes,
the number of constraints on the model is $N_c=14$ if the galaxy
position is known, or $N_c=12$ if not.  With $N_p=10$ free parameters,
we can relax one or two flux constraints and still have a model that
is overconstrained or at worst has $\nu=0$ degrees of freedom.

Once we have found a macromodel that reproduces all of the observed
positions and some of the observed fluxes, we interpret the remaining
(discrepant) fluxes as evidence for substructure.  We characterize the
flux perturbation by the ratio $\mm_{\rm obs} = f_{\rm obs}/f_{\rm mod}$
of the observed flux to that predicted by the macromodel.  We can then
plug this value into our substructure analysis to find the smallest
size of an SIS clump that could produce that perturbation (as discussed
in \S 3).  The substructure analysis depends on the macromodel through
the local convergence and shear, but we show below that the statistical
uncertainties are small and unimportant for our analysis.  In other
words, formally we take the substructure goodness of fit from
eq.~(\ref{eq:chisub}), hold $\kappa$ and $\gamma$ fixed from the
macromodel, and then optimize over $u_0$ and $v_0$ to trace out
$\chi^2_{\rm sub}$ as a function of $b/a$.  We then use this function
obtain a $1\sigma$ lower limit on $b/a$.  We conservatively assume
10 per cent flux uncertainties in the substructure analysis,
dominated not by measurement uncertainties (which can reach the
per cent level; e.g., \citealt{fas1608}) but by systematic effects
such as time delays.  Modifying that assumption would produce a
fairly simple change in our mass bounds (see the Appendix), but would
not affect our conclusions.

To convert the limit on $b/a$ into a minimum Einstein radius
$b_{\rm min}$ and then to a minimum mass within that Einstein radius
$M(b)_{\rm min}$, we must specify a source size $a$.  It has been
argued that a lower bound on the size of the emitting region of a
quasar in the radio is $a \ga 1$ pc \citep{wy}, and that a reasonable
size is $a \sim 10$ pc \citep[see][]{met}.  The source size does
contribute uncertainty to our analysis, but we shall see that it does
not really affect our conclusions.

\subsection{B1422+231}
\label{sec:1422}

The 4-image radio and optical lens B1422+231 was the first system
identified as likely to contain substructure based on its anomalous
flux ratios \citep{MS}.  The fluxes of images A, B, and C violate the
relation $f_A-f_B+f_C \approx 0$ generically expected for a lens in
a `cusp configuration' corresponding to a source lying near a cusp
caustic (\citealt{sch}; \citealt{mao2}; but see \citealt{cuspreln}).
To model the lens, we use the image positions and fluxes from the
8.4 GHz observations by \citet{pa}.  Their VLBA maps yield very
precise relative positions, but we conservatively inflate the
uncertainties to 5 mas because we do not study astrometric
perturbations in detail in this paper (see \refsec{pos}).  The
radio fluxes are essentially constant \citep[see][]{pat}, so we can
neglect systematics and use the flux measurement uncertainties quoted
by \citet{pa}.  We use radio rather than optical fluxes because they
should be sensitive only to dark matter substructure (not to
microlensing by stars).  For the lens galaxy, we use the position
given by CASTLES.\footnote{CfA/Arizona Space Telescope Lens Survey;
see {\tt http://cfa-www.harvard.edu/castles}.}

Table~\ref{tb:1422sietbl} shows our results for fitting the system with
an ellipsoid+shear macromodel.  Fitting the fluxes of all four images
gives a very poor fit ($\chi^2/\nu = 113/4$), so we relax the flux
constraints one at a time to consider the possibility that one of the
images might be perturbed by substructure.  Fitting the fluxes of A,
C, and D yields an equally bad fit, so we can rule out the hypothesis
that only image B is perturbed by substructure.  The same result holds
if we fit ABC or ACD.  However, if we consider A to be perturbed, then
we get a good fit with $\chi^2/\nu = 3.1/3$.

\begin{table}
\begin{center}
  \begin{tabular}{cccccrr}
\hline
\multicolumn{7}{c}{\Large{B1422+231 ellipsoid+shear}} \\
\hline \hline
 & \multicolumn{4}{c}{Normalized Magnification} & & \\
\cline{2-5}
case & Image A & Image B & Image C & Image D & $\chi^2_{\rm tot}$ & $\chi^2_{\rm flux}$ \\
\hline \hline
abcd & 1.02 & 0.993 & 0.942 & 0.746 & 112.97 & 48.08 \\
\hline
abc  & 1.02 & 0.998 & 0.941 & 0.704 &  98.34 & 34.59 \\
abd  & 1.01 & 0.986 & 0.828 & 0.704 &  35.08 & 19.81 \\
acd  & 1.02 & 0.961 & 0.938 & 0.757 & 111.19 & 49.17 \\
bcd  & 1.21 & 1.01  & 0.995 & 0.925 &   3.07 &  0.89 \\
\hline
ab   & 1.00 & 0.990 & 0.816 & 0.649 &  12.91 &  0.90 \\
ac   & 1.02 & 0.981 & 0.939 & 0.704 &  97.99 & 35.69 \\
ad   & 1.00 & 0.829 & 0.820 & 0.793 &  10.29 &  6.83 \\
bc   & 1.21 & 1.01  & 0.993 & 0.909 &   2.40 &  0.56 \\
bd   & 1.19 & 1.00  & 0.977 & 0.909 &   2.41 &  1.16 \\
cd   & 1.23 & 1.04  & 1.00  & 0.909 &   1.71 &  0.94 \\
\hline
  \end{tabular}
\caption{
Modeling results for B1422+231 using an ellipsoid+shear macromodel.
Column 1 lists the image fluxes that were used to constrain the
macromodel.  Columns 2--5 list $\mm_{\rm obs}$, or the ratio of the observed
magnification to that predicted by the macromodel, for the four images.
Columns 6--7 give the total $\chi^2$ and the contribution from the fluxes.
}\label{tb:1422sietbl}
\end{center}
\end{table}

We then apply our substructure analysis to this model to find the minimum
clump mass required to perturb image A.  The macromodel has convergence
$\kappa =  0.381$ and shear $\gamma = 0.496$ at the position of image A.   
In order to produce a perturbation of $\mm_{\rm obs} = 1.213$, an SIS clump
in this convergence and shear field must have $b/a > 0.0561$ (1$\sigma$).
Given the source redshift $z_s = 3.62$ and lens redshift $z_l = 0.34$,
this $b/a$ bound translates to a mass within the Einstein radius of
$M(b) > 2.07 \times 10^3 (a/10\mbox{ pc})^2\Msun$, or equivalently
to a velocity dispersion of $\sigma > 2.24 (a/10\mbox{ pc})$ km s$^{-1}$.
We emphasize that we are quoting the mass within the Einstein radius,
and the total clump mass may be much larger.  The large lower limit on
the perturber mass confirms and quantifies the conventional wisdom that
microlensing cannot explain radio flux ratio anomalies.

\begin{figure}
 \begin{center}
\includegraphics[width=0.45\textwidth]{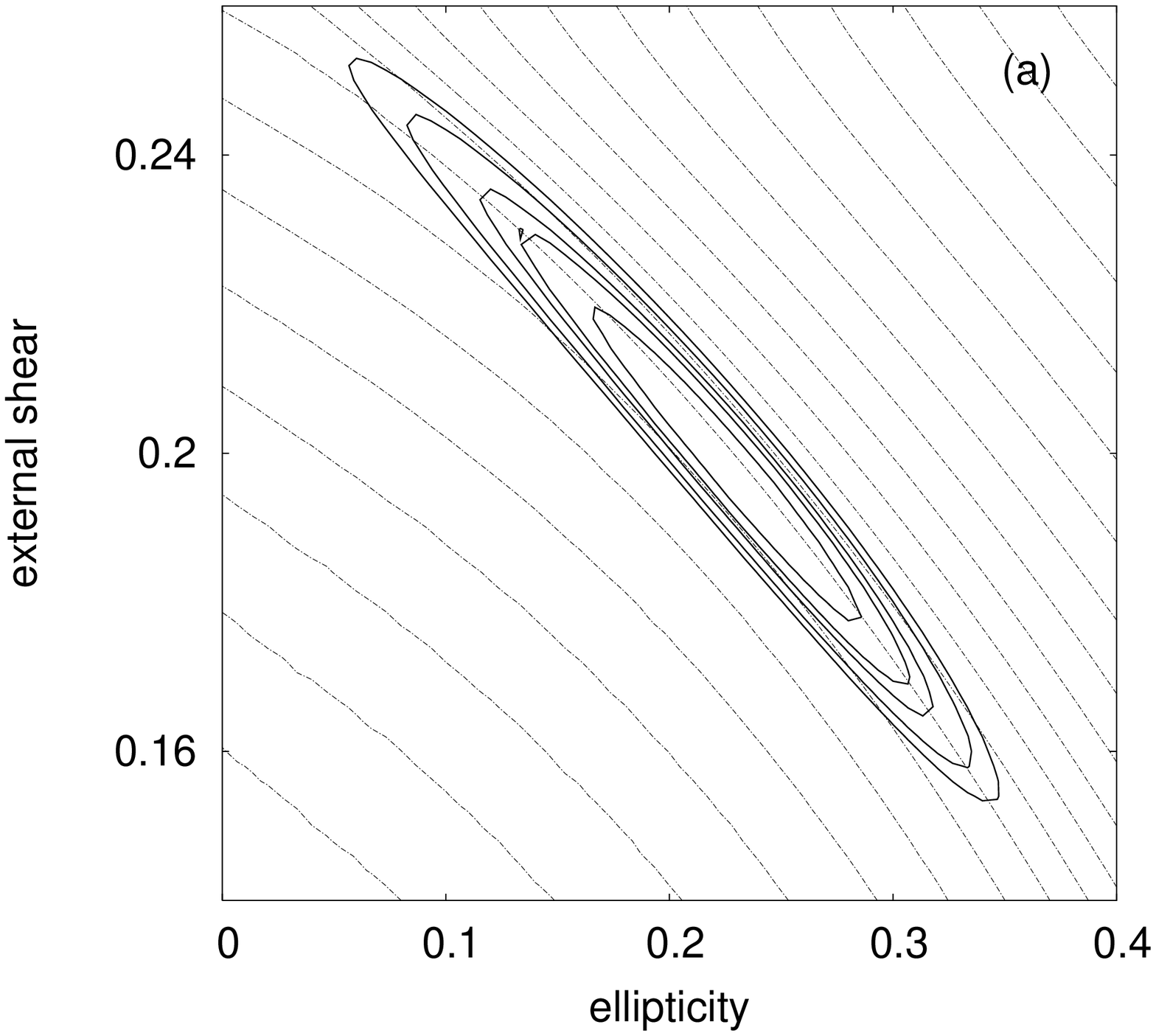}
\includegraphics[width=0.45\textwidth]{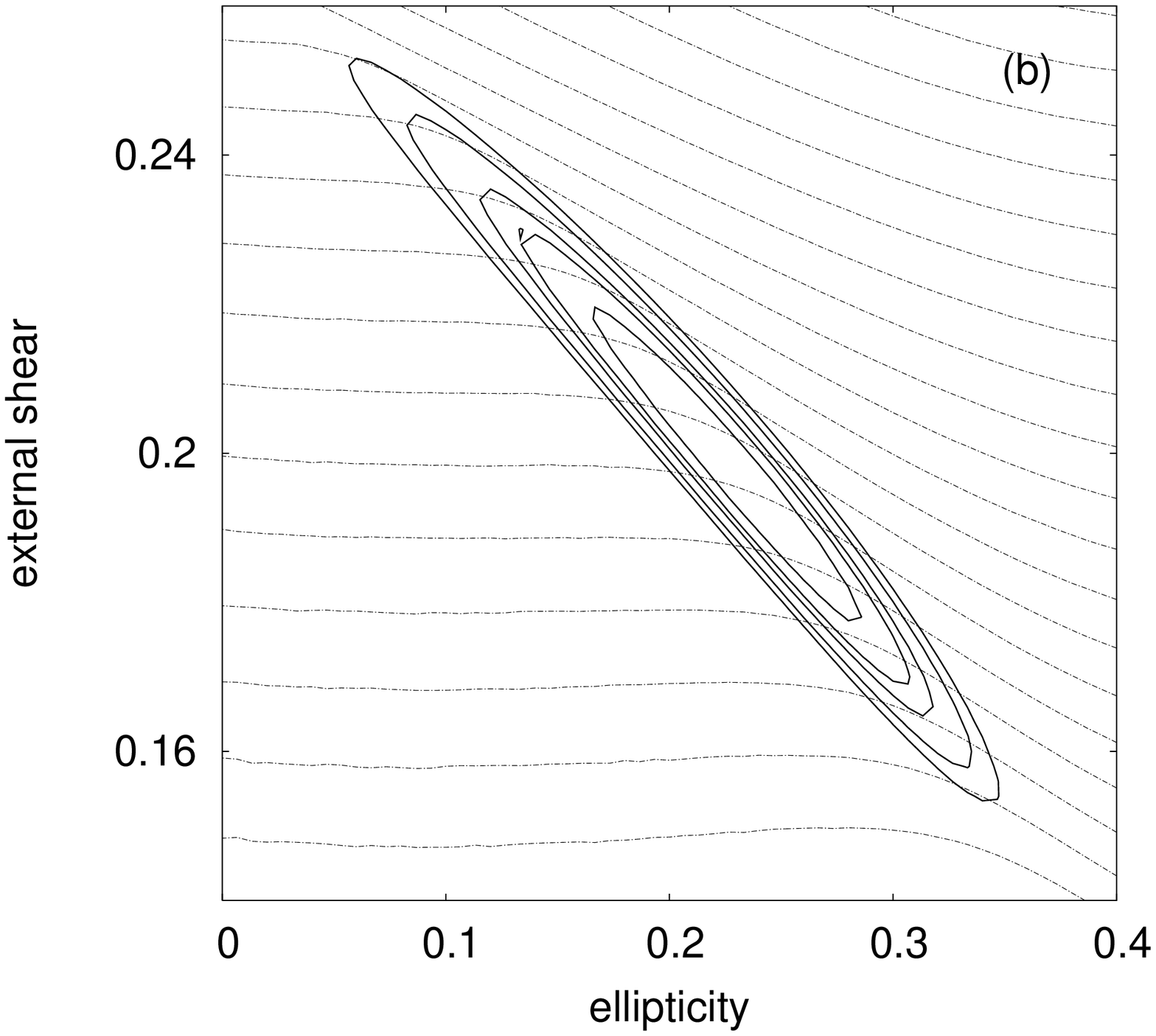}
\caption{
The ellipses show $\chi^2$ contours for B1422+231 in the ellipticity-shear
plane, showing the $1\sigma$, 90 per cent, $2\sigma$, 99 per cent, and
$3\sigma$ confidence regions.  The dotted contours are (a) $\kappa$ and
(b) $\gamma$ contours plotted in intervals of
$\delta\kappa = \delta\gamma = 0.005$.  The small variation in $\kappa$
and $\gamma$ over the ellipses implies that uncertainties in the
substructure analysis due to the macromodel are small.
}\label{fig:1422contkapgam}
 \end{center}
\end{figure}

The substructure analysis does depend on the macromodel, through the
local convergence and shear at the position of image A.  There is an
uncertainty in the macromodel between the ellipticity and external
shear which leads to an uncertainty in $\kappa$ and $\gamma$, as shown
in Fig.~\ref{fig:1422contkapgam}.  The effect is small, however:
over the $1\sigma$ confidence region of the macromodel, $\kappa$ varies
by about 0.01 and $\gamma$ varies by about 0.025.  This small variation
affects the mass bound by only $\sim$8 per cent, and the velocity
dispersion bound by even less.  Another uncertainty in the macromodel
arises from the mass sheet degeneracy, but we saw in \refsec{mag} that
this has no effect on our substructure analysis.  In other words, the
macromodel is constrained well enough for our purposes.

If we allow for the possibility of clumps in front of more than one
image, we find that there are three models that give a good fit to the
data (BC, BD, and CD; see Table \ref{tb:1422sietbl}).  All three models
still require substructure in front of image A, with mass bounds
similar to that found for the BCD model.  For each model, the $1\sigma$
lower limit on the mass in front of the other image is zero.  That is,
we can generically conclude that there must be a clump of mass
$M(b) > 2 \times 10^3\Msun$ in front of image A, and there is
no evidence of clumps in front of any other images.

\subsection{B2045+265}

B2045+265 is a 4-image radio and optical lens with the source quasar
at redshift $z_s = 1.28$ and the lens galaxy at redshift $z_l = 0.87$
\citep{fas}.  It is the tightest known cusp configuration lens, and
it exhibits a strong violation of the cusp relation in both radio and
optical bands \citep{cuspreln}.  We seek to fit the 5 GHz MERLIN
radio data from \citet{fas}, taking radio component E to indicate the
position of the lens galaxy.

Table \ref{tb:2045sietbl} shows the results of modeling this system
with an ellipsoid+shear macromodel.  Attempting to fit all of the
positions and fluxes gives a very bad fit ($\chi^2/\nu = 172/4$).
Relaxing some of the flux constraints yields somewhat better fits,
with the best case being when we only fit the fluxes of images A and C
($\chi^2/\nu = 14.4/2$).  However, all of these models underpredict
the flux of image D, by as much as a factor of 20, implying that there
must be a clump producing a very large perturbing magnification.  The
problem is that when a clump is placed in front of a negative parity
image like D, the cross section for significant magnification is very
small.  Therefore, the large magnifications shown in these models would
require a very massive clump in a very particular position.  Not only
would such a large perturbing mass almost certainly result in resolvable
splittings of image D that are not observed, it would probably affect
the positions and fluxes of the other images as well.

\begin{table}
\begin{center}
  \begin{tabular}{cccccrr}
\hline
\multicolumn{7}{c}{\Large{B2045+265 ellipsoid+shear}} \\
\hline \hline
 & \multicolumn{4}{c}{Normalized Magnification} & & \\
\cline{2-5}
case & Image A & Image B & Image C & Image D & $\chi^2_{\rm tot}$ & $\chi^2_{\rm flux}$ \\
\hline \hline
abcd & 1.80  & 0.726 & 3.14  & 17.32 & 172.17 & 171.24 \\
\hline
abc  & 1.84  & 0.744 & 3.22  & 17.59 &  81.80 &  80.48 \\
abd  & 1.99  & 0.805 & 3.48  & 18.75 & 124.29 & 122.21 \\
acd  & 0.796 & 0.321 & 1.39  &  7.46 &  93.05 &  90.94 \\
bcd  & 2.05  & 0.828 & 3.58  & 19.33 & 150.05 & 148.05 \\
\hline
ab   & 2.06  & 0.829 & 3.59  & 20.00 &  31.55 &  30.77 \\
ac   & 0.844 & 0.339 & 1.47  &  8.21 &  14.43 &  13.66 \\
ad   & 0.912 & 0.367 & 1.59  &  8.54 &  82.57 &  80.45 \\
bc   & 2.12  & 0.856 & 3.71  & 20.67 &  56.97 &  56.19 \\
bd   & 2.38  & 0.960 & 4.15  & 22.41 &  95.25 &  93.28 \\
cd   & 0.497 & 0.200 & 0.869 &  4.63 &  67.29 &  65.02 \\
\hline
  \end{tabular}
\caption{
Modeling results for B2045+265 using an ellipsoid+shear macromodel.
}\label{tb:2045sietbl}
\end{center}
\end{table}

\begin{table}
\begin{center}
  \begin{tabular}{cccccrr}
\hline
\multicolumn{7}{c}{\Large{B2045+265 internal+external shear}} \\
\hline \hline
 & \multicolumn{4}{c}{Normalized Magnification} & & \\
\cline{2-5}
case & Image A & Image B & Image C & Image D & $\chi^2_{\rm tot}$ & $\chi^2_{\rm flux}$ \\
\hline \hline
abcd & 1.88  & 0.743 & 3.31  &   1.00 & 89.57 & 82.72 \\
\hline
abc  & 1.72  & 0.741 & 2.81  & 354.38 & 71.89 & 71.40 \\
abd  & 2.09  & 0.827 & 3.68  &   1.00 & 38.66 & 31.74 \\
acd  & 0.841 & 0.333 & 1.48  &   1.00 & 20.25 & 14.09 \\
bcd  & 2.17  & 0.856 & 3.81  &   1.00 & 64.26 & 57.31 \\
\hline
ab   & 1.93  & 0.829 & 3.15  & 381.13 & 28.03 & 27.53 \\
ac   & 0.852 & 0.365 & 1.39  & 166.94 & 11.42 & 10.92 \\
ad   & 0.995 & 0.393 & 1.75  &   1.00 &  6.31 &  0.00 \\
bc   & 1.97  & 0.846 & 3.21  & 396.07 & 51.22 & 50.72 \\
bd   & 2.52  & 0.996 & 4.44  &   1.00 &  7.05 &  0.00 \\
cd   & 0.565 & 0.224 & 0.994 &   1.00 &  5.74 &  0.01 \\
\hline
  \end{tabular}
\caption{
Modeling results for B2045+265 using an internal+external shear macromodel.
}\label{tb:2045intshrtbl}
\end{center}
\end{table}

These problems lead us to consider the alternate internal+external
shear macromodel, whose results are given in Table
\ref{tb:2045intshrtbl}.  Trying to fit all of the fluxes still gives
a poor fit ($\chi^2/\nu = 90/4$).  Dropping one of the flux constraints
improves the fit, but even the best case is still not an acceptable fit
(the ACD model has $\chi^2/\nu = 20/3$).  Dropping two flux constraints,
however, yields three acceptable models.  Fitting A and D gives
$\chi^2/\nu = 6.31/2$, fitting B and D gives $\chi^2/\nu = 7.05/2$,
and fitting C and D gives $\chi^2/\nu = 5.74/2$; all three models
differ from the data by only $\sim\!2\sigma$.  Although the CD case
has the lowest $\chi^2$, it requires the positive parity A image to
be demagnified by about a factor of two, which is not possible with
an SIS clump.  So we are only left with the AD and BD cases as viable
models for the system.  The AD model has substructure perturbing masses
in front of images B ($\mm_{\rm obs} = 0.3938$) and C ($\mm_{\rm obs} = 1.7518$)
while the BD model has perturbers in front of images A
($\mm_{\rm obs} = 2.5283$) and C ($\mm_{\rm obs} = 4.4474$).

The results of applying our substructure analysis are shown in Table
\ref{tb:massestbl}.  In the BD model, the substantial magnifications
of images A and C require large clumps: the mass in front of A must be 
larger than
$2.29 \times 10^6 \Msun$ ($\sigma > 15.51$ km s$^{-1}$),
and the mass in front of image C must be larger than
$1.58 \times 10^7 \Msun$ ($\sigma > 25.12$ km s$^{-1}$).  In the
AD model, the minimum masses necessary to reproduce the anomalous
fluxes are somewhat smaller:
$M(b) > 4.77 \times 10^5 \Msun$ ($\sigma > 10.78$ km s$^{-1}$) for
image B, and
$M(b) > 3.71 \times 10^5 \Msun$ ($\sigma > 9.84$ km s$^{-1}$) for
image C.
(We have again assumed a source size $a = 10$ pc.)  Although we cannot
identify a unique model, the important points are that the clump masses
are $\gg \Msun$ and therefore exclude microlensing as an explanation
for the observed flux ratio anomalies, and also that they agree well
with the sizes of clumps predicted by a $\Lambda$CDM cosmology
\citep[e.g.,][]{klypin,moore}.

\begin{figure}
 \begin{center}
\includegraphics[width=0.45\textwidth]{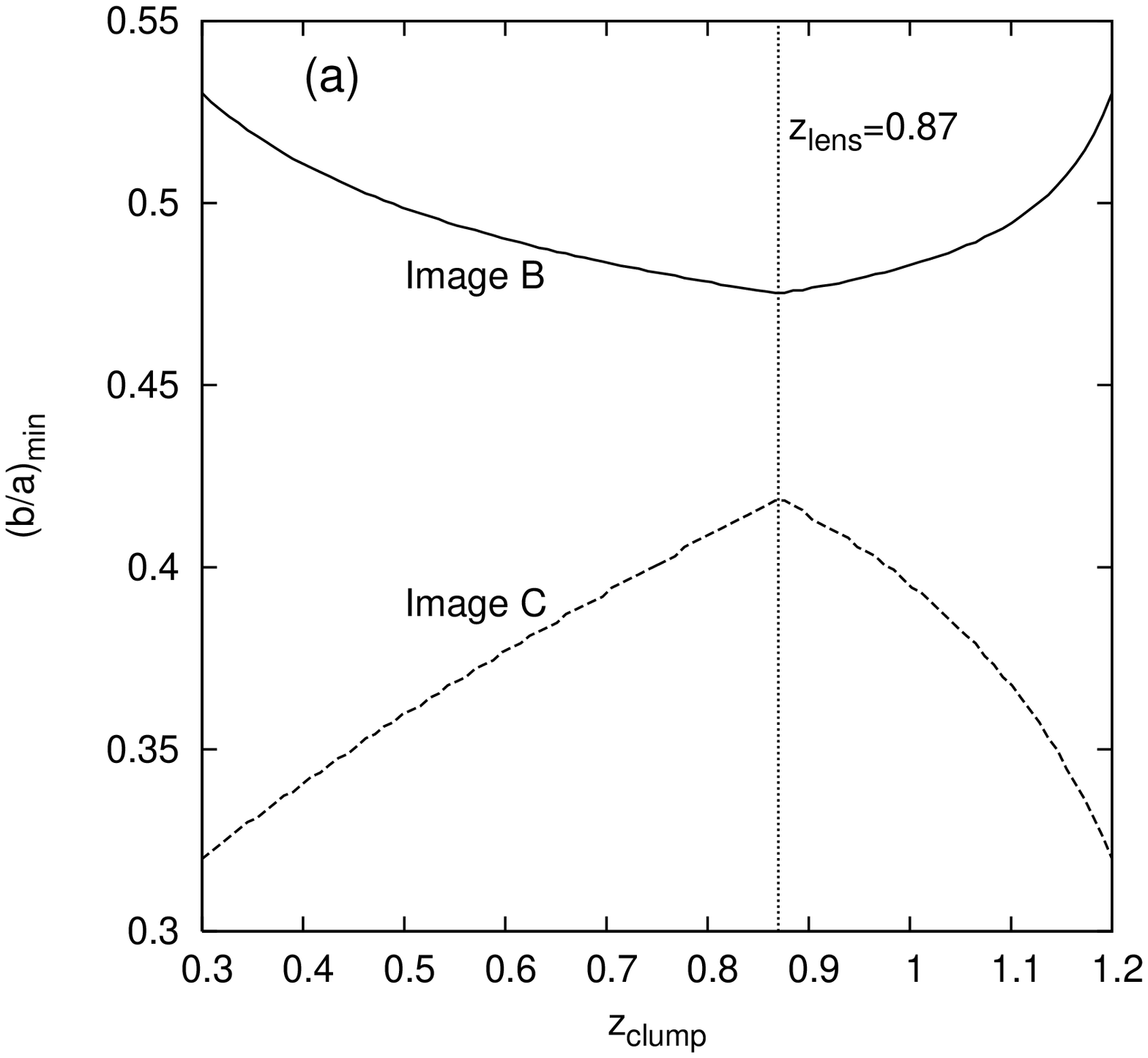}
\includegraphics[width=0.45\textwidth]{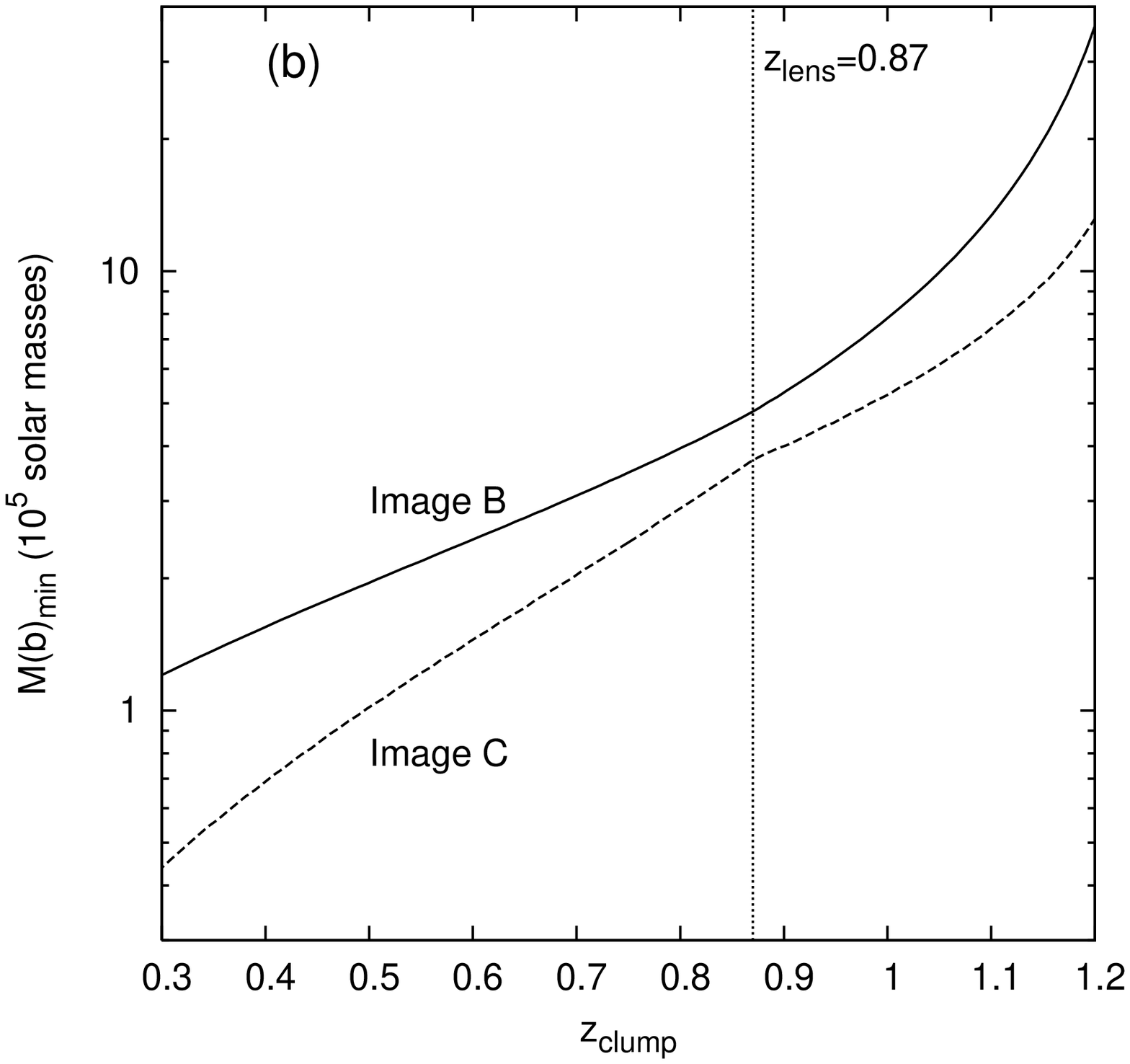}
\caption{
Dependence of the clump bound on the clump redshift, for the two
perturbed images in the AD internal+external shear model for B2045+265.
(a) The lower bound on $b/a$.
(b) The lower bound on the mass within the Einstein radius, $M(b)$.
}\label{fig:losbounds}
 \end{center}
\end{figure}

The mass bounds given above were calculated for a clump lying within
halo of the main lens galaxy.  It is possible, though, that the clump
lies elsewhere along the line of sight \citep{analytics,metcalfLOS};
the likelihood depends on the relative abundances of embedded and
isolated clumps \citep[see][]{chen}.  As discussed in \S 2, our
formalism can easily accommodate the hypothesis that the clump lies
at redshift $z_c \ne z_l$ (see eq.~\ref{eq:LOS}).
Fig.~\ref{fig:losbounds} shows how the clump bounds for images B
and C vary with clump redshift for the AD internal+external shear
model.  Moving the clump in redshift away from the lens galaxy
increases the lower bound on $b/a$ for negative parity image B, and
decreases it for positive parity image C.  (These dependences can be
understood in terms of eq.~\ref{eq:LOS} and Fig.~\ref{fig:fmaxkapgam}.)
However, the effect is only tens of percent over a wide range in
redshift.  A stronger variation is seen in the mass bound, because
of the additional redshift dependence in the lensing critical density
($\Sigma_{\rm cr} \propto D_{os}/D_{ol} D_{os}$).  Even so, the change is
a factor of a few, so uncertainty in the location of the clump along
the line of sight does not significantly affect order of magnitude
conclusions.

\subsection{B1555+375}

B1555+375 is a faint 4-image radio lens discovered by \citet{mar},
whose fluxes violate the relation $f_A - f_B \approx 0$ expected for
a lens in a `fold configuration' \citep{foldreln}.  We fit the 5 GHz
data from Marlow et al.  The position of the lens galaxy with respect
to the images has not been measured.  The lens and source redshifts
are not known, but Marlow et al.\ estimate them to lie in the ranges
$1.0 \la z_s \la 3.0$ and $0.5 \la z_l \la 1.0$.

Our attempts to fit this system with an ellipsoid+shear macromodel
result in models with very large and perpendicularly aligned
ellipticities and shears ($e \sim 0.9$, $\gamma_{\rm ext} \sim 0.3$, and
$\Delta\theta = 90^\circ$).  These models are highly contrived, and
have extremely large and implausible magnifications.  We consider
them to be unphysical, and turn instead to internal+external shear
macromodels.  

Fitting all four images yields a model with $\chi^2/\nu = 45.6/2$ that
reproduces the image positions well but not the flux ratios.  Relaxing
the flux constraints in front of one image can improve the fit, but
the resulting models are unacceptable in that they require the positive
parity image A to be demagnified or the negative parity image D to
be highly amplified by a clump.  The only acceptable models are found
when we relax two flux constraints, and in fact there are two good
cases (see Table \ref{tb:massestbl}).  One possibility is to have a
clump in front of image B with $b/a > 0.328$, plus a clump in front of
image C with $b/a > 0.214$.  This model fits the data perfectly, which
is not surprising because it has $\nu = 0$ degrees of freedom.  Assuming
redshifts of $z_s = 2.0$ and $z_l = 0.75$, the $b/a$ bounds 
translate into clump mass limits of 
$M > 1.19 \times 10^5 \Msun$ ($\sigma > 6.30$ km s$^{-1}$) and
$M > 5.08 \times 10^4 \Msun$ ($\sigma > 5.09$ km s$^{-1}$) for
images B and C, respectively.  Varying the redshifts can change the
mass limits by a factor of a few up to $\sim$10, but does not affect
the conclusion that the fluxes cannot be explained by microlensing.
The other possibility is to have a clump in front of image A with
$b/a > 0.420$ ($M > 1.96 \times 10^5 \Msun$, or
$\sigma > 7.14$ km s$^{-1}$), plus a clump in front of image C with
$b/a > 1.783$ ($M > 3.54 \times 10^6 \Msun$, or
$\sigma > 14.72$ km s$^{-1}$).  This model gives $\chi^2 = 1.045$ for
$\nu = 0$, which is formally unacceptable.  However, as an exercise
we added random noise to the data and refit.  A substantial fraction
of these cases yielded $\chi^2 = 0$, which suggests that the model is
in fact consistent with the data given the measurement uncertainties.

\begin{table}
  \begin{center}
  \begin{tabular}{ccccccc}
\hline
\multirow{2}{*}{Lens} & \multirow{2}{*}{Macromodel} & Perturbed & \multirow{2}{*}{$\chi^2/\nu$} & \multirow{2}{*}{Image} & $M(b)_{\rm min}$  & $\sigma_{\rm min}$ \\
                      &                             & Image(s)  &                               &                        & (M$_{\odot}$) & (km s$^{-1}$)  \\
\hline \hline
B1422+231                  & ellipsoid+shear                          & A                        & $3.073/3$                  & A & $2.07 \times 10^3$ & $ 2.24$ \\
\hline \hline
\multirow{4}{*}{B2045+265} & \multirow{4}{*}{internal+external shear} & \multirow{2}{*}{A and C} & \multirow{2}{*}{$7.051/2$} & A & $2.29 \times 10^6$ & $15.51$ \\
                           &                                          &                          &                            & C & $1.58 \times 10^7$ & $25.12$ \\
\cline{3-7}
                           &                                          & \multirow{2}{*}{B and C} & \multirow{2}{*}{$6.320/2$} & B & $4.77 \times 10^5$ & $10.48$ \\
                           &                                          &                          &                            & C & $3.71 \times 10^5$ & $ 9.84$ \\
\hline \hline
\multirow{4}{*}{B1555+375} & \multirow{4}{*}{internal+external shear} & \multirow{2}{*}{A and C} & \multirow{2}{*}{$1.045/0$} & A & $1.96 \times 10^5$ & $ 7.14$ \\
                           &                                          &                          &                            & C & $3.54 \times 10^6$ & $14.72$ \\
\cline{3-7}
                           &                                          & \multirow{2}{*}{B and C} & \multirow{2}{*}{$0.000$/0} & B & $1.19 \times 10^5$ & $ 6.30$ \\
                           &                                          &                          &                            & C & $5.08 \times 10^4$ & $ 5.09$ \\
\hline
  \end{tabular}
\caption{
The $1\sigma$ lower limits on the mass within the Einstein radius
and the velocity dispersion of perturbing clumps.  There is only one
acceptable model for B1422+231, but there are two possibilities for
B2045+265 and B1555+375.  The bounds scale with the assumed source
size as $M(b)_{\rm min} \propto (a/10\mbox{ pc})^2$ and
$\sigma_{\rm min} \propto (a/10\mbox{ pc})$.
}\label{tb:massestbl}
  \end{center}
\end{table}

\subsection{B0712+472}

B0712+472 is a 4-image lens with an image configuration intermediate
between a cusp and fold \citep{jack}.  The optical flux ratios
strongly violate the cusp and fold relations, but at radio wavelengths
the violation is marginal \citep{cuspreln,foldreln}.  The difference
suggests that the optical flux ratios are affected by microlensing.
We focus on the radio flux ratios, and fit the data given by
\citet{jack00}.  As shown in Table \ref{tb:0712tbl}, attempting to
fit all four radio fluxes yields $\chi^2/\nu = 18.77/4$ for an
ellipsoid+shear macromodel, or $\chi^2/\nu = 7.29/4$ for an
internal+external shear macromodel.  The internal+external shear
model differs from the data at only 88 per cent confidence, so it is
a reasonably good fit.  Nevertheless, it is interesting to consider
whether the fit can be improved by relaxing the flux constraints,
as shown in Table \ref{tb:0712tbl}.  Among three-flux models, only
ABD yields a noticeably better fit.  Among two-flux models, only
the AD case yields a reasonable model that gives a better fit.
(The AB model can be ruled out because it requires the negative
parity image D to be magnified by a factor of $\sim$20 relative to
the macromodel; while the CD model can be ruled out because it
requires the positive parity image A to be demagnified by a factor
of $\sim$2.)

Although we see that relaxing flux constraints lowers the $\chi^2$,
we must ask whether that really provides evidence for substructure,
or whether it just indicates that we are using fewer constraints.
The test for statistical significance when removing degrees of freedom
is called the F-test \citep[e.g.,][]{bev}.  The F-test returns a
probability that the change in $\chi^2$ is due simply to the change
in the number of degrees of freedom -- so a \emph{low} value of the
probability indicates that the fit really has improved.  The test
results are given in Table \ref{tb:0712tbl}, and they confirm our
intuition that many of the three-flux and two-flux models are not
significantly better than the ABCD model.  However, the ABD and AD
models have relatively low F-test values (0.08 and 0.05, respectively),
so we conclude that there is marginal evidence for a radio flux
ratio anomaly, and if real it is probably in image C.

\begin{table}
\begin{center}
  \begin{tabular}{crrrcrrr}
\hline
\multicolumn{8}{c}{\Large{B0712+472}} \\
\hline \hline
& \multicolumn{3}{c}{internal+external shear} && \multicolumn{3}{c}{ellipsoid+shear}\\
\cline{2-4} \cline{6-8}
case & $\chi^2_{\rm tot}$ & $\chi^2_{\rm flux}$ & F-test && $\chi^2_{\rm tot}$ & $\chi^2_{\rm flux}$ & F-test \\
\hline 
abcd & 7.29 & 1.49 & $\equiv 1.00$ && 18.83 & 9.37 & $\equiv 1.00$ \\
\hline
abc  & 7.06 & 1.41 &         0.85  && 14.59 & 7.94 &         0.20  \\
abd  & 4.51 & 0.98 &         0.08  && 10.34 & 7.00 &         0.06  \\
acd  & 6.59 & 2.96 &         0.50  &&  9.42 & 2.26 &         0.04  \\
bcd  & 6.48 & 0.63 &         0.44  &&  6.06 & 0.03 &         0.02  \\
\hline
ab   & 2.88 & 1.20 &         0.06  &&  2.22 & 1.63 &         0.01  \\
ac   & 6.48 & 2.95 &         0.49  &&  3.46 & 0.01 &         0.02  \\
ad   & 2.42 & 0.73 &         0.05  &&  2.02 & 0.02 &         0.01  \\
bc   & 6.44 & 0.65 &         0.47  &&  4.10 & 0.00 &         0.03  \\
bd   & 4.48 & 0.99 &         0.14  &&  2.88 & 0.02 &         0.02  \\
cd   & 0.72 & 0.11 &         0.01  &&  0.98 & 0.02 &         0.01  \\
\hline
  \end{tabular}
\caption{
Modeling results for B0712+472 using both macromodels.  The F-test
gives the probability that $\chi^2$ has decreased (relative to the
abcd model) only because the model has fewer constraints, rather than
because the fit is significantly better.
}\label{tb:0712tbl}
\end{center}
\end{table}

\section{Conclusions}

We have developed a semi-analytic formalism for computing the effects
of substructure on the lensed images of a finite-size source.  By
considering the local effects of a clump modeled as an isothermal
sphere, we can solve analytically for the perturbed micro-image(s),
and then compute numerically the change in the position and
magnification of the macro-image.  While this is a simplified toy
model, it yields valuable insight into the general features of finite
source effects in substructure lensing:
\begin{itemize}

\item
The perturbations do not have a simple dependence on source size,
but are related to intersections of the source with micro-caustics.

\item
Positive parity images are always amplified by isothermal clumps,
but negative parity images may be either amplified or suppressed
depending on the source position and size.

\item
Sources that are more than an order of magnitude larger than the
clump Einstein radius can still be perturbed at the percent level,
which mildly contradicts conventional wisdom that a source cannot
`feel' lensing structure on scales smaller than itself.

\item
Statistical uncertainties in the macromodel do not significantly
affect the substructure analysis.  Remarkably, the mass sheet
degeneracy in the macromodel has \emph{no effect} on the
substructure analysis, at least for isothermal clumps.

\item
Astrometric perturbations could be at the few milli-arcsec level,
and could be identified by comparing the relative image positions
at different wavelengths.

\end{itemize}
The bottom line is that there is a tremendous amount to be learned
from high-resolution observations at a variety of wavelengths that
correspond to different source sizes.  The promising possibilities
are observations of the optical, mid-IR, and radio continua, and
the optical emission lines.  (Detailed X-ray observations seem less
valuable, because the source will be much smaller than the caustics
for millilensing.)  The first steps in this direction have been taken
\citep{agol,0435b,met}, but a more concerted effort to do this for
flux ratio anomaly lenses is called for.

It is already possible to place limits on the substructure mass scale:
since there is a finite range of magnifications possible for a given
ratio of the clump Einstein radius $b$ to the source size $a$, an
observed flux perturbation leads directly to a lower bound on $b/a$
(even with no knowledge of the relative positions of perturber and
source).  Adding knowledge of (or assumptions about) the source size
then leads to a lower bound on the clump mass.  These substructure
bounds do depend on our assumption that each flux ratio anomaly is
caused by a single, isolated, isothermal clump; how they change for 
different clump models and for the limit of moderate or high optical 
depth will be the subject of a follow-up study.  Still, the principle
that finite source effects permit simple \emph{lower} bounds on
the substructure mass scale should be general.

With this background, we have sought to understand three known
lensing systems with strong flux ratio anomalies at radio wavelengths
(B1422+231, B1555+375, and B2045+265), plus one system with marginal
evidence for a radio flux ratio anomaly (B0712+472).  We carefully
examined macromodels consisting of an isothermal lens galaxy with
different types of angular structure, in order to determine which
of the lensed images are perturbed and by how much.  Assuming
isothermal clumps, we could then use our substructure analysis to
place lower bounds on the clump masses.  For B1422+231, we find
strong evidence for a clump in front of image A, and the mass
within the clump Einstein radius must be
$M > 2 \times 10^3\,(a/10\mbox{ pc})^2 \Msun$.  For B2045+265 and
B1555+375, we find strong evidence for clumps in front of \emph{two}
images in each systems.  The masses within the Einstein radii are
$M \ga 10^4$--$10^7 \Msun$, which generally agreees with $\Lambda$CDM
predictions, although it is important to consider whether $\Lambda$CDM
predicts enough clumps to explain the presence of multiple anomalies
in multiple lenses.  In B0712+472, there is marginal evidence for a
clump in front of image C.  We emphasize that our identification of
the images that are perturbed is independent of assumptions about
the nature of substructure; those assumptions enter only when we
derive quantitative clump mass bounds.

To round out our analysis, we have considered several systematic
effects in the substructure mass bounds.  As noted above, statistical
uncertainties in the macromodel propagate into the substructure
analysis, but their effects are small.  In many lensing applications
the main problem is the mass sheet degeneracy in the macromodel, but
we have shown that it has no effect on the substructure analysis,
at least for isothermal clumps.  Thus, it turns out that the main
systematics are the uncertainty in the source size, and lack of
knowedlge about whether the clump lies within the main lens galaxy
(a standard assumption) or elsewhere along the line of sight.
Varying the clump redshift over a reasonable range can change the
clump mass bounds by a factor of a few.  So it is certainly important
for detailed quantitative results, but not so important for order
of magnitude reasoning.

Ultimately, using flux ratio anomalies to test $\Lambda$CDM and draw
conclusions about the nature of dark matter relies on sophisticated
statistical analyses with realistic clump models
\citep[e.g.,][]{DK,KD,met}.  Toy models like ours are still valuable,
though, because they reveal and clarify the general principles on
which the sophisticated analyses are based.  For example, our results
suggest that looking for effects requiring comparable scales for the
source size and the lensing substructure will be the best way to
distinguish the substructure explanation for flux ratio anomalies
from competing hypotheses that may be disfavored but not yet ruled
out.  Furthermore, we believe that a detailed understanding of flux
ratio anomalies in individual lenses will always be an important
complement to `black box' statistical machinery.

\appendix

\section{Behavior at large source size}

\subsection{Taylor series expansion of the magnification}

We noted in \refsec{mag} that at large $a$ the magnification appeared
to be roughly independent of source position.  We now confirm rigorously
that it is independent of source position to first order in $1/a$.
First, we define some quantities to simplify the notation:
\begin{eqnarray*}
c_1 & = & 2u_0(1-\kappa-\gamma)\cos\theta + 2b(1-\kappa-\gamma\cos 2\theta)
  + 2v_0(1-\kappa+\gamma)\sin\theta \\
c_2 & = & b^2 + u_0^2 + v_0^2 +2\,b\,u_0 \cos\theta + 2\,b\,v_0\sin\theta \\
c_3 & = & 4[u_0(1-\kappa-\gamma)\cos\theta+b(1-\kappa-\gamma\cos 2\theta)
  + v_0(1-\kappa+\gamma)\sin\theta]^2 \\
d_1 & = & 4[(1-\kappa)^2+\gamma^2-2\gamma(1-\kappa)\cos 2\theta]
\end{eqnarray*}
Note that $c_1$, $c_2$, and $c_3$ all depend on the source position,
while $d_1$ does not.  With these definitions, we can write
\begin{equation}
  \frac{r_{\pm}(\theta)}{a} = \frac{2}{d_1} \left( \xi c_1
    \pm \sqrt{(c_3-c_2 d_1)\xi^2 + d_1} \right)\,,
\end{equation}
where $\xi \equiv 1/a$.  We immediately see that as $a \to \infty$
($\xi \to 0$), $r_{-}/a$ is negative while $r_{+}/a$ is positive and
finite, so the image boundary is formed only by $r_{+}$, and the
magnification $\mm \propto \int [r_{+}(\theta)/a]^2\,d\theta$ remains
finite.  Expanding in $\xi$, we find:
\begin{eqnarray}
  \mm &=& \frac{(1-\kappa)^2-\gamma^2}{2\pi}
    \int \left(\frac{r_{+}(\theta)}{a}\right)^2\,d\theta \nonumber \\
  &=& \frac{(1-\kappa)^2-\gamma^2}{2\pi}
    \int \left[ \frac{4}{d_1} + \frac{8 c_1}{d_1^{3/2}}\ \xi
    + \mathcal{O}(\xi^2) \right]\,d\theta \nonumber \\
  &=& 1 + \xi\ \frac{(1-\kappa)^2-\gamma^2}{2\pi}
    \int \frac{2b(1-\kappa-\gamma\cos 2\theta)}
      {[(1-\kappa)^2+\gamma^2-2\gamma(1-\kappa)\cos 2\theta]^{3/2}}\ d\theta
    + \mathcal{O}(\xi^2) \label{eq:Mser}
\end{eqnarray}
where in the last step we used the fact that any periodic function
whose period is an odd multiple of $\pi$ integrates to zero.  The zeroth
order term shows that a sufficiently large source is insensitive to the
perturber; it has a normalized magnification $\mm \approx 1$,
meaning that it only feels the convergence and shear field.  In the
first order term, the integral cannot be evaluated analytically, but
the important result is that it does not depend on the source position
$(u_0,v_0)$.  Thus, to first order in $1/a$, the magnification is
independent of source position.  Carrying the expansion further reveals
that $u_0$ and $v_0$ enter only at second order in $1/a$.

\subsection{Substructure limits at large source size}

\begin{figure}
 \begin{center}
\includegraphics[width=0.35\textwidth]{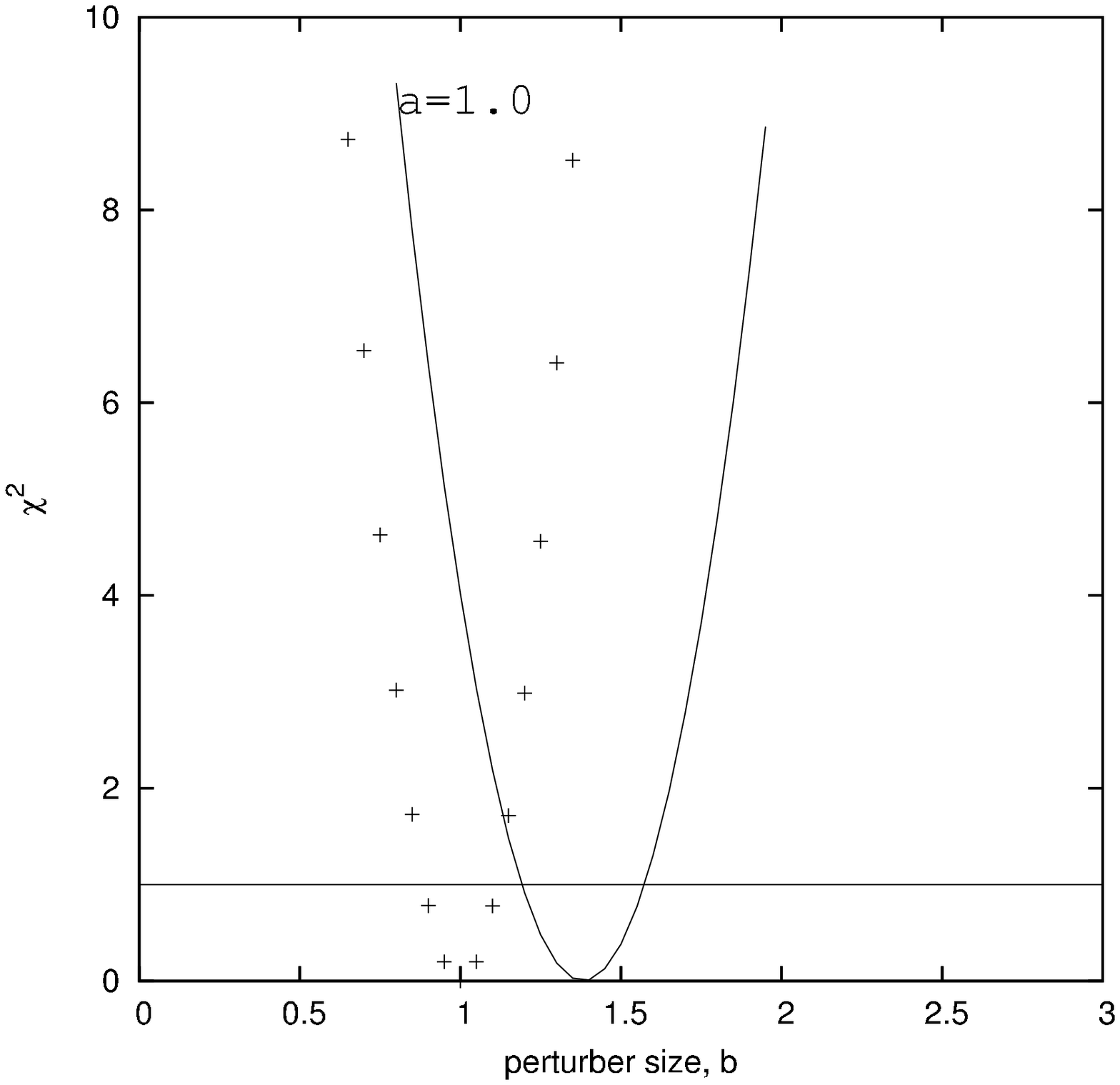}
\includegraphics[width=0.35\textwidth]{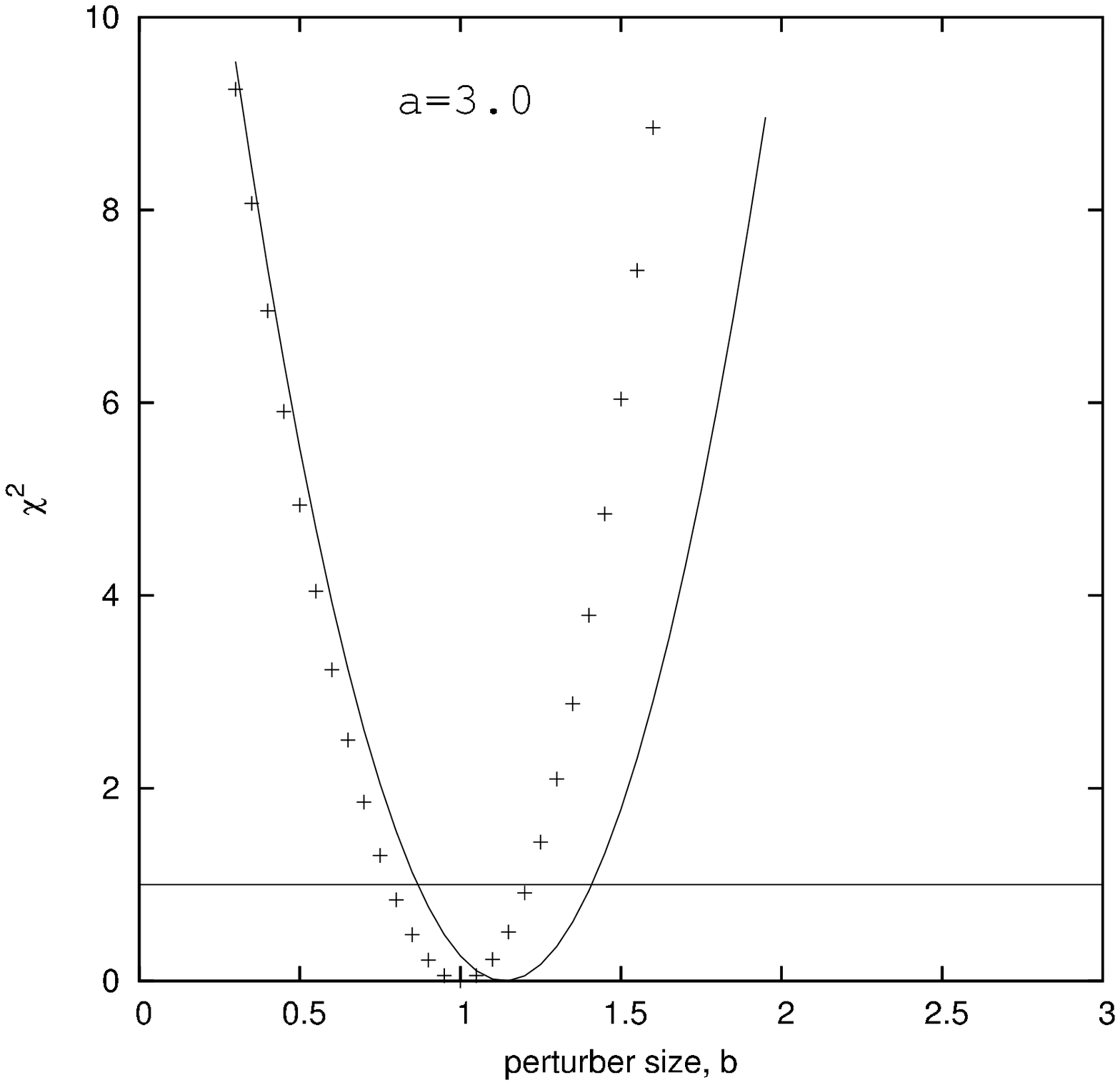}
\linebreak
\includegraphics[width=0.35\textwidth]{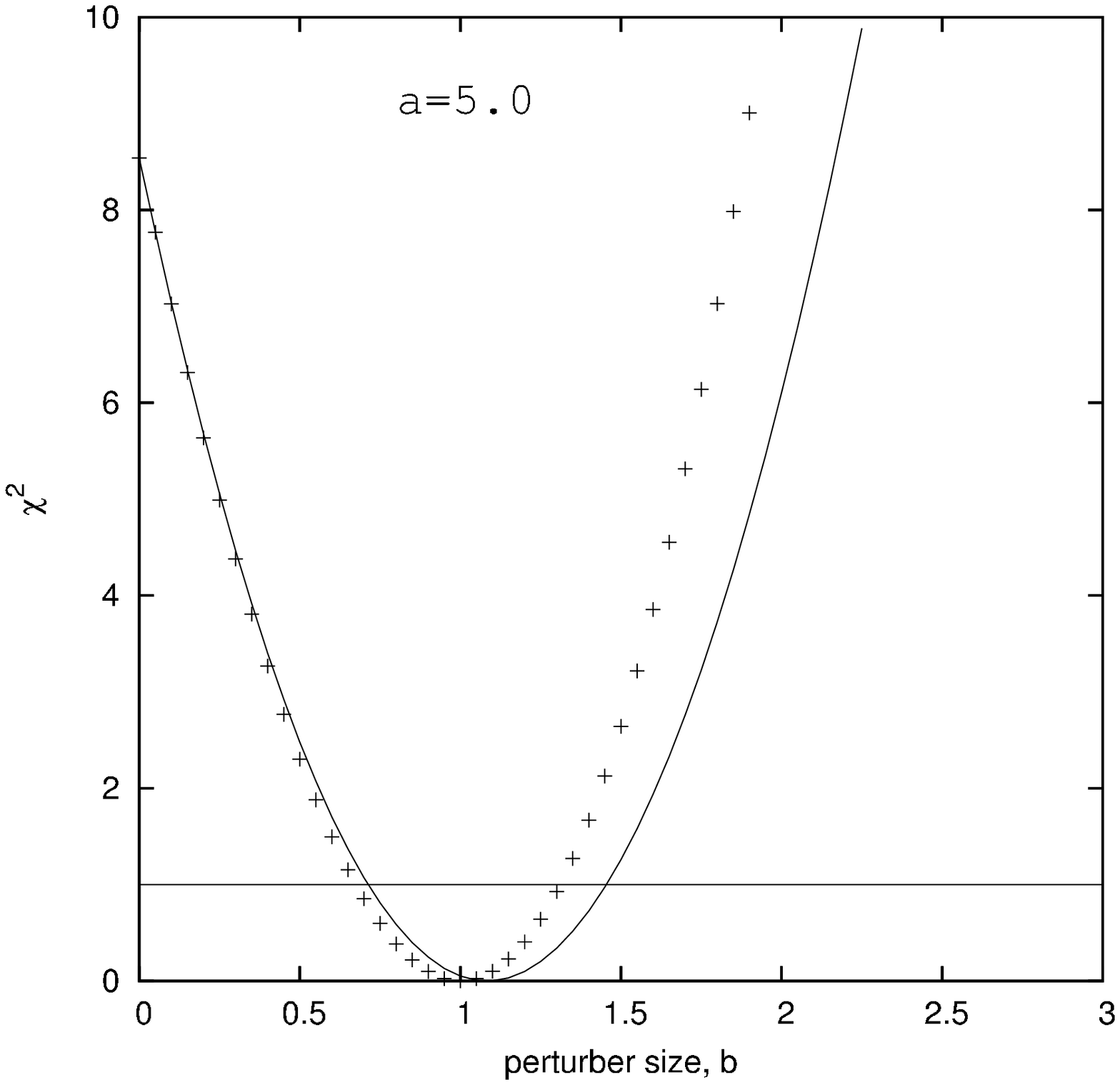}
\includegraphics[width=0.35\textwidth]{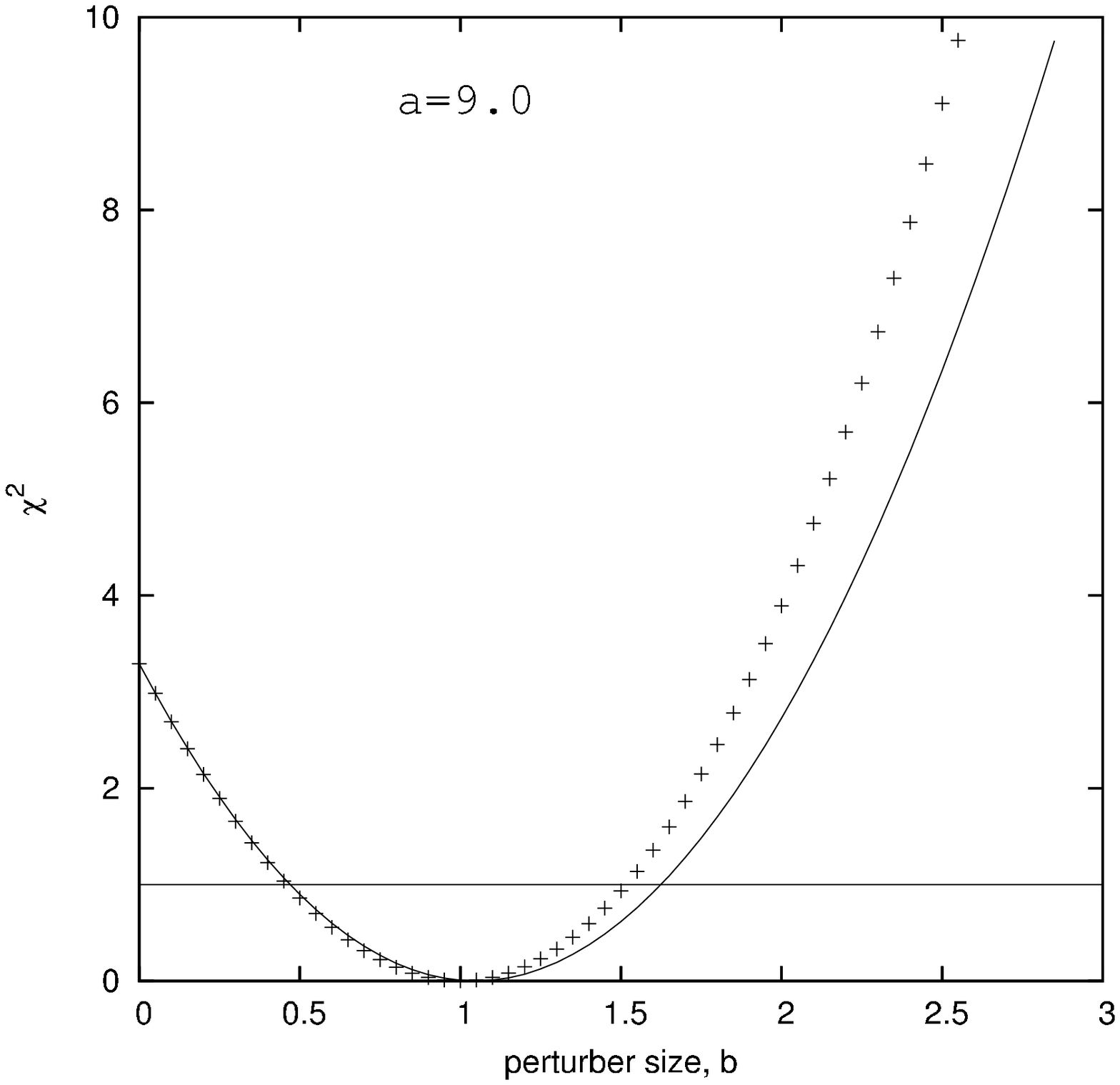}
\linebreak
\includegraphics[width=0.35\textwidth]{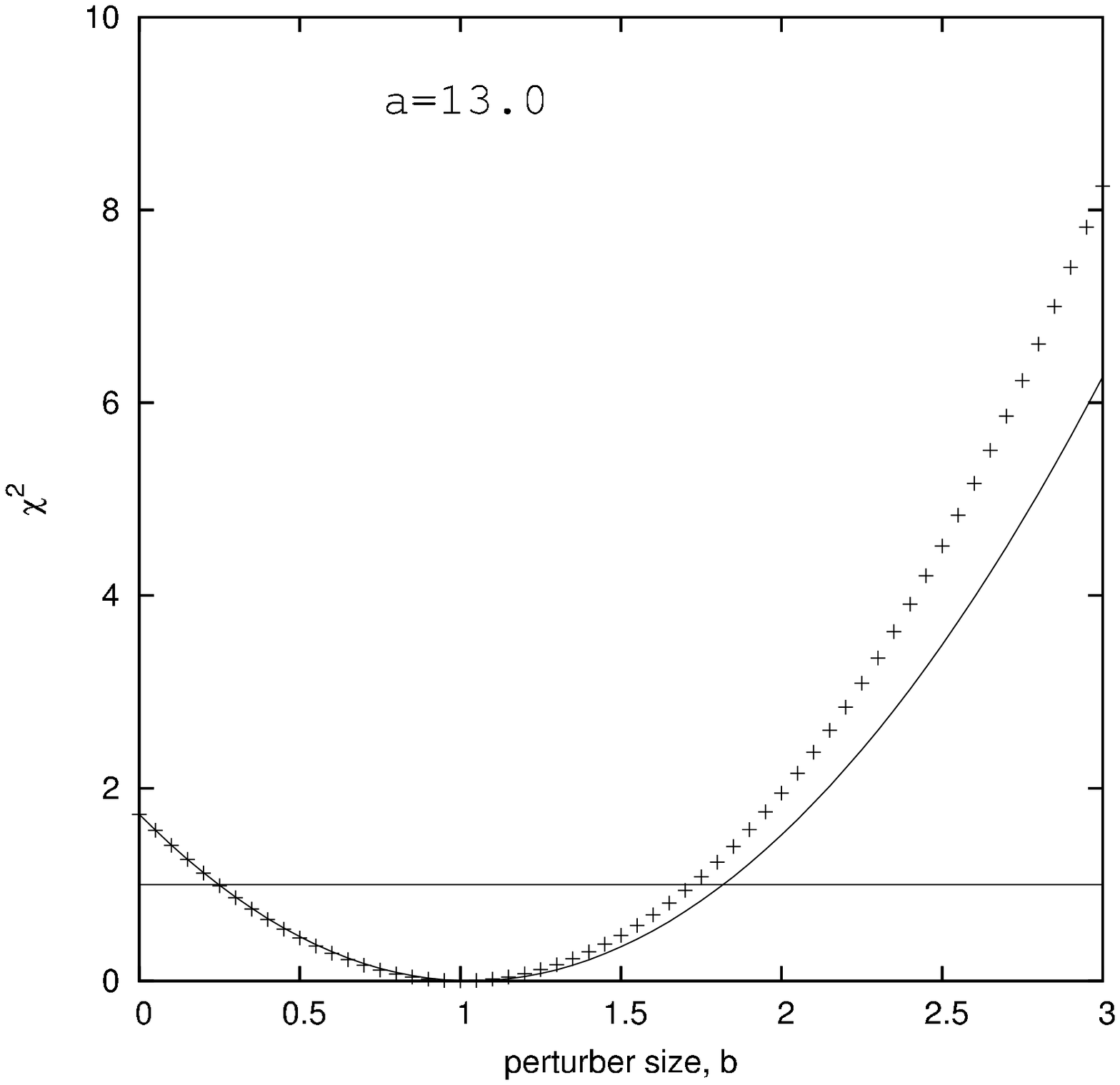}
\includegraphics[width=0.35\textwidth]{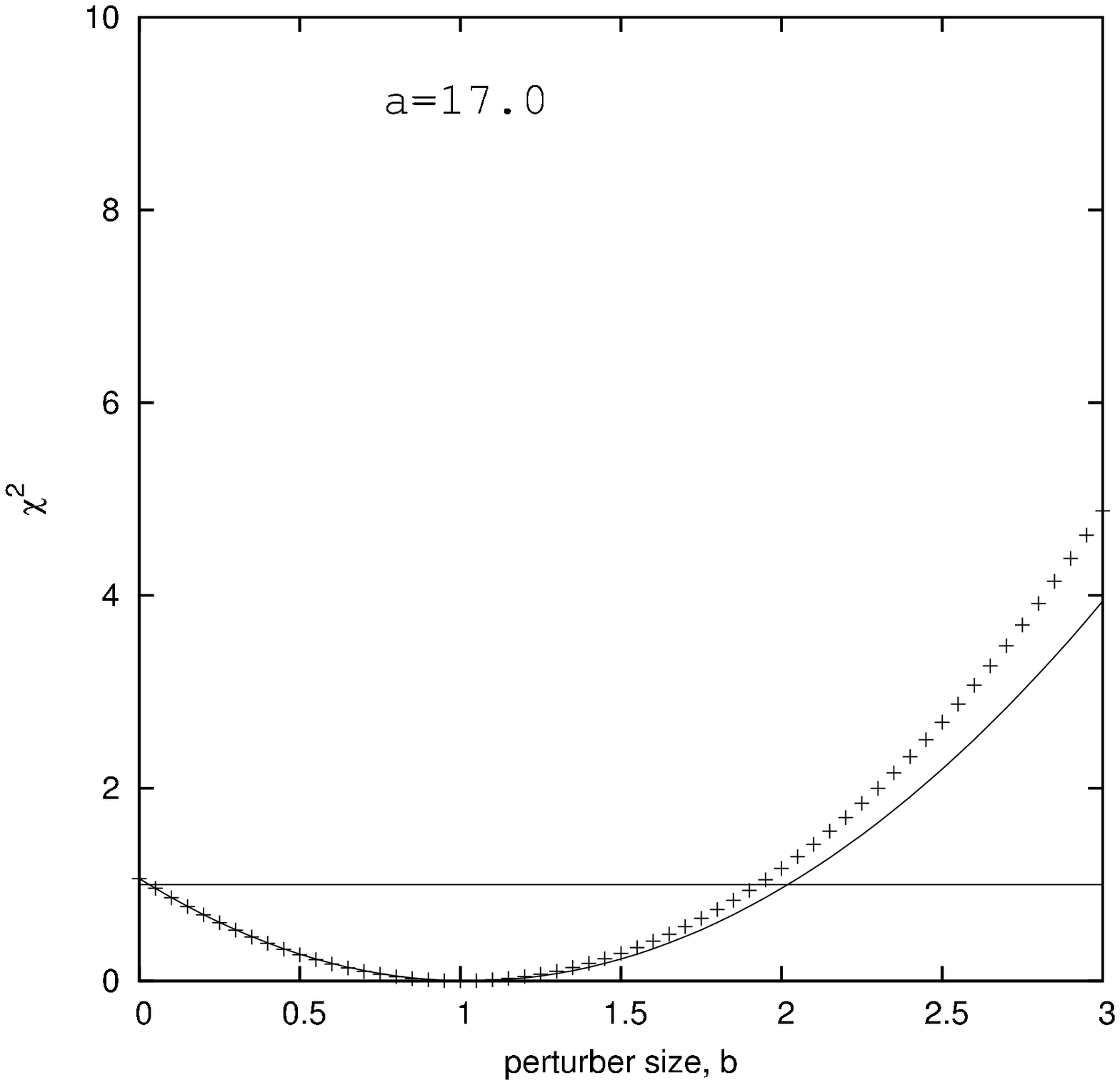}
\caption{
$\chi^2$ as a function of perturber size for various $a$.  The crosses
represent the exact analysis (holding the source position fixed), while
the solid lines represent the analysis using the first order Taylor
series expansion of the model magnification.  The expansion analysis
gives approximately the correct $1\sigma$ lower limit on $b$ for
$a \ga 5$.
}\label{fig:appendixfig}
 \end{center}
\end{figure}

The series expansion of the normalized magnification can be combined
with eq.~(\ref{eq:chisub}) to obtain an analytic result for the
minimum size of the perturbing mass from a single flux measurement.
The expansion has the form $\mm_{\rm obs} \approx 1 + C(b/a)$
where $C$ is a number that depends on $\kappa$ and $\gamma$.  Using
this in the definition of the substructure $\chi^2$, we find that
the upper and lower bounds on $b$ can be written as
\begin{equation}
  b_{\pm} \approx \frac{a}{C}\,\left(\mm_{\rm obs} - 1
    \pm \sigma_{\rm obs} \sqrt{\chi^2} \right)\,,
\end{equation}
where $\sqrt{\chi^2}$ indicates the confidence level desired
(e.g., $\sqrt{\chi^2}=N$ for $N$-$\sigma$).
Fig.~\ref{fig:appendixfig} compares this analysis to the exact
$\chi^2$ analysis used in the main paper, and shows that it recovers
reasonably accurate lower limits on $b$ even for sources as small as
$a/b \sim 5$.

\end{document}